\documentclass[twocolumn,superscriptaddress,showpacs,preprintnumbers,amsmath,amssymb]{revtex4}

\usepackage{booktabs}
\usepackage{graphicx,color}
\usepackage{dcolumn}
\usepackage{bm}
\usepackage{CJK}
\usepackage{mathrsfs}
\usepackage[dvipdfm,
            pdfstartview=FitH,
            CJKbookmarks=true,
            bookmarksnumbered=true,
            bookmarksopen=true,
            colorlinks,
            pdfborder=001,
            linkcolor=blue,
            anchorcolor=blue,
            citecolor=blue
            ]{hyperref}

\newcommand{\la}{\langle}
\newcommand{\ra}{\rangle}
\newcommand{\lb}{\left(}
\newcommand{\rb}{\right)}
\newcommand{\ls}{\left[}
\newcommand{\rs}{\right]}
\newcommand{\Lb}{\left\{}
\newcommand{\Rb}{\right\}}

\begin{document}
\begin{CJK}{GBK}{song}

\title{Fully self-consistent relativistic Brueckner-Hartree-Fock theory for finite nuclei}

\author{Shihang Shen}
\affiliation{State Key Laboratory of Nuclear Physics and Technology, School of Physics,
Peking University, Beijing 100871, China}
\affiliation{RIKEN Nishina Center, Wako 351-0198, Japan}

\author{Haozhao Liang}
 \affiliation{RIKEN Nishina Center, Wako 351-0198, Japan}
 \affiliation{Department of Physics, Graduate School of Science, The University of Tokyo, Tokyo 113-0033, Japan}
\author{Jie Meng\footnote{Email: mengj@pku.edu.cn}}
 \affiliation{State Key Laboratory of Nuclear Physics and Technology, School of Physics,
Peking University, Beijing 100871, China}
 \affiliation{School of Physics and Nuclear Energy Engineering, Beihang University,
              Beijing 100191, China}
 \affiliation{Department of Physics, University of Stellenbosch, Stellenbosch 7602, South Africa}
\author{Peter Ring}
 \affiliation{State Key Laboratory of Nuclear Physics and Technology, School of Physics,
Peking University, Beijing 100871, China}
 \affiliation{Physik-Department der Technischen Universit\"at M\"unchen, D-85748 Garching, Germany}
\author{Shuangquan Zhang}
 \affiliation{State Key Laboratory of Nuclear Physics and Technology, School of Physics,
Peking University, Beijing 100871, China}

\date{\today}

\begin{abstract}
Starting from the relativistic form of the Bonn potential as a bare nucleon-nucleon interaction, the full Relativistic Brueckner-Hartree-Fock (RBHF) equations are solved for finite nuclei in a fully self-consistent basis. This provides a relativistic \textit{ab initio} calculation of the ground state properties of finite nuclei without any free parameters and without three-body forces. The convergence properties for the solutions of these coupled equations are discussed in detail at the example of the nucleus $^{16}$O. The binding energies, radii, and spin-orbit splittings of the doubly magic nuclei $^{4}$He, $^{16}$O, and $^{40}$Ca are calculated and compared with the earlier RBHF calculated results in a fixed Dirac Woods-Saxon basis and other non-relativistic \textit{ab initio} calculated results based on pure two-body forces.
\end{abstract}


\maketitle

\section{Introduction}

To understand the nuclear system from the underlying interaction between nucleons has been one of the central problems in nuclear physics.
Because of the strong repulsive core at short distance \cite{Jastrow1951}, the realistic nucleon-nucleon ($NN$) interaction is notoriously difficult to be solved in the usual many-body framework.
Many methods have been proposed in the past to treat this singular behavior, such as Brueckner theory \cite{Brueckner1954}, variational method \cite{Jastrow1955},  Lee-Suzuki method \cite{Suzuki1980}, the unitary correlation operator method \cite{Feldmeier1998}, the low momentum effective $NN$ interaction $V_{low-k}$ \cite{Bogner2002}, and the similarity renormalization group \cite{Bogner2007}.
Recently, with the great progress of the high-precision $NN$ interactions, such as Reid93 \cite{Stoks1994}, AV18 \cite{Wiringa1995}, CD Bonn \cite{Machleidt2001}, or chiral potentials \cite{Epelbaum2009, Machleidt2011}, and with the rapid increase of computational power, more and more \textit{ab initio} methods have been developed to study the nuclear many-body system.
Celebrated examples include the quantum Monte Carlo method \cite{Carlson2015}, the coupled-cluster method \cite{Hagen2014}, the no core shell model \cite{Barrett2013}, the self-consistent Green's function method \cite{Dickhoff2004}, the lattice chiral effective field theory \cite{Lee2009}, the in-medium similarity renormalization  group \cite{Hergert2016}, the Monte Carlo shell model \cite{Liu2012}, or the Brueckner-Hartree-Fock (BHF) theory \cite{Day1967}.

Among all \textit{ab initio} methods, the Brueckner-Hartree-Fock theory is one of the most promising theories for an extension to heavy nuclei.
Historically, the Brueckner theory was introduced to deal with the hard core of the nuclear force in nuclear many-body calculations \cite{Brueckner1954}.
The basic idea is to describe nuclear structure in a mean-field approximation, but replacing the bare nuclear force by an effective interaction in the medium. This effective interaction is the reaction $G$-matrix, which takes into account the two-nucleon short range correlations by summing up all the ladder diagrams in the nuclear medium.
In this way, the saturation property of nuclear matter can be obtained qualitatively \cite{Brueckner1954}.
A formal derivation of Brueckner theory was provided by Goldstone \cite{Goldstone1957}, and substantial progress has been made by Bethe and co-workers \cite{Bethe1957, Bethe1963}.
This was one of the breakthroughs in microscopic nuclear many-body theory, and many developments have been done along this direction.
The readers are referred to the review papers \cite{Day1967, Rajaraman1967, Baranger1969} for the basic idea of Brueckner theory, the three hole-line expansion beyond BHF, and BHF for finite nuclei, respectively.

However, in the 1970s it was realized that all non-relativistic potentials failed to reproduce the saturation properties of infinite nuclear matter in detail.
The saturation points obtained with various forces are distributed along the so-called Coester line \cite{Coester1970}, which systematically deviates from the empirical value.
It is the general opinion that this discrepancy was caused by the missing of the three-body forces \cite{Fujita1957, Brown1969a}, which have been used in all the modern non-relativistic investigations phenomenologically.
In this way, at the cost of additional phenomenological parameters, one was able to reproduce the saturation properties of nuclear matter \cite{Zuo2002} as well as the ground states and a few excited states of light nuclei \cite{Pieper2001}.

On the other hand, nuclear structure has also been investigated in a relativistic framework.
Johnson and Teller showed already in 1955 that proper nuclear saturation properties can be obtained provided that the potential depends on the velocity of the nucleons~\cite{Johnson1955}.
Later this theory was reformulated by Duerr in a relativistically invariant way~\cite{Duerr1956}. In this way, the collapse of the nucleus occurring in the non-relativistic theories for high kinetic energies was avoided, and at the same time an extremely strong spin-orbit coupling was found~\cite{Duerr1956}, in agreement with experimental data on the magic numbers~\cite{GoeppertMayer1955}.
The Hamiltonian proposed by Duerr was then applied to finite nuclei \cite{Rozsnyai1961}.
After the one-boson-exchange potential was established gradually \cite{Green1967}, Miller and Green developed a Relativistic Hartree-Fock (RHF) approach based on attractive scalar and repulsive vector meson-exchange forces~\cite{Miller1972}.
The relativistic approach became popular when Walecka established the $\sigma+\omega$ model and applied it successfully to highly condensed matter~\cite{Walecka1974}.
See Ref.~\cite{Meng2016} for a recent review.

Inspired by the success of early phenomenological relativistic investigations for nuclear structure, several groups proposed the relativistic versions of BHF theory and developed the Relativistic Brueckner-Hartree-Fock (RBHF) method.
In the pioneering works by the Brooklyn group~ \cite{Anastasio1980, Anastasio1983}, the wave function of nucleon was chosen as a Dirac spinor in free space and the relativistic effects were taken into account in the first-order perturbation theory.
Further developments were advanced by Horowitz and Serot \cite{Horowitz1984,Horowitz1987}, Brockmann and Machleidt \cite{Brockmann1984,Brockmann1990}, and ter Haar and Malfliet \cite{TerHaar1986,Haar1987}.
In these studies, the relativistic effects produced a strong density-dependent repulsion and therefore the saturation point was shifted remarkably close towards the empirical value.
Using perturbation theory, it was found that relativistic effects lead to three-body forces through virtual nucleon-antinucleon excitations in the intermediate states (the so-called Z diagrams) \cite{Brown1987}.
Later, using the newly developed Bonn A/B/C potentials \cite{Machleidt1989}, the nuclear matter saturation points obtained within the RBHF theory were located on a new Coester line improving significantly the old one \cite{Brockmann1990}.
The study of RBHF theory in nuclear matter has also been extended to the investigations of optical potential \cite{Nuppenau1990,Xu2016}, asymmetric nuclear matter \cite{Huber1995,VanDalen2004}, or neutron stars \cite{Huber1996,Katayama2013}. See Ref.~\cite{Muther2017} for a recent review.

Even though the RBHF theory has achieved great success in the study of nuclear matter properties, the corresponding progress in finite nuclei was rather slow.
Due to its enormous computational requirement, for a long time, RBHF theory for finite nuclei has been available only with certain approximations, such as the effective density approximation (EDA) \cite{Muether1988}, or the local density approximation (LDA) \cite{Celenza1990,Marcos1991,Gmuca1991,Brockmann1992,Fritz1993}.
In the LDA approach, the density-dependence of the effective interaction, i.e. of the $G$-matrix, in nuclear matter is mapped onto a density-dependent relativistic Hartree or Hartree-Fock (DDRH or DDRHF) model, which is easy to be solved for finite nuclei. However, this mapping is far from unique and therefore this method suffers from large ambiguities as discussed in detail in Ref.~\cite{VanGiai2010}. As a consequence, different LDA approaches lead to rather different results.
Only very recently,  for the first time the RBHF equations were solved directly for finite nuclei \cite{Shen2016}.
The adopted Dirac Woods-Saxon (DWS) basis, which is obtained by solving the Dirac equation with a Woods-Saxon potential \cite{Zhou2003}, guarantees the full relativistic structure of the Dirac spinors.
Furthermore, the angle averaging \cite{Schiller1999,Suzuki2000} is avoided by solving the Bethe-Goldstone (BG) equation in the rest frame.
Taking the nucleus $^{16}$O as an example, convergence has been achieved with an energy cut-off close to 1.1 GeV and good descriptions of binding energy and radius have been obtained without any adjustable parameters.

In this work, we will adopt the self-consistent RHF basis in solving the BG equation and go beyond the previous work \cite{Shen2016}, where the BG equation was solved in a fixed DWS basis. All the cut-offs in the calculation will be checked and presented in detail. We will discuss the self-consistent single-particle potential and its uncertainties.
The center-of-mass motion will be treated in both projection before and after variation (PBV and PAV) methods.
Finally, as examples, the doubly magic nuclei $^{4}$He, $^{16}$O, and $^{40}$Ca will be calculated in the RBHF framework.
The ground state properties will be studied and compared with other \emph{ab initio} calculations in the literature.

In Sec. \ref{sec:theory}, the RBHF framework will be given. All the numerical details will be discussed at length using $^{16}$O as an example in Sec. \ref{sec:nd}. Results for $^{4}$He, $^{16}$O, and $^{40}$Ca will be presented in Sec. \ref{sec:res}. A summary and perspectives for future investigations will be given in Sec. \ref{sec:sum}.


\section{Theoretical Framework}

\label{sec:theory}

In this Section, we will outline the theoretical framework of the relativistic Brueckner-Hartree-Fock theory for finite nuclei. In particular, the relevant formulas will be shown explicitly with spherical symmetry.

The RBHF theory leads to a system of coupled non-linear equations.
The relativistic Hartree-Fock equations produce an optimized self-consistent
single-particle potential $U$ with the single-particle wave functions $|a\rangle$, and the single-particle energies $\varepsilon_a$. Filling them up to the Fermi
surface leads to a product state $|\Phi\rangle$ and the corresponding mean-field energy $E_{\rm RHF}$.
These results depend on the effective interaction $V_{\rm eff}$ used
in the mean-field equations. As compared to conventional
mean-field theory, where the effective interaction $V_{\rm eff}$ is phenomenological, in the RBHF theory this effective interaction is replaced by
the $G$-matrix, derived in an \emph{ab initio} calculation from the bare nucleon-nucleon interaction
using the Bethe-Goldstone equation. Its solution depends in a self-consistent
way on the mean-field potential $U$ and its single-particle wave functions and energies.
Therefore, RBHF theory presents a coupled system of equations which has to be solved by iteration.
Its starting point is a relativistic form of the bare nucleon-nucleon interaction.

\subsection{Relativistic Brueckner-Hartree-Fock theory for finite nuclei}

\subsubsection{Realistic one-Boson-exchange Lagrangian}

We start with a relativistic one-boson-exchange $NN$ interaction which describes the $NN$ scattering data~\cite{Machleidt1989}:
\begin{align}
\mathscr{L}_{NNpv} &= -\frac{f_{ps}}{m_{ps}} \bar{\psi}\gamma^5\gamma^\mu
\psi \partial_\mu \varphi^{(ps)},  \notag \\
\mathscr{L}_{NNs} &= g_s \bar{\psi}\psi\varphi^{(s)}, \\
\mathscr{L}_{NNv} &= -g_v \bar{\psi}\gamma^\mu\psi\varphi_\mu^{(v)} - \frac{%
f_v}{4M}\bar{\psi}\sigma^{\mu\nu}\psi \left( \partial_\mu\varphi_\nu^{( v)}
- \partial_\nu \varphi_\mu^{(v)} \right),  \notag
\end{align}
where $\psi$ denotes the nucleon field. The bosons to be exchanged are
characterized by the index $\alpha$ and
include the pseudoscalar mesons ($\eta,\pi$) with pseudovector ($pv$)
coupling, the scalar ($s$) mesons ($\sigma,\delta$), and the vector ($v$) mesons
($\omega,\rho$). For each pair, e.g., ($\eta,\pi$), the first (second) meson has
isoscalar (isovector) character. For the isovector mesons, the
field operator $\varphi_\alpha$ will be replaced by $\vec{\varphi}%
_\alpha\cdot\vec{\tau}$ with $\vec{\tau}$ being the usual Pauli matrices.

\subsubsection{Hamiltonian}

The Hamiltonian density is obtained using the Legendre transformation,
\begin{equation}
\mathscr{H}=\sum_{i}\frac{\partial \mathscr{L}}{\partial (\partial ^{0}\phi
_{i})}\partial ^{0}\phi _{i}-\mathscr{L},
\end{equation}%
where $\phi_{i}$ represent the nucleon field $\psi $, the meson fields $\varphi _{\alpha }$,
and the photon field.

In the stationary case the Hamiltonian is found as a three-dimensional integral over the
Hamiltonian density:
\begin{equation}
H=\int \mathrm{d}^{3}r\mathscr{H}(\mathbf{r}).
\end{equation}%
Eliminating the meson fields one finds the following many-body Hamiltonian for
the nucleons~\cite{Brockmann1978},
\begin{widetext}
\begin{equation}
H= \int \mathrm{d}^{3}r\bar{\psi}\left( -i\bm{\gamma }\cdot \bm{\nabla}+M\right)\psi
 +\frac{1}{2}\sum_{\alpha }\int \mathrm{d}^{3}r_{1}\mathrm{d}^{3}r_{2}
 \bar{\psi}(\mathbf{r}_{1})\Gamma _{\alpha }^{(1)}\psi(\mathbf{r}_{1})
 D_{\alpha }(\mathbf{r}_{1},\mathbf{r}_{2})
 \bar{\psi}(\mathbf{r}_{2})\Gamma _{\alpha }^{(2)}\psi (\mathbf{r}_{2}),
\end{equation}%
\end{widetext}
where $\Gamma _{\alpha }^{(1)},\Gamma _{\alpha }^{(2)}$ are the interaction
vertices for particles 1 and 2, with the coordinates $\mathbf{r}_{1}$ and $\mathbf{r}%
_{2}$, respectively:
\begin{subequations}\label{eq:gamma12}
\begin{align}
\Gamma _{s}=& g_{s}, \\
\Gamma _{pv}=& \frac{f_{ps}}{m_{ps}}\gamma ^{5}\gamma ^{i}\partial _{i}, \\
\Gamma _{v}^{\mu }=& g_{v}\gamma ^{\mu }+\frac{f_{v}}{2M}\sigma ^{i\mu
}\partial _{i}.
\end{align}%
\end{subequations}
For the Bonn interaction \cite{Machleidt1989}, a form factor of monopole-type is attached to each
vertex. In momentum space it has the form:
\begin{equation}
\frac{\Lambda _{\alpha }^{2}-m_{\alpha }^{2}}{\Lambda _{\alpha }^{2}+\mathbf{q}%
^{2}},  \label{eq:form}
\end{equation}%
where $\Lambda _{\alpha }$ is the cut-off parameter for meson $\alpha $ and $%
\mathbf{q}$ is the momentum transfer.

In Minkowski space, the meson propagators $D_{\alpha }(x_1,x_2)$ are the retarded solutions of the
Klein-Gordon equations,
\begin{equation}
D_{\alpha }(x_1,x_2)=\pm \int \frac{d^{4}q}{(2\pi )^{4}}\frac{1}{m_{\alpha}^{2}-q^{2}}e^{-iq\cdot (x_1-x_2)},
\end{equation}%
where $q$ is the four-momentum transfer between the two particles. The sign $-$
holds for the scalar (and pseudoscalar) mesons and the sign $+$ holds for the
vector fields. The dependence on the zero-component momentum transfer $q_0$ (energy) reflects the retardation of the interaction between the two particles. In the Bonn interaction of Ref.~\cite{Machleidt1989}, this effect was deemed to
be small and was ignored from the beginning. In this way, the meson propagators are just Yukawa
functions:

\begin{eqnarray}
D_{\alpha }(\mathbf{r}_1,\mathbf{r}_2)&=&
\pm \int \frac{d^{3}q}{(2\pi )^{3}}\frac{1}{m_{\alpha }^{2}+\mathbf{q}^{2}}e^{i\mathbf{q}\cdot(\mathbf{r}_1-\mathbf{r}_2)}\notag\\
&=&\pm \frac{1}{4\pi}\frac{e^{-m_{\alpha }|\mathbf{r}_1-\mathbf{r}_2|}}{|\mathbf{r}_1-\mathbf{r}_2|}.
\label{eq:propa}
\end{eqnarray}%
Note, however, that with the form factor in Eq.~(\ref{eq:form}) the meson
propagators are no longer simple Yukawa functions. In practice, the relevant
matrix elements are calculated numerically in the momentum space, see Appendix~\ref{sec:app1} for details.

Now we expand the nucleon-field operators $\psi (\mathbf{r}),
$ $\psi ^{\dag }(\mathbf{r})$ in terms of a complete orthonormal static relativistic basis $|k\rangle $:
\begin{equation*}
\psi ^{\dag }(\mathbf{r})=\sum_{k}\psi _{k}^{\dag }(\mathbf{r})b_{k}^{\dag },\text{
\ \ \ \ }\psi (\mathbf{r})=\sum_{k}\psi _{k}(\mathbf{r})b_{k}^{{}},
\end{equation*}%
where $b_{k}^{\dagger }$ and $b_{k}$ form a complete set of creation and
annihilation operators for nucleons in the state $|k\rangle $, which can be
of positive energy or of negative energy. Here $\psi _{k}(\mathbf{r})$ is the
corresponding Dirac spinor. {The quantum number $k$ characterizing the state $|k\rangle$ contains also the
isospin $\tau=n,\,p$ for neutrons and protons}. We then have the Hamiltonian for nuclear
system in the second quantized form as:
\begin{equation}
H=\sum_{kk^{\prime }}\langle k|T|k^{\prime }\rangle b_{k}^{\dagger
}b_{k^{\prime }}^{{}}+\frac{1}{2}\sum_{klk^{\prime }l^{\prime }}\langle
kl|V|k^{\prime }l^{\prime }\rangle b_{k}^{\dagger }b_{l}^{\dagger
}b_{l^{\prime }}^{{}}b_{k^{\prime }}^{{}},  \label{eq:hami}
\end{equation}%
where the matrix elements are given by
\begin{align}
\langle k|T|k^{\prime }\rangle & =\int d^{3}r\,\bar{\psi}_{k}(\mathbf{r})\left( -i%
\bm{\gamma}\cdot \nabla +M\right) \psi _{k^{\prime }}(\mathbf{r}), \\
\langle kl|V_{\alpha }|k^{\prime }l^{\prime }\rangle & =\int
d^{3}r_{1}d^{3}r_{2}\,\bar{\psi}_{k}(\mathbf{r}_{1})\Gamma _{\alpha }^{(1)}\psi
_{k^{\prime }}(\mathbf{r}_{1})  \notag \\
&~~~~~~~~\times D_{\alpha }(\mathbf{r}_{1},\mathbf{r}_{2})\bar{\psi}_{l}(\mathbf{r}_{2})\Gamma
_{\alpha }^{(2)}\psi _{l^{\prime }}(\mathbf{r}_{2}).
\end{align}%
The two-body interaction $V$ contains contributions from the different mesons $\alpha$.

The indices $k,l$ run over an arbitrary complete basis of Dirac spinors with
positive and negative energies, as, for instance, over plane wave states $u(%
\mathbf{k},s)$ and $v(\mathbf{k},s)$ in the momentum space~\cite{Itzykson1980} or over the eigensolutions of a
Dirac equation with potentials of Woods-Saxon shapes discussed in Refs. \cite{Koepf1991,Zhou2003}.

\subsubsection{Bethe-Goldstone equation}

As is well known, the matrix elements of the bare nucleon-nucleon interaction
$\langle ab|V|cd\rangle $ are very large and difficult to be used directly in nuclear
many-body theory. Within the Brueckner theory, one takes into account
the fact that nucleons in the nuclear medium do not feel the same
interaction as that in free space. All the states below the Fermi surface are
occupied and therefore the Pauli principle allows only scattering processes
into intermediate states above the Fermi surface. The $T$-matrix, which
describes scattering processes in free space is therefore, in the nuclear
medium, replaced by the $G$-matrix~\cite{Brueckner1954}. It sums up all the
ladder diagrams with two particles in intermediate states above the
Fermi surface. It is deduced from the Bethe-Goldstone equation~\cite{Bethe1957},
\begin{widetext}
\begin{equation}
\langle ab|\bar{G}(W)|a^{\prime }b^{\prime }\rangle =\langle ab|\bar{V}%
|a^{\prime }b^{\prime }\rangle +\frac{1}{2}\sum_{cd}\langle ab|\bar{V}%
|cd\rangle \frac{Q(c,d)}{W-\varepsilon _{c}-\varepsilon _{d}}\langle cd|\bar{%
G}(W)|a^{\prime }b^{\prime }\rangle,
\label{eq:BG}
\end{equation}%
\end{widetext}
where $\langle ab|\bar{V}|a^{\prime }b^{\prime }\rangle=
\langle ab|V|a^{\prime }b^{\prime }-b^{\prime }a^{\prime }\rangle$
 is the antisymmetrized two-body matrix element, $W$ is the starting
energy, and $\varepsilon _{c}$, $\varepsilon _{d}$ are the single-particle
energies of the two particles in the intermediate states. The Pauli operator
$Q(c,d)$ allows the scattering only to states $c$ and $d$ above the Fermi
surface. It is defined as
\begin{equation}
Q(c,d)=\begin{cases} 1, & \rm{for}~~\varepsilon_{c}>\varepsilon_{F}~and~
\varepsilon_{d}>\varepsilon_{F}, \\ 0, & \rm{otherwise}. \end{cases}
\label{eq:Q}
\end{equation}%
In this paper, we use the convention that indices $a,b,c,\dots$ run over the
single-particle states being the solutions of RBHF equation, whereas the indices $k,l,m,\dots$
run over an arbitrary complete single-particle basis. Of course, the single-particle energies
$\varepsilon_{c}$ and $\varepsilon_{d}$ depend on the solution of the corresponding Hartree-Fock equation and
therefore we are left with a coupled system of equations, which has to be
solved by iteration.

\subsubsection{Relativistic Hartree-Fock equation}

The relativistic Hartree-Fock equation reads
\begin{equation}
(T+U)|a\rangle =e_{a}|a\rangle ,
\label{eq:rhf}
\end{equation}%
where $e_{a}=\varepsilon _{a}+M$ is the single-particle energy with the rest
mass of nucleon $M$, and $U$ is the self-consistent single-particle potential.
In conventional relativistic Hartree-Fock theory~\cite{Brockmann1978,Long2006} based on an effective interaction
$V_{\rm eff}$, this potential is defined as
\begin{equation}
U_{ab}=\frac{1}{2}\sum_{c=1}^{A}\langle ac|\bar{V}_{\rm eff}|bc\rangle.
\label{eq:UHF}
\end{equation}%
According to the no-sea approximation, the index $c$ runs over all the occupied states
in the Fermi sea. In the relativistic Brueckner-Hartree-Fock framework,
the interaction $V_{\rm eff}$ in the definition of $U$ is replaced by the $G$-matrix. Because of the fact
that the effective interaction in the medium $G(W)$ depends on the starting
energy, this definition is not as straightforward as in the conventional
Hartree-Fock case, where the effective interaction does not depend on
energy. The connection between the matrix element $U_{ab}$ and $G(W)$ was
first discussed in Ref.~\cite{Bethe1963} in nuclear matter. In the
framework of perturbation theory, it was shown that, according to the
Bethe-Brandow-Petschek (BBP) theorem~\cite{Bethe1963}, a specific choice of the starting energy
in terms of the single-particle energies causes a large set of diagrams beyond the Hartree-Fock level to vanish.
The extension to finite nuclei gives the following results \cite{Baranger1969,Davies1969}:
\begin{equation}
U_{ab}=\frac{1}{2}\sum_{c=1}^{A}\langle ac|\bar{G}(\varepsilon
_{a}+\varepsilon _{c})+\bar{G}(\varepsilon _{b}+\varepsilon _{c})|bc\rangle ,
\label{eq:Uhh}
\end{equation}%
if $|a\rangle $ and $|b\rangle $ are both hole (i.e. occupied) states, and
\begin{equation}
U_{ab}=\sum_{c=1}^{A}\langle ac|\bar{G}(\varepsilon _{a}+\varepsilon
_{c})|bc\rangle ,
\label{eq:Uph}
\end{equation}%
if $|a\rangle $ is a hole state and $|b\rangle $ is a particle (i.e. unoccupied) state, and
\begin{equation}
U_{ab}=\frac{1}{2}\sum_{c=1}^{A}\langle ac|\bar{G}(\varepsilon _{a}^{\prime
}+\varepsilon _{c})+\bar{G}(\varepsilon _{b}^{\prime }+\varepsilon
_{c})|bc\rangle ,
\label{eq:Upp}
\end{equation}%
if $|a\rangle $ and $|b\rangle $ are both particle states.

In the above expressions, $\varepsilon$ labels the self-consistent
single-particle energy, while $\varepsilon^{\prime }$ is somewhat uncertain
\cite{Davies1969}. This means the matrix elements of the self-consistent potential
$U_{ab}$ with both states $|a\rangle$ and $|b\rangle$ above the Fermi level are not well defined in
the Brueckner-Hartree-Fock theory. Different choices have been proposed in the literature \cite{Rajaraman1967,Davies1969}.
One extreme is to set the potential $U_{ab}=0$ in Eq.~(\ref{eq:Upp}). This is known as the {\it gap choice} in nuclear
matter. Another extreme is to set $\varepsilon_a^{\prime }= \varepsilon_a$,
which is known as the {\it continuous choice} in nuclear matter. The BHF
theory can be viewed as the first-order approximation of the so-called
hole-line expansion \cite{Day1967}, which orders the
Bethe-Brueckner-Goldstone expansion diagrams according to the number of
independent hole lines. In principle, the final result will not depend on
different choices of the single-particle potential $U$ if this hole-line expansion
is taken into account up to high orders. This has already been confirmed in
a non-relativistic calculation for nuclear matter up to the three-hole-line
level: as shown in Ref.~\cite{Song1998} the resulting equation of state does not depend much on the choice of $U$. Meanwhile, it was found in Ref.~\cite{Song1998}
on the BHF level, i.e., on the two-hole-line level, that near the saturation density
the {\it continuous choice} produces several MeV$/A$ more binding than the {\it gap choice}.
The results in the three-hole-line level agree with each other for these two
different choices, and lie in between the BHF results.

Following the discussions in Ref.~\cite{Davies1969}, we choose in the present investigation
a prescription in between the above two extremes. Precisely,
$\varepsilon^{\prime }_a=\varepsilon^{\prime }_b=\varepsilon^{\prime }$ is fixed as an energy among the occupied states and
we discuss the difference of the results by fixing $\varepsilon^{\prime }$ as the highest and as the lowest energy of the occupied states in the Fermi sea.

\subsubsection{Solution in the RHF basis}

The coupled RBHF equations in finite nuclei are solved within a complete
Dirac basis $\{|k\rangle\}$, i.e. in a basis in Dirac space with states of positive
and negative energies. This means that the RBHF single-particle states $|a\rangle$
are expressed as linear combinations
\begin{equation}
|a\rangle = \sum_k D_{ka} |k\rangle.
\end{equation}
Because of the requirement of the completeness of the basis, $|k\rangle$
runs not only over the positive-energy states in and above the Fermi sea but also over
the negative-energy states in the Dirac sea~\cite{Zhou2003}, even though the no-sea
approximation is adopted for the RBHF or RHF calculations, which means that the sums in the evaluation of various densities run only over the occupied states in the Fermi sea. As a consequence, also the index $c$ in Eqs. (\ref{eq:Uhh}--\ref{eq:Upp}) is restricted to the $A$ occupied states in the Fermi sea. The
RHF equation in the basis $\{|k\rangle\}$ reads
\begin{equation}  \label{eq:rhf2}
\sum_{l} \left( T_{kl} + U_{kl} \right) D_{la} = e_{a} D_{ka}.
\end{equation}
It is solved by diagonalization in the calculations of the present work.

It should be noticed that both the BG equation~(\ref{eq:BG}) and
the single-particle potential (\ref{eq:Uhh}--\ref{eq:Upp}) are defined in the
RBHF single-particle basis $\{|a\rangle\}$. Therefore, it requires a
double-iteration procedure.

We start with a set of trial single-particle states, a discrete DWS basis \cite{Zhou2003} with the single-particle wave functions $|k\rangle$ and the corresponding energies $\varepsilon_k$, and we evaluate the matrix elements of the kinetic energy $T_{kl}$ and the antisymmetrized two-body matrix elements of the bare interaction  $\bar{V}_{klmn}$ in this basis. This is a rather lengthy process, but it is done only once. Details are given in Appendix~\ref{sec:app1}.

Then we start the iteration:
\begin{enumerate}
\item We solve the BG equation (\ref{eq:BG}) in this basis by matrix inversion. This yields a first set of $G$-matrix elements in the DWS basis.

\item These $G$-matrix elements are used to evaluate the single-particle potentials $U_{kl}$ in Eqs.~(\ref{eq:Uhh}--\ref%
{eq:Upp}).

\item The RHF equation (\ref{eq:rhf}) is solved as
\begin{equation}
\sum_{l} \left( T_{kl} + U_{kl} \right) D_{lk^{\prime }} = e_{k^{\prime }}
D_{kk^{\prime }},
\label{eq:rhf3}
\end{equation}
and a new set of single-particle states $\{|k^{\prime }\rangle\}$ with single-particle energies
$\varepsilon_{k^{\prime }}$ is obtained. The iteration has converged if $\{|k^{\prime}\rangle\}=\{|k\rangle\}\equiv\{|a\rangle\}$. Otherwise the set $\{|k^{\prime}\rangle\}$
forms a new RHF basis and we continue.

\item The single-particle matrix elements of the kinetic energy and the two-body matrix elements of the bare interaction
are transformed to the new RHF basis
\begin{align}
T^{}_{k^{\prime}l^{\prime}} &= \sum_{kl} D_{kk^{\prime}}^* D^{}_{ll^{\prime }} T^{}_{kl}, \\
\bar{V}^{}_{k^{\prime }l^{\prime }m^{\prime }n^{\prime }} &=
\sum_{klmn}D_{kk^{\prime }}^* D_{ll^{\prime }}^* D^{}_{mm^{\prime }}D^{}_{nn^{\prime }} \bar{V}^{}_{klmn} .
\label{eq:transV}
\end{align}
Then we go back to Step 1 and solve the BG equation~(\ref{eq:BG}) in this new RHF basis. This yields a second set of $G$-matrix elements and we continue with Step 2.
\end{enumerate}

In practice, we found that performing an RHF iteration in Step 3 in each step of the RBHF iteration can speed up the convergence, and avoid the very time consuming Step 1, i.e. the solution of the BG equation.

This complicated iteration scheme allows the fully self-consistent solution of the RBHF problem. In Ref. \cite{Shen2016} a simplified version has been applied. In that case Step 4 was omitted and for each step of the iteration the BG equation was solved in the original DWS basis.  Only the changes in the single-particle energies $\varepsilon_{k^{\prime}}$ in the propagator were taken into account. In practice this means, that the self-consistent single-particle wave functions in the Pauli operator in Eq.~(\ref{eq:Q}) were replaced by the corresponding one DWS basis.

\subsubsection{Center-of-mass motion}
\label{sec:center_of_mass}
In the above formulation, the spurious center-of-mass (c.m.) motion is included in
the total Hamiltonian in Eq.~(\ref{eq:hami}). It is not of interest and should be
removed. Since the Hamiltonian is invariant against translations, the exact many-body
eigenstates of the system should be eigenfunctions of the total momentum
$ \mathbf{P} = \sum_i^A \mathbf{p}_i$. The spurious center-of-mass motion is
therefore removed by projection onto the eigenspaces with vanishing eigenvalues of this operator. It has been
shown in Refs. \cite{Beck1970,Ring1980} that, for large values of the particle number $A$, the projected energy is obtained in a good approximation by removing the center of mass energy
\begin{equation}
\label{eq:Ecm}
E_{cm} =  \frac{{\langle\bf{P}\rangle}^2}{2AM}
\end{equation}
from the total energy $\langle H\rangle=\langle T+V\rangle$.

In most of the (R)HF calculations the variation is carried out without projection, i.e.,
the (R)HF equations are solved for the total Hamiltonian $H$ and
the spurious center of mass energy in Eq.~(\ref{eq:Ecm}) is removed after the variation, as for instance in
Ref. \cite{Long2004}. This is a projection after variation (PAV).
A more strict treatment~\cite{Zeh1965} would be to exclude this term also in the (R)HF equation, i.e., to carry out a projection before variation (PBV), as it has been done for instance in
Ref.~\cite{Chabanat1998}. We will discuss these two different choices in the
RBHF framework as it has been done in Ref. \cite{Becker1974} for the non-relativistic BHF framework. We also should mention that the two-body part of the center of mass correction $\langle {\bf{P}} \rangle^2/2AM$  as well as the Coulomb force (see Appendix \ref{sec:app3}) has only been taken into account in the RHF equation (\ref{eq:rhf3}) and not in the BG equation for the calculation of the $G$-matrix.

\subsection{RBHF theory for spherical nuclei}

\subsubsection{Spherical DWS basis}

The eigenfunctions of a Dirac equation with spherical symmetry can be written as (for simplicity we neglect here the isospin indices):
\begin{equation}  \label{eq:wf}
|a\rangle = \frac{1}{r} \left(%
\begin{array}{c}
F_{n_a\kappa_a}(r) \Omega_{j_am_a}^{l_a}(\theta,\varphi) \\
iG_{n_a\kappa_a}(r) \Omega_{j_am_a}^{\tilde{l}_a}(\theta,\varphi)%
\end{array}%
\right),
\end{equation}
where $\Omega_{jm}^l(\theta, \varphi)$ are the spinor spherical harmonics.
The radial, orbital angular momentum, total angular momentum, and magnetic
quantum numbers are denoted by $n,\, l,\, j,$ and $m$, respectively, while the
quantum number $\kappa$ is defined as $\kappa = \pm(j+1/2)$ for $j=l\mp1/2$.
$\tilde{l} = 2j - l$ is the orbital angular momentum for the lower component.
$F(r),\, G(r)$ are the radial wave functions which satisfy the radial Dirac equation:
\begin{equation}
\left(
\begin{array}{cc}
M+\Sigma(r) & -\frac{d}{dr}+\frac{\kappa}{r} \\
\frac{d}{dr}+\frac{\kappa}{r} & -M+\Delta(r)
\end{array}
\right) \left(
\begin{array}{c}
F_a(r) \\
G_a(r) \\
\end{array}
\right) =e_a \left(
\begin{array}{c}
F_a(r) \\
G_a(r) \\
\end{array}
\right),
\label{eq:direq}
\end{equation}
where $\Sigma=V+S$ and $\Delta=V-S$ are the sum and the difference of vector and
scalar potentials. For a DWS basis \cite{Zhou2003}, $\Sigma(r)$ and $%
\Delta(r)$ are potentials with Woods-Saxon shape parameterized as
\begin{align}
\Sigma(r) =& \frac{V_0}{1+\exp((r-R)/a)}, \label{eq:dws1}\\
\Delta(r) =& \frac{W_0}{1+\exp((r-R^{ls})/a^{ls})}.\label{eq:dws2}
\end{align}

\subsubsection{Symmetries of the BG equation}

The BG equation has to be solved in the space of particle pairs. The quantum numbers of each particle are labeled as
$a=(n_a,j_a,\pi_a,t_a)$ with the parity $\pi=(-)^l$ and the isospin quantum number $t=n,\,p$.
Symmetries including rotation, parity, and charge can be used to reduce the dimension of this space. We therefore consider particle pairs with angular momentum ${\bf{J}}={\bf{j}}_a+{\bf{j}}_b$, parity $P=\pi_a\pi_b$, and $z$-component of the isospin (charge) $T=t_a+t_b$:
\begin{equation}
|ab\rangle_{pp}^{JPT} = \sum_{m_am_b}C_{j_am_aj_bm_b}^{JM}|a,m_a\rangle\,|b,m_b\rangle.
\label{eq:pairs}
\end{equation}

Introducing the particle-particle ($pp$) $jj$-coupled matrix elements in Eq.~(\ref{eq:VppJ}),
the BG equation can be reduced into sub-systems (channels) with the channel quantum numbers $JPT$:
\begin{align}
\label{eq:BG2}
\langle ab|\bar{G}|a^{\prime }b^{\prime }\rangle_{pp}^{JPT} ~=&
~\langle ab|\bar{V}|a^{\prime }b^{\prime }\rangle_{pp}^{JPT}~ +\\
+~\frac{1}{2}{\sum_{cd}}\langle ab&|\bar{V}|cd\rangle_{pp}^{JPT}
\frac{Q(c,d)}{W-\varepsilon_c-\varepsilon_d} \langle cd|\bar{G}|a^{\prime }b^{\prime }\rangle_{pp}^{JPT}
\notag,
\end{align}
where the $m$ degeneracy has been summed up by Eq.~(\ref{eq:pairs}). The starting energy $W$ is chosen according to Eqs.~(\ref{eq:Uhh}--\ref%
{eq:Upp}).

In practice, the BG equation has to be solved independently for each channel with the quantum numbers $JPT$. For the isospin, there are three channels ($T=-1,0,+1$), which can be solved independently as
\begin{equation}
\langle nn|\bar{G}|nn\rangle_{pp}^J,~~
\left(
\begin{array}{cc}
\langle np|\bar{G}|np\rangle_{pp}^J & \langle np|\bar{G}|pn\rangle_{pp}^J \\
\langle pn|\bar{G}|np\rangle_{pp}^J & \langle pn|\bar{G}|pn\rangle_{pp}^J
\end{array}
\right),~~
\langle pp|\bar{G}|pp\rangle_{pp}^J.
\end{equation}

The RHF equation is a single-particle equation of the structure $b^\dag b$. Therefore we need the antisymmetrized particle-hole (ph) $jj$-coupled matrix elements as given in Eqs. (\ref{eq:VphI}) and (\ref{eq:VphIe}). In spherical nuclei, only $I=0$ matrix elements are relevant \cite{Tarbutton1968}:
\begin{align}
\label{eq:spp0}
\langle a|U|b\rangle =
~\sum_{c}^A \frac{\hat{j}_c}{\hat{j}_a}
\langle a c|\bar{G}(W)|bc\rangle_{ph}^{I=0}
\delta_{t_a t_b}\delta_{\kappa_a\kappa_b},
\end{align}
where $\hat j=\sqrt{2j+1}$. The antisymmetrized $ph$-coupled matrix elements $\langle a c|\bar{G}(W)|bc\rangle_{ph}^{I=0}$ are derived from the $pp$-coupled matrix elements, i.e. from the solutions of the BG equation (\ref{eq:BG2}) using Eq. (\ref{eq:Vph-pp}):
\begin{align} \label{eq:Gph-pp}
\langle 12|\bar{G}|34\rangle_{ph}^I =& \sum_{J} (2J+1)(-1)^{j_3+j_4+J}
\notag \\
&\times \left\{%
\begin{array}{ccc}
j_1 & j_3 & I \\
j_4 & j_2 & J%
\end{array}%
\right\} \langle 12|\bar{G}|34\rangle_{pp}^J.
\end{align}

The maximum $J$ in $pp$ coupling used in the BG equation is determined by
the single-particle angular momentum cut-off $j_{\mathrm{cut}}$ of the
basis. In practice, the high $J$ matrix elements give relatively small
contributions to $\langle 12|\bar{G}|34\rangle_{ph}^{I=0}$, thus we introduce a
cut-off $J_{\mathrm{cut}}$ and solve the $G$-matrix for $0\leq J\leq J_{%
\mathrm{cut}}$.

In practice, we first construct the $ph$ coupled matrix elements $\bar{V}_{ph}^I$ (for details see Appendix~\ref{sec:app1}). From those we obtain the $pp$ coupled matrix elements $\bar{V}_{pp}^J$ by an inverse recoupling in Eq. (\ref{eq:Vpp-ph}).
They are used to solve the BG equation (\ref{eq:BG2}) and to obtain $\bar{G}_{pp}^J$ for all $J$ values satisfying
$0\leq J\leq J_{\mathrm{cut}}$. Finally using Eq. (\ref{eq:Gph-pp}) and recoupling again to the $ph$ channel, the RHF equation can be solved for the next step.

\subsubsection{Observables}

After getting the solution of the coupled system of RBHF equations, the total energy is expressed as
\begin{align}  \label{eq:finalE}
E = \sum_{a}^A \langle a|T|a\rangle + \frac{1}{2}\sum_{ab}^A \langle ab|\bar{%
G}(W)|ab\rangle + E_{C} - E_{cm},
\end{align}
with the starting energy $W=\varepsilon_a+\varepsilon_b$. In the present
framework, the Coulomb energy $E_C$ is not included in the evaluation of the $G$-matrix
and is calculated separately. $E_{cm}$ is the center of mass energy in
Eq.~(\ref{eq:Ecm}). For details in the calculation of $E_{cm}$ and $E_{C}$ see Appendices~%
\ref{sec:app2} and \ref{sec:app3}.

The charge density distribution is obtained from  \cite{Campi1972,Long2004}:
\begin{align}
\rho_c(r) =& \frac{1}{\sqrt{\pi}ar} \int r^{\prime }dr^{\prime
}\rho_p(r^{\prime }) \left[ e^{-(r-r^{\prime})^2/\lambda^2} - e^{-(r+r^{\prime})^2/\lambda^2}\right], \label{eq:rhoc}
\end{align}
where $\lambda^2 = 1/(a^2-B^2)$, with $a = \sqrt{\frac{2}{3}}\times 0.8$ fm the correction due to the finite proton size, and $B^2 = \frac{3}{2}\la \mathbf{P}^2\ra^{-2}$ the correction due to the center-of-mass motion.
For PBV, $B^2 = 0$ since the center-of-mass correction has already been considered in the single-particle wave functions. The charge radius is calculated as
\begin{align}
\langle r_c^2 \rangle=& \frac{1}{Z} 4\pi \int r^4dr \rho_c(r).
\end{align}

\section{Numerical details}
\label{sec:nd}

As an example we consider the nucleus $^{16}$O.
We use the realistic $NN$ interaction Bonn A which has been adjusted to the $NN$ scattering data~\cite{Machleidt1989}.
The Woods-Saxon potential in Eqs.~(\ref{eq:dws1},\ref{eq:dws2}) is taken from Ref.~\cite{Koepf1991}.
The DWS basis is then obtained by solving the spherical Dirac equation~(\ref{eq:direq}) in a box with the box size $R$ and mesh size $dr=0.05$~fm.

The RHF equation~(\ref{eq:rhf}) can be solved in either the DWS basis or the obtained RHF basis.
The validity of this RHF code is confirmed by reproducing the
results of other RHF codes
in the oscillator basis \cite{Lalazissis2009} and in coordinate space \cite{Long2006}.

The BG equation~(\ref{eq:BG2}) is solved by matrix inversion \cite{Haftel1970} in the space of pair states $|ab\ra$ given in Eq. (\ref{eq:pairs}).
These are pairs of Dirac spinors with the full relativistic structure coupled to good angular momentum $J$ ($pp$ coupling).
The indices $a$ and $b$ run over all solutions of the Dirac equation (with positive and negative energies).
The BG equations are solved for each of the channels characterized by the quantum numbers $J$, parity $\pi$, and
$z$-component of the isospin $T=t_a+t_b$. This leads, for various $J$ values, to a set of $pp$-coupled matrix elements of the $G$-matrix.
Because we work always in the RHF basis during the iteration, the Pauli operator in Eq.~(\ref{eq:Q}) and all its relativistic structure is here fully taken into account.
In particular, there is no angle averaging involved as it is the case in most Brueckner calculations in nuclear matter \cite{Schiller1999,Suzuki2000}.
The fully self-consistent RHF basis is used in comparison with previous investigation with a fixed
DWS basis~\cite{Shen2016}.
The BG equation~(\ref{eq:BG2}) is solved for four different values of the starting energy $W$, equally distributed between the lowest and the highest single-particle energies in the Fermi sea.
The $G$-matrix with specific starting energy $W$ is obtained by a four point Lagrange polynomial interpolation \cite{Davies1969}.

In the following convergence check, $\varepsilon'$ in Eq.~(\ref{eq:Upp}) is chosen as the minimum single-particle energy of the hole states ($\varepsilon'=\varepsilon_{\nu 1s1/2}$).
The center-of-mass correction is not considered.
After the convergence check, discussion will be focused on the single-particle potential of particle states in Eq.~(\ref{eq:Upp}) as well as PAV and PBV.
Without explicitly stating, all the calculations are performed in the self-consistent RHF basis.

\subsection{Convergence check}

It is well known that the bare $NN$ interaction contains a repulsive core and a strong tensor part connecting the nucleons below the Fermi surface to the states with high momentum in the continuum.
In order to take this coupling fully into account, one needs a relatively large basis space and it is crucial to investigate the corresponding convergence of the RBHF calculations.
In our previous investigation with a fixed basis \cite{Shen2016}, a reasonable convergence is achieved near an energy cut-off $\varepsilon_{\rm cut} = 1.1$ GeV.
In this work, we will carry out the same check for the fully self-consistent RBHF calculation.
Moreover, we will give a convergence check for the other cut-offs, i.e., the single-particle orbital angular momentum cut-off $l_{\rm cut}$, the single-particle energy cut-off in the Dirac sea $\varepsilon_{\rm Dcut}$, and the total angular momentum cut-off $J_{\rm cut}$ in the derivation of the $ph$ matrix elements of the G-matrix in Eq. (\ref{eq:Gph-pp}).

\begin{figure}
\includegraphics[width=8cm]{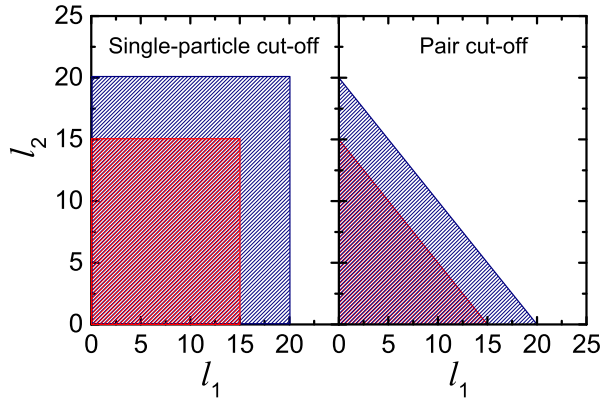}
\caption{(Color online) The difference between the single-particle cut-off and the pair cut-off is illustrated at the example of orbital angular momentum coupling for $l_{\rm cut}=15~\hbar$ or $l_{\rm cut}=20~\hbar$ (left panel) and for $L_{\rm cut}=15~\hbar$ or $L_{\rm cut}=20~\hbar$ with $L=l_1+l_2$ (right panel).
\label{fig1}}
\end{figure}

Except these single-particle cut-offs for the basis space, we have
introduced pair cut-offs with $\varepsilon_1+\varepsilon_2\leq E_{\rm cut}$ and $l_1+l_2\leq L_{\rm cut}$.
The difference between the single-particle cut-off and the pair cut-off is illustrated in Fig.~\ref{fig1}.
Since the matrix elements with small total angular momentum $J$ in the $pp$ coupling have a larger contribution to the matrix elements of total angular momentum $I=0$ in the $ph$ coupling, the pair cut-offs will be introduced for $J>J_h$ with $J_h$ defined as the largest total angular momentum that the hole states can couple to. For $^{16}$O, we have $J_h=3~\hbar$ coupled from two particles in the $1p_{3/2}$ orbit. In present work, we set $E_{\rm cut} = \varepsilon_{\rm cut},\, L_{\rm cut} = l_{\rm cut}$, and the convergence in the pair cut-off will be achieved automatically in the convergence of the single-particle cut-off.

\begin{figure}
\includegraphics[width=8cm]{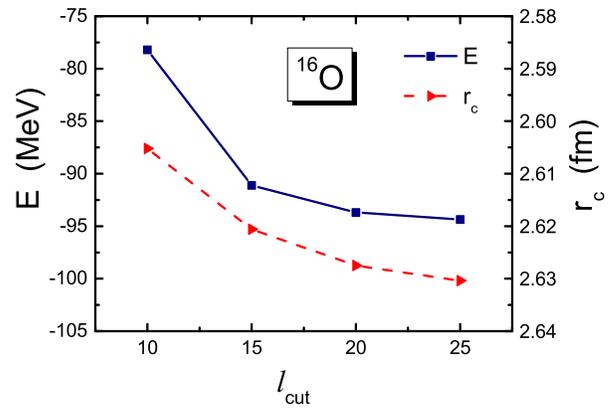}
\caption{(Color online) Total energy $E$ and charge radius $r_c$ of $^{16}$O as a function of angular momentum cut-off $l_{\rm cut}$ calculated in RBHF theory.
\label{fig2}}
\end{figure}

First, we give the convergence check for single-particle angular momentum cut-off $l_{\rm cut}$ in Fig.~\ref{fig2}.
In this check, the other cut-offs are chosen as $J_{\rm cut} = 6~\hbar,\,\varepsilon_{\rm cut} = 1100$ MeV, and $\varepsilon_{\rm Dcut} = -1700$ MeV.
With $\varepsilon_{\rm Dcut} = -1700$ MeV, single-particle states with high angular momentum ($l>7~\hbar$) in the Dirac sea are not included in the basis, this will be discussed later in the check of $\varepsilon_{\rm Dcut}$.
At $l_{\rm cut} = 20~\hbar$, the convergence is achieved. Increasing $l_{\rm cut}$ to $25~\hbar$ will change the energy by $0.6$ MeV and charge radius by $0.003$ fm.

\begin{figure}
\includegraphics[width=8cm]{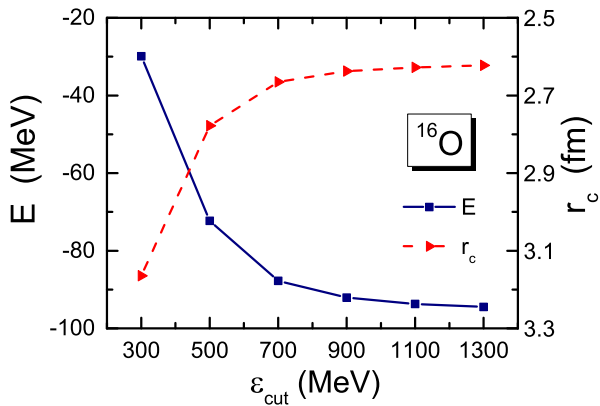}
\caption{(Color online) Total energy $E$ and charge radius $r_c$ of $^{16}$O as a function of the energy cut-off $\varepsilon_{\rm cut}$ calculated in RBHF theory.
\label{fig3}}
\end{figure}

As shown in the RBHF calculation with fixed DWS basis~\cite{Shen2016}, one needs a very large energy cut-off to take into account the short-range correlation of the strong repulsive core.
The general feature of this convergence still holds in the self-consistent RBHF calculation as shown in Fig.~\ref{fig3}.
The other cut-offs are $J_{\rm cut} = 6~\hbar,\,l_{\rm cut} = 20~\hbar$, and $\varepsilon_{\rm Dcut} = -1700$ MeV.
In comparison with RBHF calculation with fixed DWS basis~\cite{Shen2016}, we have included a pair cut-off for the energy, that is, only pair states $|ab\ra$ with $\varepsilon_a+\varepsilon_b\leq\varepsilon_{\rm cut}$ are included in the intermediate states in the BG equation~(\ref{eq:BG2}) for the channels with $J> J_h$. As in the case with the fixed DWS basis, we find good convergence at $\varepsilon_{\rm cut}=1.1$ GeV. Increasing $\varepsilon_{\rm cut}$ to 1.3 GeV will change the energy by $0.8$ MeV and charge radius by $0.005$ fm.

\begin{figure}
\includegraphics[width=8cm]{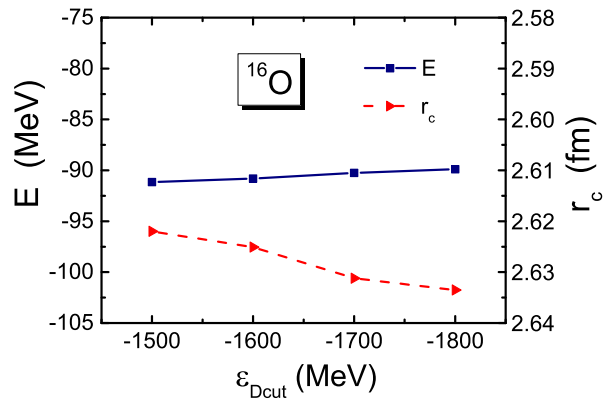}
\caption{(Color online) Total energy $E$ and charge radius $r_c$ of $^{16}$O as a function of energy cut-off in the Dirac sea $\varepsilon_{\rm Dcut}$ calculated in RBHF theory.
\label{fig4}}
\end{figure}

As has been pointed out in Ref.~\cite{Zhou2003}, the single-particle states with negative energy in the Dirac sea must be included in the basis space for completeness. Because of spherical symmetry, the Dirac equation is solved independently in different blocks with the quantum numbers $\kappa=(l,j)$.
We include at least 2 states in the Dirac sea for each block, and then we add more states by gradually decreasing the energy cut-off in the Dirac sea $\varepsilon_{\rm Dcut}$ to check the convergence.
The other cut-offs are $J_{\rm cut} = 6~\hbar,\,l_{\rm cut} = 20~\hbar$, and $\varepsilon_{\rm cut} = 1100$ MeV.
The result is shown in Fig.~\ref{fig4}. Convergence is achieved at $\varepsilon_{\rm Dcut} = -1700$ MeV.
Decreasing $\varepsilon_{\rm Dcut}$ to $-1800$ MeV will change the energy by $0.4$ MeV and charge radius by $0.002$ fm.

Up to now, convergence has been checked with respect to the single-particle basis space and the final choice is $l_{\rm cut} = 20~\hbar,\,\varepsilon_{\rm cut} = 1100$ MeV, and
we include at least 2 states in the Dirac sea for each block, and then add more states by choosing the energy cut-off in the Dirac sea $\varepsilon_{\rm Dcut} = -1700$ MeV.
For a given DWS potential with a box size $R=7$ fm, there will be 686 single-particle states distributed among 41 blocks (i.e. $0\leq l \leq 20$), where 92 are states with negative energy and 594 are states with positive energy.
The dimension of pair states $|ab\ra$ is different for different channels ($J$, parity $P$, and isospin $T$), for example, for ($0,-,1$) it is 5834, and for ($2,+,0$) it is 54654.

\begin{figure}
\includegraphics[width=8cm]{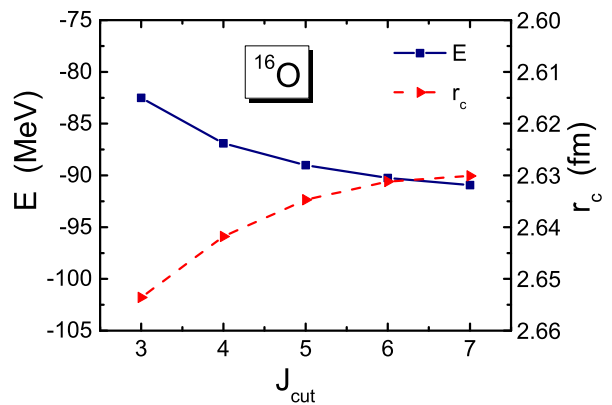}
\caption{(Color online) Total energy $E$ and charge radius $r_c$ of $^{16}$O calculated in RBHF theory as a function of total angular momentum cut-off $J_{\rm cut}$ introduced in Eq.~(\ref{eq:Gph-pp}).
    \label{fig5}}
\end{figure}

In Fig.~\ref{fig5}, we show the convergence with total angular momentum cut-off $J_{\rm cut}$ introduced in Eq.~(\ref{eq:Gph-pp}).
Increasing $J_{\rm cut}$ from $6$ to $7$ will change the energy by $0.3$ MeV and charge radius by $0.0007$ fm.
Therefore $J_{\rm cut} = 6$ will be used in the following calculations.

\begin{figure}
\includegraphics[width=8cm]{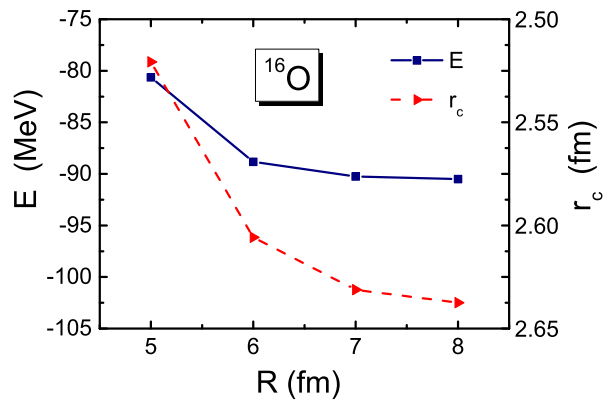}
\caption{(Color online) Total energy $E$ and charge radius $r_c$ of $^{16}$O as a function of box size $R$ calculated in RBHF theory.
    \label{fig6}}
\end{figure}

All the above convergence checks were performed at box size $R = 7$ fm.
In Fig.~\ref{fig6}, we show that this is enough for $^{16}$O.
Further increasing the box size to 8 fm will cause changes by $0.3$ MeV in the energy and by $0.006$ fm in the charge radius.

\subsection{Center-of-mass motion}

In the above convergence check, the center-of-mass motion in Eq.~(\ref{eq:Ecm}) has not been corrected.
Usually there are two ways to treat this term:
\begin{enumerate}
  \item Treat it as a first-order correction after the solution of the RHF equation. As discussed in Sec~\ref{sec:center_of_mass}, we will label this method as a projection after variation (PAV).
  \item Exclude this term in the RHF equation similarly as in Ref.~\cite{Chabanat1998}, which we will label as a projection before variation (PBV).
\end{enumerate}
In none of these cases the center-of-mass term is included in the solution of the BG equation (\ref{eq:BG2}).

\begin{table}
\caption{Total energy, charge radius, and c.m. correction for $^{16}$O calculated by RBHF for PBV and PAV.}
\label{tab1}
\centering
\begin{ruledtabular}
\begin{tabular}{lD{.}{.}{3.3}D{.}{.}{3.3}}
& $PBV$ & $PAV$ \\
\hline
$E$ (MeV)  &  $-110.1$ & $-101.4$ \\
$r_c$ (fm) &  $2.566$   & $2.577$   \\
$E_{cm}$ (MeV)&  $-11.83$ & $-11.12$ \\
\end{tabular}
\end{ruledtabular}
\end{table}

\begin{figure}[!htbp]
\includegraphics[width=8cm]{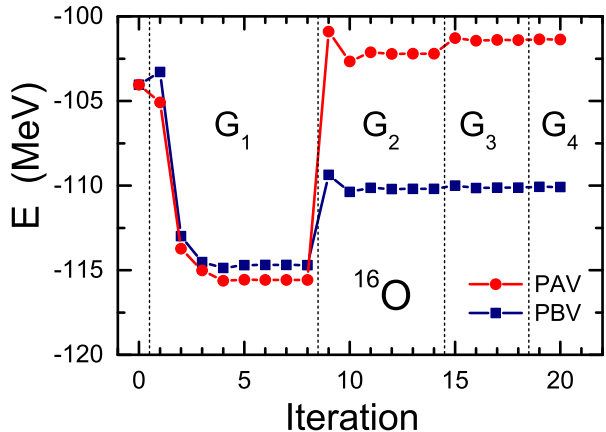}
\caption{(Color online) Total energies at each RBHF iteration step for PAV and PBV.
\label{fig7}}
\end{figure}

The results are listed in Table \ref{tab1}.
It can be seen that the total energy given by PBV is about $9$ MeV smaller than PAV, while the charge radii and center-of-mass correction energy $E_{cm}$ are almost the same for both cases.
In order to understand it more clearly, we plot out the total energy at each iteration step in Fig.~\ref{fig7}.
As can be seen from the figure, there is not so much difference between PBV and PAV during the first RBHF iteration step, where the $G$-matrix $G_1$ is calculated from the initial DWS basis.
Accordingly, the $G$-matrices $G_2,\,G_3,$ and $G_4$ are respectively calculated from the convergent RHF basis.
One may say that with the same interaction, PAV can be viewed as a good approximation of PBV.
But in the next RBHF iteration step, the energy of PBV suddenly becomes smaller than that of PAV.

\begin{figure}[!htbp]
\includegraphics[width=8cm]{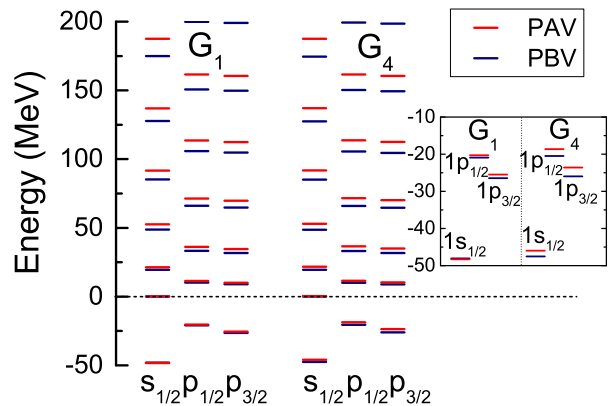}
\caption{(Color online) Single-particle spectra in $s$ and $p$ blocks at each RBHF iteration step for PAV and PBV.
    \label{fig8}}
\end{figure}

The reason for this sudden change can be understood from Fig.~\ref{fig8}, where the single-particle spectra in $s$ and $p$ blocks are given after the first ($G_1$) and last ($G_4$) RBHF iteration.
It can be seen, the single-particle energies in the RBHF calculation with PBV are generally lower than those of PAV, especially for those high-lying states.
In the conventional RHF calculations, the results for PAV and PBV are similar because only the occupied states are concerned and they are similar for PAV and PBV.
But for RBHF calculations, the difference in single-particle spectra, in particular high-lying states, will lead to different $G$-matrices in next iteration.
Generally, the lower the unoccupied states, the more attractive the $G$-matrix elements of occupied states.
As a result, the $G$-matrix in PBV is more attractive and the total binding energy is larger.

PBV and PAV have already been discussed in the non-relativistic BHF calculation \cite{Becker1974}.
As the single-particle energies of BHF basis states are not fed back into the $G$-matrix,
the conclusion in Ref.~\cite{Becker1974} was that the two methods give almost the same results for $^{16}$O.
This is the case in Fig.~\ref{fig7} calculated with the $G$-matrix $G_1$.
In the fully self-consistent RBHF calculation, however, the total energy given by PBV is
about 9 MeV smaller and the single-particle energies are generally
lower than those by PAV.

\subsection{Self-consistent basis and fixed basis}

To take into account the relativistic structure of the Pauli operator in Eq.~(\ref{eq:BG2}),
we adopt a relativistic basis in our calculation.
The relativistic DWS basis~\cite{Zhou2003} has advantages in comparison with the harmonic oscillator basis~\cite{Gambhir1990} like a proper asymptotic behavior of nuclear density distribution, which is crucial for describing, e.g., halo nuclei.
More important here is that the nucleon single-particle potential is close to the DWS shape, which serves as a good approximation for the final converged RBHF single-particle states.

The DWS basis is obtained by solving the radial Dirac equation ~(\ref{eq:direq}) in potentials of Woods-Saxon shape given in  Eqs.~(\ref{eq:dws1}) and (\ref{eq:dws2}).
The parameters are chosen with reference to~\cite{Koepf1991} for the neutron potential, with potential depths $V_0$ from $-60$~MeV to $-80$~MeV.
They are listed in Table \ref{tab2}.

The RBHF calculation has been previously carried out for finite nuclei with a fixed DWS basis in Ref.~\cite{Shen2016}.
We will compare the results with fixed basis and results with the self-consistent RBHF basis in the following.

\begin{table}[!htbp]
\caption{Parameters for DWS potentials in Eqs.~(\ref{eq:dws1}) and (\ref{eq:dws2}).}
\label{tab2}
\centering
\begin{ruledtabular}
\begin{tabular}{cccccc}
  \toprule
  $V_0$ (MeV) & $R$ (fm) & $a$ (fm) & $W_0$ (MeV) & $R^{ls}$ (fm) & $a^{ls}$ (fm) \\
  \hline
  $-60$ to $-80$ & 3.10697 & 0.615 & 725.9136 & 2.8827 & 0.648 \\
  \bottomrule
\end{tabular}
\end{ruledtabular}
\end{table}

\begin{figure}[!htbp]
\includegraphics[width=8cm]{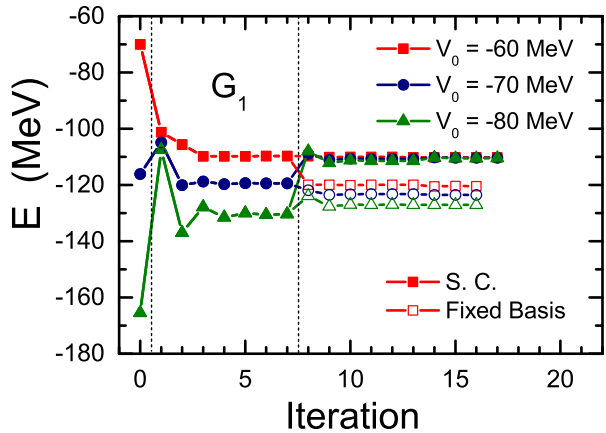}
\caption{(Color online) Total energies at each iteration step calculated by RBHF in the self-consistent RHF basis (solid symbols) and in the fixed DWS basis (open symbols).
 \label{fig9}}
\end{figure}

The total energies at each iteration step are plotted in Fig.~\ref{fig9}.
Solid symbols stand for the self-consistent RBHF calculations, while open symbols stand for fixed DWS basis calculations.
Different shapes represent different DWS potential depths $V_0$.
Similar to Fig.~\ref{fig7}, the first RBHF iteration is represented by $G_1$.

We mention that in Fig.~\ref{fig7} or Fig.~\ref{fig9}, different calculations may have different number of iteration steps.
We have adjusted them slightly in the figures to make them the same without losing too much precision.

For the first RBHF iteration the results are identical to those of the DWS basis calculation for each value of $V_0$, because in both cases the same basis is used.
After the first RBHF iteration, there appears a difference between the self-consistent calculation and the fixed basis one. In the end, the results of the self-consistent calculation do not depend on the initial DWS basis, while the results of a fixed basis one do. As expected, self-consistency is very important to get an unambiguous result.

\subsection{Single-particle potential $U$ of particle states}

One uncertainty in (R)BHF theory is the choice of the single-particle potential for particle states in Eq. (\ref{eq:Upp}).
Different choices have been discussed in BHF for finite nuclei and the readers are referred to Ref. \cite{Davies1969} for more details. Here we choose Eq. (\ref{eq:Upp}) for $U_{ab}$ where
$\varepsilon^{\prime }_a$ is different from the single-particle energy $\varepsilon_a$,
with $\varepsilon^{\prime }_a=\varepsilon^{\prime }_b=\varepsilon^{\prime }$ as the single-particle energy of hole states, and compare the results between the lowest value $\varepsilon^{\prime }=\varepsilon_{\nu 1s1/2}$ and the highest value $\varepsilon^{\prime }=\varepsilon_{\pi 1p1/2}$.
We also give as a comparison the results for the {\it gap choice}: $U_{ab}=0$.
These results are presented in Table~\ref{tab3}.

\begin{table}[!htbp]
\caption{Binding energy per nucleon, charge radius, and proton $1p$ spin-orbit splitting of $^{16}$O in the RBHF calculations with different choices for $U_{ab}$ in Eq. (\ref{eq:Upp}).}
\label{tab3}
\centering
\begin{ruledtabular}
\begin{tabular}{lcccc}
  \toprule
    & $\varepsilon'=\varepsilon_{\pi 1p1/2}$ & $\varepsilon'=\varepsilon_{\nu 1s1/2}$ & Gap & Exp. \\
  \hline
  $E/A$ (MeV) & $-7.10$ & $-6.88$ & $-5.41$ & $-7.98$ \\
  $r_c$ (fm)&  ~~2.56  & ~~2.57   & ~~2.64  & ~~2.70   \\
  $\Delta E_{\pi1p}^{ls}$ (MeV) & ~5.4 & ~5.4 & ~5.4 & ~6.3 \\
  \bottomrule
\end{tabular}
\end{ruledtabular}
\end{table}

It can be seen that the {\it gap choice} gives the least bound system.
While fixing $\varepsilon'$ as an energy among the occupied states gives $1.4$ to $1.7$ MeV per nucleon more binding.

In the framework of Bethe-Brueckner-Goldstone (BBG) (see for instance Ref.~\cite{Day1967}), the {\it gap choice} and the {\it continuous choice} have been discussed in nuclear matter in Ref.~\cite{Song1998}. The {\it continuous choice} gives more binding than the {\it gap choice} at the BHF level. By performing the BBG expansion up to the three hole-line level,
the above two choices give similar results, and lie between the results of these two choices at the BHF level~\cite{Song1998}.
From this point of view, the choice of $\varepsilon'$ as an energy among the occupied states is a reasonable choice.

\begin{figure}[!htbp]
\includegraphics[width=8cm]{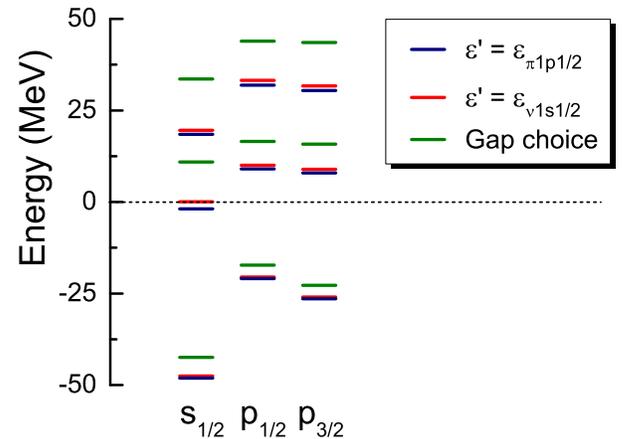}
\caption{(Color online) Single-particle spectrum of $^{16}$O calculated by RBHF theory with the interaction Bonn A for different choices of the single-particle potential $U_{ab}$ of particle states in Eq. (\ref{eq:Upp}).
\label{fig10}}
\end{figure}

In Fig.~\ref{fig10} we present the single-particle spectrum of RBHF calculations with different choices and we find:
First, in the {\it gap choice} there is a larger gap between the unoccupied states and occupied states as compared to other choices. This has been observed in earlier non-relativistic calculations and therefore this name was chosen~\cite{Day1967}.
Second, as compared to the {\it gap choice}, the choices of $\varepsilon'$ fixed as an energy among the occupied states give more attraction to low-lying states.
This can be understood because the single-particle energies $\varepsilon$ of low-lying particle states are close to the chosen $\varepsilon'$.
Therefore it can be viewed as the {\it continuous choice} and the $G$-matrix is attractive.
Third, the single-particle energies of low-lying states with choices of $\varepsilon'=\varepsilon_{\pi 1p1/2}$ and $\varepsilon'=\varepsilon_{\nu 1s1/2}$ are close to each other.

Summing up the previous discussions, the numerical details for the solution of RBHF equations in the following applications include
$l_{\rm cut}=20~\hbar$ (Fig.~\ref{fig2}),
$\varepsilon_{\rm cut}=-1.1 $ GeV (Fig.~\ref{fig3}),
$\varepsilon_{\rm Dcut}=-1700 $ MeV (Fig.~\ref{fig4}),
$J_{\rm cut}= 6~\hbar$ (Fig.~\ref{fig5}), and
$R_{\rm box}= 7 $ fm (Fig.~\ref{fig6}).
The center-of-mass term is treated with PBV.
For the $pp$-potential $U_{ab}$ in Eq. (\ref{eq:Upp}), we consider a {\it continuous choice} with $\varepsilon'$ as the last occupied state in the Fermi sea. This means  $\varepsilon'=\varepsilon_{\pi 1p1/2}$ for $^{16}$O.
For the nucleus $^{40}$Ca we use  $\varepsilon^\prime = \varepsilon_{\pi1d3/2}$, $l_{\rm cut} = 25~\hbar, J_{\rm cut} = 9~\hbar$, and the same values for the remaining quantities.

\section{Results and discussion}
\label{sec:res}

\subsection{The nucleus $^{16}$O}

The total energy, charge radius $r_c$, matter radius $r_m$,
and proton spin-orbit splitting for the $1p$ shell of $^{16}$O calculated by RBHF with the interaction Bonn A \cite{Machleidt1989} are listed in Table~\ref{tab4}, in comparison with experimental data~\cite{Wang2012,Angeli2013,Ozawa2001,Coraggio2003}.
The corresponding results from RBHF in fixed DWS basis \cite{Shen2016}, Density-Dependent Relativistic Hartree-Fock (DDRHF) with
PKO1~\cite{Long2006} and PKA1~\cite{Long2007}, non-relativistic BHF~\cite{Hu2017a} with $V_{{\rm low}-k}$ derived from Argonne $v_{18}$~\cite{Wiringa1995}, Coupled-Cluster (CC) method~\cite{Hagen2009} and No Core Shell Model (NCSM) \cite{Roth2011} with N$^3$LO~\cite{Entem2003}, and Nuclear Lattice Effective Field Theory (NLEFT)~\cite{Lahde2014} with N$^2$LO~\cite{Epelbaum2009a} are also included.

The phenomenological functional PKO1 includes tensor correlations induced by the $\pi$-exchange and PKA1 includes in addition tensor correlations by the $\rho$-exchange. As expected, the phenomenological results, which are both fitted to the experimental data \cite{Long2006,Long2007}, are in much better agreement with data than all the \emph{ab initio} results.

The total energy of our self-consistent RBHF calculation is underbound by 14.1 MeV (or by 11\%) and the charge radius is smaller by 0.14 fm (or by 5\%) as compared to the experimental values. This is similar to the fixed basis calculation~\cite{Shen2016} and consistent with the infinite nuclear matter results~\cite{Brockmann1990}.
In the self-consistent non-relativistic BHF calculations~\cite{Hu2017a}, the interaction $V_{{\rm low}-k}$ derived from the Argonne interaction $v_{18}$ was used.
The result shows an overbinding by $6.6$ MeV (or by 5\%) and the rms radius is too small.
In comparison, the values of
$E=-120.9$ MeV is obtained with the CC~\cite{Hagen2009} using the chiral $NN$ interaction N$^{3}$LO, $E=-119.7(6)$ MeV obtained with the NCSM~\cite{Roth2011} using the same interaction N$^{3}$LO, and $E=-121.4(5)$ obtained with the NLEFT~\cite{Lahde2014} using the interaction N$^2$LO.

The experimental matter radius $r_m=2.54$ fm is larger than $r_m=2.30$ fm calculated with the CC method. It has been shown that this is a general feature of all the \emph{ab initio} calculations with conventional forces based on chiral effective field theory~\cite{Lapoux2016}. Only modern chiral forces~\cite{Ekstrom2015}, which include also radii in the adjustment of the parameters for the bare forces, are able to cure this problem.

The spin-orbit splittings of $1p$ proton shell in RBHF theory $\Delta E_{\pi1p}^{ls}=5.4$ MeV is smaller than the previous RBHF results with fixed DWS basis $\Delta E_{\pi1p}^{ls}=6.0$ MeV, and the deviation with the data is respectively $14\%$ and $5\%$.

\begin{table}
\caption{Total energy, charge radius, matter radius, and $\pi1p$ spin-orbit splitting of $^{16}$O calculated by RBHF theory with the interaction Bonn A \cite{Machleidt1989}, in comparison with experimental data. The corresponding results from RBHF in fixed DWS basis \cite{Shen2016}, DDRHF with PKO1 \cite{Long2006} and PKA1 \cite{Long2007}, non-relativistic BHF~\cite{Hu2017a} with $V_{{\rm low}-k}$ derived from Argonne $v_{18}$, Coupled-Cluster (CC) method \cite{Hagen2009} and No Core Shell Model (NCSM) \cite{Roth2011} with N$^3$LO~\cite{Entem2003}, and Nuclear Lattice Effective Field Theory (NLEFT)~\cite{Lahde2014} with N$^2$LO~\cite{Epelbaum2009a} are also included.}
\centering
\begin{ruledtabular}
\begin{tabular}{lcccc} 
&\multicolumn{1}{c}{$E$ (MeV)} & \multicolumn{1}{c}{$r_c$ (fm)} & \multicolumn{1}{c}{$r_m$ (fm)} &
\multicolumn{1}{c}{$\Delta E_{\pi1p}^{ls}$ (MeV)} \\
\hline
Exp. \cite{Wang2012,Angeli2013,Ozawa2001,Coraggio2003} &
$-127.6$ & $2.70$ & $2.54(2)$ & $6.3$ \\
RBHF, Bonn A & $-113$.$5$ & $2.56$ & $2.42$ & $5$.$4$ \\
RBHF (DWS) \cite{Shen2016}& $-120$.$7$ & $2.52$ & $2.38$ & $6$.$0$ \\
DDRHF, PKO1 \cite{Long2006} & $-128.3$ & $2.68$ & $2.54$ & $6.4$ \\
DDRHF, PKA1 \cite{Long2007} & $-127.0$ & $2.80$ & $2.67$ & $6.0$ \\
BHF \cite{Hu2017a}, AV18  & $-134.2$ &  & $1.95$ & $13.0$ \\
CC \cite{Hagen2009}, N$^3$LO  & $-120.9$ &  & 2.30 &  \\
NCSM \cite{Roth2011}, N$^3$LO  & $-119.7(6)$ &  &  &  \\
NLEFT \cite{Lahde2014}, N$^2$LO  & $-121.4(5)$ &  &   \\
\end{tabular}
\end{ruledtabular}
\label{tab4}
\end{table}

\begin{figure}[!htbp]
\includegraphics[width=8cm]{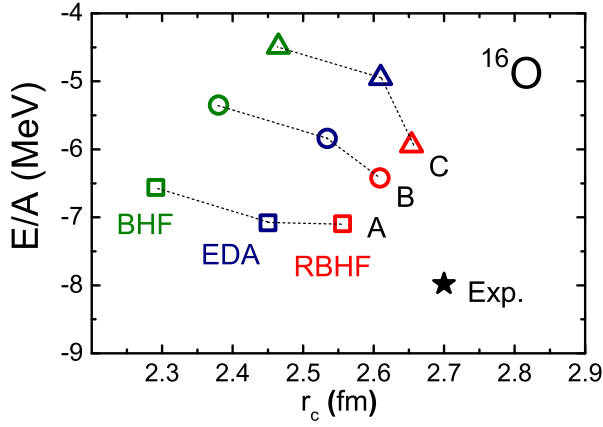}
\caption{(Color online) Energy per nucleon $E/A$ for $^{16}$O as a function of the charge radius $r_c$ calculated in RBHF using the interactions Bonn A, B, and C, in comparison with BHF and the relativistic effective density approximation (EDA) \cite{Muether1990}.
\label{fig11}}
\end{figure}

Figure~\ref{fig11} shows the energy per nucleon $E/A$ for $^{16}$O as a function of the charge radius $r_c$ calculated in RBHF using the interactions Bonn A, B, and C, in comparison with BHF and the relativistic effective density approximation (EDA) \cite{Muether1990}.
It can be seen in all cases that relativistic effects improve the results considerably.
By comparing EDA with RBHF, one can see that self-consistency has important effect.

In Table~\ref{tab5}, the total energy, charge radius, single-particle energies, and $\pi1p$ spin-orbit splitting of $^{16}$O calculated by RBHF theory with the interactions Bonn A, B, and C are listed, and in comparison with the experimental data.
The experimental single-particle energies are taken from Ref.~\cite{Coraggio2003}.

\begin{table}[h]
\caption{Total energy, charge radius, single-particle energies, and $\pi1p$ spin-orbit splitting of $^{16}$O calculated by RBHF theory with the interactions Bonn A, B, and C. All energies are in MeV and radius in fm.}\label{table:all}
\centering
\begin{ruledtabular}
\begin{tabular}{lcccc}
& Exp. & Bonn A & Bonn B & Bonn C \\
\hline
$E$                          &  $-127.6$ & $-113.5$ & $-102.7$ & $-95.0$  \\
$r_c$                        &    $2.70$ &  $2.56$  & $2.61$   & $2.65$   \\
$\varepsilon_{\nu 1s_{1/2}}$ & $-47$     & $-48.1$  & $-45.2$  & $-43.1$  \\
$\varepsilon_{\nu 1p_{3/2}}$ & $-21.8$   & $-26.4$  & $-24.9$  & $-23.7$  \\
$\varepsilon_{\nu 1p_{1/2}}$ & $-15.7$   & $-21.0$  & $-20.2$  & $-19.7$  \\
$\varepsilon_{\pi 1s_{1/2}}$ & $-44\pm7$ & $-43.9$  & $-41.1$  & $-39.1$  \\
$\varepsilon_{\pi 1p_{3/2}}$ & $-18.5$   & $-22.5$  & $-21.1$  & $-20.0$  \\
$\varepsilon_{\pi 1p_{1/2}}$ & $-12.1$   & $-17.1$  & $-16.5$  & $-16.0$  \\
$\Delta E_{\pi1p}^{ls}$      &    $6.3$  &   $5.4$  &  $4.6$   &  $4.0$   \\
\end{tabular}
\end{ruledtabular}\label{tab5}
\end{table}

\begin{figure}[!htbp]
\includegraphics[width=8cm]{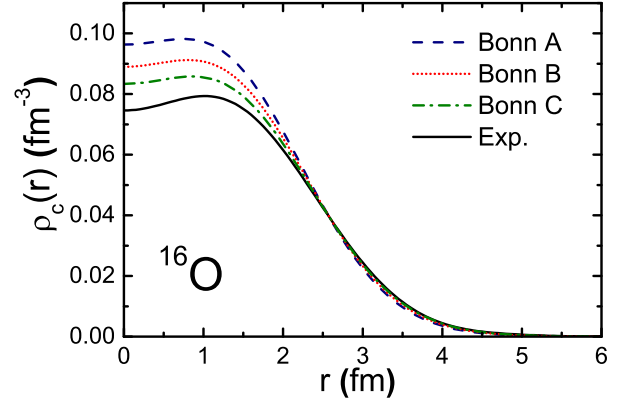}
\caption{(Color online) Charge density distributions of $^{16}$O calculated by RBHF theory with the interactions Bonn A, B, and C, in comparison with experimental data~\cite{DeVries1987}.
\label{fig12}}
\end{figure}

Figure~\ref{fig12} shows the charge density distributions of $^{16}$O calculated by RBHF theory with the interactions Bonn A, B, and C, in comparison with experimental data.
The experimental data is from Ref.~\cite{DeVries1987}.
It can be seen Bonn A, B, and C interactions give too large central distribution, as a result the charge radius is smaller than the experimental value.

\subsection{The nuclei $^{4}$He and $^{40}$Ca}

\begin{table}[h]
\caption{Total energy, charge radius, and proton radius of $^{4}$He calculated by RBHF theory using the interaction Bonn A~\cite{Machleidt1989} with PBV and PAV, in comparison with experimental data~\cite{Wang2012,Angeli2013,Lu2013}. The corresponding results from DDRHF with PKO1 \cite{Long2006} and PKA1 \cite{Long2007}, solution of the Faddeev-Yakubovsky (FY) equation~\cite{Nogga2000} with CD-Bonn, FY~\cite{Binder2016} with N$^4$LO, NCSM~\cite{Navratil2007a} with N$^3$LO, NLEFT~\cite{Lahde2014} with N$^2$LO, and BHF~\cite{Hu2017a} with $V_{{\rm low}-k}$ derived from Argonne $v_{18}$ are also included.}
\centering
\begin{ruledtabular}
\begin{tabular}{llll} 
&\multicolumn{1}{c}{$E$ (MeV)} & \multicolumn{1}{c}{$r_c$ (fm)} & \multicolumn{1}{c}{$r_p$ (fm)} \\
\hline
Exp. \cite{Wang2012,Angeli2013,Lu2013} &
$-28.30$ & $1.68$ & $1.46$  \\
RBHF (PBV), Bonn A & $-35.05$ & $1.83$ & $1.64$  \\
RBHF (PAV), Bonn A & $-26.31$ & $1.90$ & $1.73$  \\
DDRHF, PKO1 \cite{Long2006} & $-28.45$ & $1.90$ & $1.72$ \\
DDRHF, PKA1 \cite{Long2007} & $-28.28$ & $2.06$ & $1.90$ \\
FY \cite{Nogga2000}, CD-Bonn  & $-26.26$ &  &   \\
FY \cite{Binder2016}, N$^4$LO  & $-24.27(6)$ &  & 1.547(2)  \\
NCSM \cite{Navratil2007a}, N$^3$LO  & $-25.39(1)$ &  & $1.515(2)$  \\
NLEFT \cite{Lahde2014}, N$^2$LO  & $-25.60(6)$ &  &   \\
BHF \cite{Hu2017a}, AV18  & $-25.90$ &  &   \\
\end{tabular}
\end{ruledtabular} \label{tab6}
\end{table}

The total energy, charge radius, and proton radius of $^{4}$He calculated by RBHF theory using the interaction Bonn A~\cite{Machleidt1989} with PBV and PAV are listed in Table~\ref{tab6}, in comparison with experimental data~\cite{Wang2012,Angeli2013,Lu2013}. The corresponding results from DDRHF with PKO1 \cite{Long2006} and PKA1 \cite{Long2007}, solution of the Faddeev-Yakubovsky (FY) equation~\cite{Nogga2000} with CD-Bonn~\cite{Machleidt1996}, FY~\cite{Binder2016} with N$^4$LO~\cite{Epelbaum2015}, NCSM~\cite{Navratil2007a} with N$^3$LO~\cite{Entem2003}, NLEFT~\cite{Lahde2014} with N$^2$LO, and BHF~\cite{Hu2017a} with $V_{{\rm low}-k}$ derived from Argonne $v_{18}$~\cite{Wiringa1995} are also included.

It can be seen that by considering only the two-body interaction, all the non-relativistic calculations predicated underbinding for $^4$He. In contrast, RBHF with PBV gives too much binding and a too large radius.
As expected, the center-of-mass correction plays an important role for such a light nucleus: PBV and PAV give very different results.

\begin{table}[b]
\caption{Total energy, charge radius, matter radius,
and proton spin-orbit splitting for the $1d$ shell of $^{40}$Ca calculated by RBHF with the interaction Bonn A \cite{Machleidt1989}, in comparison with experimental data~\cite{Wang2012,Angeli2013,Coraggio2003}.
The corresponding results from DDRHF with PKO1~\cite{Long2006} and PKA1~\cite{Long2007}, BHF~\cite{Hu2017a}, NCSM \cite{Roth2007}, and CC \cite{Hagen2007} with $V_{{\rm low}-k}$ derived from Argonne $v_{18}$~\cite{Wiringa1995}, and CC \cite{Hagen2010} with N$^3$LO~\cite{Entem2003} are also included.}
\centering
\begin{ruledtabular}
\begin{tabular}{lcccc} 
&\multicolumn{1}{c}{$E$ (MeV)} & \multicolumn{1}{c}{$r_c$ (fm)} & \multicolumn{1}{c}{$r_m$ (fm)} & $\Delta E_{\pi1d}^{ls}$ (MeV) \\
\hline
Exp. \cite{Wang2012,Angeli2013,Coraggio2003} &
$-342.1$ & $3.48$ &   & $6.6\pm 2.5$ \\
RBHF , Bonn A & $-290.8$ & $3.23$ & $3.11$ & $5.8$  \\
DDRHF, PKO1 \cite{Long2006} & $-343.3$ & $3.44$ & $3.33$ & $6.6$ \\
DDRHF, PKA1 \cite{Long2007} & $-341.7$ & $3.53$ & $3.41$ & $7.2$ \\
BHF \cite{Hu2017a}, AV18  & $-552.1$ &  & $2.20$ & $24.9$ \\
NCSM \cite{Roth2007}, AV18  & $-461.8$ &  & $2.27$ &  \\
CC \cite{Hagen2007}, AV18  & $-502.9$ &  &  &  \\
CC \cite{Hagen2010}, N$^3$LO  & $-345.2$ &  &  &  \\
\end{tabular}
\end{ruledtabular}
\label{tab7}
\end{table}

In Table~\ref{tab7}, the total energy, charge radius, matter radius,
and proton spin-orbit splitting for the $1d$ shell of $^{40}$Ca calculated by RBHF with the interaction Bonn A \cite{Machleidt1989} are listed, in comparison with experimental data~\cite{Wang2012,Angeli2013,Coraggio2003}.
The corresponding results from DDRHF with PKO1~\cite{Long2006} and PKA1~\cite{Long2007}, BHF~\cite{Hu2017a}, NCSM \cite{Roth2007}, and CC \cite{Hagen2007} with $V_{{\rm low}-k}$ derived from Argonne $v_{18}$~\cite{Wiringa1995}, and CC \cite{Hagen2010} with N$^3$LO~\cite{Entem2003} are also included.

Similar as for $^{16}$O, the total energy of $^{40}$Ca calculated by RBHF with the interaction Bonn A is underbound by 51.3 MeV (or by 15\%) and the charge radius is smaller by 0.25 fm (or by 7\%) as compared to the experimental values.
For the proton $1d$ spin-orbit splitting, RBHF with Bonn A gives a very good description for the data. Most of the non-relativistic results, because of the missing three-body force, give too large binding energy and too small radius, except the CC method with N$^3$LO which reproduces the experimental binding energy well.

\section{Summary}\label{sec:sum}

The relativistic Brueckner-Hartree-Fock equations have been solved for finite nuclei in a self-consistent relativistic Hartree-Fock basis. For this purpose a basis transformation is performed in each step of the iteration, and the $G$-matrix is obtained by solving the Bethe-Goldstone equation in this self-consistent RHF basis.
The Pauli operator has been taken into account fully self-consistently without any approximation.
Relativistic versions of the bare $NN$ interactions Bonn A, B, and C~\cite{Machleidt1989} have been used.

Taking $^{16}$O as an example, the relevant convergence
properties in the RBHF calculation have been checked, including a single-particle angular
momentum cut-off $l_{\rm cut}$, a single-particle energy cut-off
$\varepsilon_{\text{cut}}$, a single-particle energy cut-off
in the Dirac sea $\varepsilon_{\text{Dcut}}$, a total angular momentum
cut-off $J_{\rm cut}$, and a box size $R$.

The binding energy calculated by the self-consistent RBHF does not depend on the initial basis,
and is generally smaller than the results of RBHF calculations in the fixed DWS basis.
This underbinding is, at a first glance, unexpected, because in conventional non-relativistic Hartree-Fock theory the variational principle yields the lowest energy for the fully self-consistent calculation.
However, this can be understood by the well known fact that the Brueckner theory does not obey the variational principle: the matrix-elements of the $G$-matrix depend on the Pauli operator and on the single-particle energies, and the single-particle potential is not defined by the variational principle as the Hartree-Fock theory does.

Two different approaches for the treatment of the center-of-mass motion have been discussed: approximate projection before variation (PBV) and after variation (PAV).
While the two approaches give similar results for RHF calculations, the total energy of $^{16}$O given by PBV is about 9 MeV smaller than that of PAV for RBHF calculations.
This is due to the fact that the single-particle energies are different in the two approaches, especially for those high-lying states, which have a strong impact on the self-consistent $G$-matrix.
Thus, the more consistent approach PBV should be used in the RBHF framework.

We have also discussed different choices for the single-particle potential of particle states, which brings a major ambiguity in the (R)BHF framework.
This ambiguity, in the more general framework of the hole-line expansion, is connected to the limitation of two hole-lines used here.
We have chosen Eq. (\ref{eq:Upp}) for the single-particle potential of particle states, and
found that the choice of $\varepsilon'$ in Eq.~(\ref{eq:Upp}) as an energy among the occupied states is a reasonable choice.

We have performed RBHF calculations for the doubly closed shell nuclei $^4$He, $^{16}$O, and $^{40}$Ca,
and the results are compared with experimental data, phenomenological DDRHF calculations, and other state-of-the-art \emph{ab initio} calculations.
For $^4$He, because of its very light nature,
the treatment of the center-of-mass motion is essential.
The total energy of $^{16}$O and $^{40}$Ca calculated by RBHF with the interaction Bonn
A is underbound by 1 MeV per nucleon and the charge radius is smaller by 5\% to 7\% as compared to the experimental values.
The spin-orbit splittings for both $^{16}$O and $^{40}$Ca are very well reproduced.

Except $^4$He,
we find considerable underbinding by RBHF for the heavier nuclei. There remains the open question of the origin of this underbinding.

First, we cannot exclude, at present, the importance of additional three-body forces, even in a relativistic description.
In the relativistic framework, three-body forces resulting from virtual nucleon-antinucleon excitations ($Z$-diagram) are included~\cite{Brown1987}.
There are also other non-relativistic origins for three-body forces in nuclei, as for instance, the Fujita-Miyazawa force~\cite{Fujita1957}. Microscopic investigations of various contributions to the three-body force have shown that for lower densities, up to the saturation density in nuclear matter, the relativistic $Z$-diagrams, which are included in our RBHF calculations, play a major role~\cite{Zuo2002}.
However, they produce too much repulsion, i.e., one should expect underbinding in finite nuclei, as we find in $^{16}$O and $^{40}$Ca. As shown in Ref.~\cite{Zuo2002} additional three-body forces of non-relativistic nature produce attraction and their contributions are not negligible. For lower densities they play a less important role, but details and more quantitative conclusions are still open questions for the future.

Second, we stayed in these calculations completely on the Brueckner-Hartree-Fock level, i.e. we did not include higher-order diagrams with more than two hole-lines in the hole-line expansion. They are known to have certain contributions in non-relativistic nuclear matter investigations~\cite{Rajaraman1967}. Among them, the third-order saturation-potential diagrams (or rearrangement diagrams) can be taken into account by the so-called renormalized BHF approach \cite{Becker1970} and it is definitely interesting to investigate its relativistic extension in the future.

The current work is intended to establish a firm ground for a relativistic \emph{ab initio} framework in finite nuclei.
With recent progress in covariant chiral nucleon-nucleon interaction \cite{Ren2016} and hyperon-nucleon interaction \cite{Li2016}, we are looking forward to study nuclear many-body problem with RBHF based on chiral effective field theory.
The ultimate goal is to extend it to heavy nuclei and help us to understand nuclear structure in a microscopic way.
We hope to learn from such \emph{ab initio} calculations and to settle down some of the open questions in modern nuclear energy density functional theory, such as the isospin dependence or the importance of the tensor terms~\cite{Lalazissis2009,Long2007}.

\section*{ACKNOWLEDGMENTS}

We thank Jinniu Hu for discussions and Wenhui Long for providing his RHF code.
This work was partly supported by the Major State 973 Program of China No.~2013CB834400, Natural Science Foundation of China under Grants No.~11335002, No.~11375015, and No.~11621131001,  the Overseas Distinguished Professor Project from Ministry of Education No.
MS2010BJDX001, the Research Fund for the Doctoral Program of Higher Education under Grant No.~20110001110087, and the DFG cluster of excellence \textquotedblleft Origin and Structure of the Universe\textquotedblright\ (www.universe-cluster.de). SS would like to thank the short-term PhD student exchange program of Peking University and the RIKEN IPA project, and HL would like to thank the RIKEN iTHES project and iTHEMS program.

\appendix

\newcommand{\pt}{\partial}
\newcommand{\nb}{\nabla}
\newcommand{\ti}{\times}
\newcommand{\lx}{\left.}
\newcommand{\rx}{\right.}
\newcommand{\ala}[1]{\begin{align}#1\end{align}}
\newcommand{\jb}[1]{\left\{\begin{array}{ccc}#1\end{array}\right\}}
\newcommand{\nt}{\notag}
\renewcommand{\a}{\alpha}
\renewcommand{\b}{\beta}
\newcommand{\C}{\Gamma}
\renewcommand{\c}{\gamma}
\newcommand{\D}{\Delta}
\renewcommand{\d}{\delta}
\newcommand{\e}{\varepsilon}
\newcommand{\s}{\sigma}
\renewcommand{\k}{\kappa}
\renewcommand{\L}{\Lambda}
\renewcommand{\l}{\lambda}
\renewcommand{\r}{\rho}
\newcommand{\w}{\omega}
\newcommand{\W}{\Omega}
\renewcommand{\t}{\theta}
\newcommand{\p}{\varphi}
\newcommand{\z}{\zeta}

\begin{widetext}
\section{Two-body matrix elements}
\label{sec:app1}
Here we give the details for the calculation of relativistic two-body matrix elements for the Bonn interaction~\cite{Machleidt1989}, which is defined in terms of meson-exchange terms:
\begin{equation}
\la ab|V_\alpha|cd\ra = \int d^{3}r_{1}d^{3}r_{2}
 \bar{\psi}_a(\mathbf{r}_{1})\Gamma _{\alpha }^{(1)}\psi_c(\mathbf{r}_{1})
 D_{\alpha }(\mathbf{r}_{1},\mathbf{r}_{2})
 \bar{\psi}_b(\mathbf{r}_{2})\Gamma _{\alpha }^{(2)}\psi_d (\mathbf{r}_{2}),
\end{equation}%
with $\alpha$ for scalar ($\sigma,\delta$), vector ($\omega,\rho$), and pseudoscalar ($\eta,\pi$) mesons.
The interacting vertices are given in Eq.~(\ref{eq:gamma12}).
Here we discuss the isoscalar mesons only since the isovector mesons are identical except for the isospin matrix elements, which can be considered separately.

Using the propagator in Eq.~(\ref{eq:propa}) in momentum space, we can rewrite the two-body matrix elements as
\begin{equation}
\la ab|V_\alpha|cd\ra =\int\frac{d^3q}{(2\pi)^3} \frac{1}{m_\alpha^2+q^2}
\la a|\c^0\C_\alpha(1) e^{i\mathbf{q\cdot r}_1} |c\ra \la b|\c^0\C_\alpha(2) e^{-i\mathbf{q\cdot r}_2} |d\ra.
\end{equation}
The form factor in Eq.~(\ref{eq:form}) depends only on the meson $\alpha$ and the absolute momentum transfer $|\mathbf{q}|$, so that it can be added at the last step.

The plane wave can be expanded as
\begin{equation}
e^{i\mathbf{q\cdot r}} = 4\pi \sum_{LM}i^L j_L(qr) Y_{LM}^*(\mathbf{\hat{q}})
Y_{LM}(\mathbf{\hat{r}}),
\end{equation}
where $j_L(qr)$ is the spherical Bessel function, $Y_{LM}$ is the spherical harmonic function, and $\mathbf{\hat{q}}$ represents the angular part of the vector $\mathbf{q}$. The integration over $\hat{\mathbf{q}}$ yields
\begin{equation}
\label{eq:A4}
\la ab|V_\alpha|cd\ra = \frac{2}{\pi} \int  \frac{q^2dq}{m_\alpha^2+q^2} \sum_{LM} (-1)^M
\la a|\c^0\C_\alpha j_{L}(qr)Y_{LM}(\mathbf{\hat{r}}) |c\ra
\la b|\c^0\C_\alpha j_{L}(qr)Y_{L-M}(\mathbf{\hat{r}}) |d\ra.
\end{equation}
For spherical nuclei we consider matrix elements coupled to good angular momentum. There are several ways to couple pairs of indices
\begin{itemize}
\item $pp$-coupled matrix elements:
\begin{equation}\label{eq:VppJ}
\la 12|V|34\ra_{pp}^{J} = \sum_{m_1m_2}\sum_{m_3 m_4} C_{j_1m_1j_2m_2}^{JM}C_{j_3m_3j_4m_4}^{JM}\la 12|V|34\ra.
\end{equation}
\item $ph$-coupled matrix elements (direct term):
\begin{equation}\label{eq:VphI}
\la 12|V|34\ra_{ph}^{I} = \sum_{m_1m_3}\sum_{m_2 m_4} (-1)^{j_3-m_3}C_{j_1m_1j_3-m_3}^{IM}
(-1)^{j_2-m_2}C_{j_4m_4j_2-m_2}^{IM}\la 12|V|34\ra.
\end{equation}
\item $ph$-coupled matrix elements (exchange term):
\begin{equation}\label{eq:VphIe}
\la 12|V|34\ra_{ph,e}^{I} = \sum_{m_1m_3}\sum_{m_2 m_4} (-1)^{j_3-m_3}C_{j_1m_1j_3-m_3}^{IM}
(-1)^{j_2-m_2}C_{j_4m_4j_2-m_2}^{IM}\la 12|V|43\ra.
\end{equation}
\end{itemize}
These different coupling schemes are related to each other by the following recoupling rules \cite{Shalit1963}
\begin{align}
\label{eq:Vphe-ph}
\la 12|V|34\ra_{ph,e}^{I} =&~\sum_{I'}(2I'+1)(-1)^{j_3+j_4+I+I'}
\left\{\begin{array}{ccc}
j_1 & j_3 & I \\ j_2 & j_4 & I'
\end{array}\right\}
\la 12|V|43\ra_{ph}^{I'},\\
\label{eq:Vpp-ph}
\la 12|V|34\ra_{pp}^{J} ~~=&~\sum_{I}(2I+1)(-1)^{j_3+j_4+J}
\left\{\begin{array}{ccc}
j_1 & j_2 & J \\ j_4 & j_3 & I
\end{array}\right\}
\la 12|V|34\ra_{ph}^{I},
\end{align}
and the inverse relation:
\begin{equation}\label{eq:Vph-pp}
\la 12|V|34\ra_{ph}^{I}~~=~\sum_{J}(2J+1)(-1)^{j_3+j_4+J}
\left\{\begin{array}{ccc}
j_1 & j_3 & I \\ j_4 & j_2 & J
\end{array}\right\}
\la 12|V|34\ra_{pp}^{J}.~~~
\end{equation}
The antisymmetrized $ph$-coupled matrix elements are obtained as
\begin{equation}
\label{eq:Vpha}
\la 12|\bar{V}|34\ra_{ph}^{J} = \la 12|V|34\ra_{ph}^{J} - \la 12|V|34\ra_{ph,e}^{J}.
\end{equation}
Expressing the various vertices $\c^0\C_\alpha Y_{LM}$ in Eq. (\ref{eq:A4}) in terms of spherical tensor operators $\hat{O}_{\lambda\mu}$, we can use the Wigner-Eckart theorem \cite{Varshalovich1988} and express the $ph$-coupled matrix elements by the reduced matrix elements of the corresponding tensors
\begin{equation}
\hat{J}\sum_{m_1m_2} (-1)^{j_2-m_2}C_{j_1m_1j_2-m_2}^{JM} \la j_1m_1|\hat{O}_{\lambda\mu}|j_2m_2\ra =
\delta_{J\lambda}\delta_{M\mu}\la j_1||\hat{O}_J||j_2\ra,
\end{equation}
where $\hat{J}=\sqrt{2J+1}$. Therefore we present in the following the direct terms of the $ph$-coupled matrix elements for the different mesons.

\subsection{Scalar meson}

For the scalar meson,
\begin{align}
\la ab|V_s|cd\ra_{ph}^{I} =& -g_s^2\frac{2}{\pi} \frac{(-1)^{j_b-j_d}}{\hat{I}^2} \int \frac{q^2dq}{m_s^2+q^2} \lb \frac{\L_{s}^2-m_{s}^2}{\L_{s}^2+q^2} \rb^2
\la a||\c^0j_{I}Y_{I} ||c\ra \la b||\c^0j_{I}Y_{I} ||d\ra,
\end{align}
where
\begin{equation}
\la a||\c^0j_{I}Y_{I} ||c\ra = \la j_al_a||Y_I||j_cl_c \ra
\int dr\ls F_{a}  j_I(qr)F_{c}  \rs
- \la j_a\tilde{l}_a||Y_I||j_c\tilde{l}_c \ra
\int dr\ls G_{a}  j_I(qr)G_{c} \rs,
\end{equation}
with $F_a(r)$ and $G_a(r)$ are the large and small components of the Dirac spinor in Eq.~(\ref{eq:wf}).

\subsection{Pseudoscalar meson}

For the pseudoscalar meson,
\begin{align}
\la ab|V_{ps}|cd\ra_{ph}^{I}
=& -\frac{f_{ps}^2}{m_{ps}^2}\frac{2}{\pi}
\frac{(-1)^{j_b-j_d}}{\hat{I}^4}
\int \frac{q^4dq}{m_{ps}^2+q^2} \lb \frac{\L_{ps}^2-m_{ps}^2}{\L_{ps}^2+q^2} \rb^2 \times\notag \\
&\quad\times \la a||\sqrt{I+1} j_{I+1}[Y_{I+1}\s]_{I}
+\sqrt{I} j_{I-1}[Y_{I-1}\s]_{I}||c\ra 
\la b||\sqrt{I+1} j_{I+1}[Y_{I+1}\s]_{I}
+\sqrt{I} j_{I-1}[Y_{I-1}\s]_{I}||d\ra,
\end{align}
where $[Y_{L}\s]_{I}$ is the reduced form of
\begin{equation}
[Y_{L}\s]_{IM} = \sum_{mk} C_{Lm1k}^{IM} Y_{Lm}\s_{1k}.
\end{equation}
The matrix elements read
\begin{equation}
\la a||j_{L}[Y_{L}\s]_{I}||c\ra = \la j_al_a||[Y_{L}\s]_I ||j_cl_c \ra
\int dr\ls F_{a}  j_{L}(qr)F_{c}  \rs
+ \la j_a\tilde{l}_a||[Y_{L}\s]_I||j_c\tilde{l}_c \ra
\int dr\ls G_{a}  j_{L}(qr)G_{c} \rs.
\end{equation}

\subsection{Vector meson}

The expression for the vector meson is more complicated. According to the Lorentz structure, we divide it into a time component ($v0$) and a space component ($v1$).
Take the vector-vector coupling (vv) as an example,
\begin{equation}
\C_{v}^{\rm (vv)} = g_v^2 \c^\mu\c_\mu = g_v^2 (\c_0\c_0 -\boldsymbol{\c}\cdot\boldsymbol{\c}) = \C_{v0} + \C_{v1}.
\end{equation}
The results for the time component are
\begin{align}
\la ab|V_{v0}^{(\rm vv)}|cd\ra_{ph}^{I}
=& g_{v}^2\frac{2}{\pi}\frac{(-1)^{j_b-j_d}}{\hat{I}^2}
\int \frac{q^2dq}{m_{v}^2+q^2}\lb \frac{\L_{v}^2-m_{v}^2}{\L_{v}^2+q^2} \rb^2
\la a||j_{I}Y_{I} ||c\ra \la b||j_{I}Y_{I} ||d\ra, \\
\la ab|V_{v0}^{(\rm tt)}|cd\ra_{ph}^{I}
=& -\lb \frac{f_v}{2M} \rb^2\frac{2}{\pi}\frac{(-1)^{j_b-j_d}}{\hat{I}^4}
\int \frac{q^4dq}{m_v^2+q^2} \lb \frac{\L_{v}^2-m_{v}^2}{\L_{v}^2+q^2} \rb^2\nt \times \\
&\quad\ti \la a||\sqrt{I+1} j_{I+1}[Y_{I+1}\c]_{I}
+\sqrt{I} j_{I-1}[Y_{I-1}\c]_{I}||c\ra \nt \\
&\quad\ti \la b||\sqrt{I+1} j_{I+1}[Y_{I+1}\c]_{I}
+\sqrt{I} j_{I-1}[Y_{I-1}\c]_{I}||d\ra , \\
\la ab|V_{v0}^{(\rm vt)}|cd\ra_{ph}^{I} =& -i\frac{f_v g_v}{2M} \frac{2}{\pi}
\frac{(-1)^{j_b-j_d}}{\hat{I}^3}  \int \frac{q^3dq}{m_v^2+q^2} \lb \frac{\L_{v}^2-m_{v}^2}{\L_{v}^2+q^2} \rb^2 \nt \times \\
& \quad\ti \Lb \la a||j_{I}Y_{I} ||c\ra \la b||\sqrt{I+1} j_{I+1}[Y_{I+1}\c]_{I}
+\sqrt{I} j_{I-1}[Y_{I-1}\c]_{I}||d\ra \rx \nt \\
& \quad\quad\lx +  \la a||\sqrt{I+1} j_{I+1}[Y_{I+1}\c]_{I}
+\sqrt{I} j_{I-1}[Y_{I-1}\c]_{I}||c\ra \la b||j_{I}Y_{I} ||d\ra \Rb,
\end{align}
for the vector-vector, tensor-tensor (tt), and vector-tensor (vt) components, respectively.

The results for the space component are (with $\boldsymbol{\a}=\c^0\boldsymbol{\c}$):
\begin{align}
&\la ab|V_{v1}^{(\rm vv)}|cd\ra_{ph}^{I}
= -g_{v}^2\frac{2}{\pi}\frac{(-1)^{j_b+j_d+I}}{\hat{I}^2}
\sum_L(-1)^L
\int \frac{q^2dq}{m_{v}^2+q^2} \lb \frac{\L_{v}^2-m_{v}^2}{\L_{v}^2+q^2} \rb^2
\la a||j_{L}[Y_{L}\a]_{I} ||c\ra \la b||j_{L}[Y_{L}\a]_{I} ||d\ra, \\
&\la ab|V_{v1}^{(\rm tt)}|cd\ra_{ph}^{I}
=\lb \frac{f_v}{M} \rb^2\frac{3}{\pi}\frac{(-1)^{j_b+j_d+I}}{\hat{I}^2}
\sum_L (-1)^L \int \frac{q^4dq}{m_v^2+q^2} \lb \frac{\L_{v}^2-m_{v}^2}{\L_{v}^2+q^2} \rb^2 \ti\notag \\
&~~~~~~\times \bigg\la a\bigg|\bigg| \c^0\sqrt{L+1}\jb{ 1 & L+1 & L \\ I & 1 & 1 } j_{L+1}[Y_{L+1}\s]_{I}
+\c^0\sqrt{L}\jb{ 1 & L-1 & L \\ I & 1 & 1 } j_{L-1}[Y_{L-1}\s]_{I} \bigg|\bigg|c\bigg\ra \notag \\
&~~~~~~\times \bigg\la b\bigg|\bigg| \c^0\sqrt{L+1}\jb{ 1 & L+1 & L \\ I & 1 & 1 } j_{L+1}[Y_{L+1}\s]_{I}
+\c^0\sqrt{L}\jb{ 1 & L-1 & L \\ I & 1 & 1 } j_{L-1}[Y_{L-1}\s]_{I} \bigg|\bigg|d\bigg\ra , \\
&\la ab|V_{v1}^{(\rm vt)}|cd\ra_{ph}^{I}
= \sqrt{6}i\frac{f_v g_v}{2M} \frac{2}{\pi}
\frac{(-1)^{j_b+j_d}}{\hat{I}^2} \int \frac{q^3dq}{m_v^2+q^2} \lb \frac{\L_{v}^2-m_{v}^2}{\L_{v}^2+q^2} \rb^2
\sum_{L} (V^{(\rm vt)}_1+V^{(\rm vt)}_2) ,
\end{align}
with
\begin{align*}
V^{(\rm vt)}_1 =& \la a||j_L[Y_L\a]_I||c\ra 
\bigg\la b\bigg|\bigg|\c^0\sqrt{L+1} \jb{ 1 & L+1 & L \\ I & 1 & 1 } j_{L+1}[Y_{L+1}\s]_{I}
+\c^0\sqrt{L} \jb{ 1 & L-1 & L \\ I & 1 & 1 } j_{L-1}[Y_{L-1}\s]_{I} \bigg|\bigg|d\bigg\ra,\nt\\
V^{(\rm vt)}_2 =& \bigg\la a\bigg|\bigg|\c^0\sqrt{L+1} \jb{ 1 & L+1 & L \\ I & 1 & 1 } j_{L+1}[Y_{L+1}\s]_{I}
+\c^0\sqrt{L} \jb{ 1 & L-1 & L \\ I & 1 & 1 } j_{L-1}[Y_{L-1}\s]_{I} \bigg|\bigg|c\bigg\ra 
\la b||j_{L}[Y_{L}\a]_{I} ||d\ra.
\end{align*}
The reduced matrix element reads
\begin{equation}
\la a||j_{L}[Y_{L}\a]_{I}||c\ra
= i\la j_al_a||[Y_{L}\s]_I ||j_c\tilde{l}_c \ra
\int dr\ls F_{a}  j_{L}(qr)G_{c}  \rs
- i\la j_a\tilde{l}_a||[Y_{L}\s]_I||j_cl_c \ra
\int dr\ls G_{a}  j_{L}(qr)F_{c} \rs.
\end{equation}

Notice that in the Bonn interaction \cite{Machleidt1989} the tensor part of $\w$ meson is deemed to be small and omitted, only the $\rho$ meson has a tensor part $f_\rho$.

\subsection{Isospin matrix elements}

The isospin operator has been separated in the vertex expression in Eq.~(\ref{eq:gamma12}).
For the isovector mesons ($\delta,\pi,\rho$), the full interaction vertices are accompanied with the isovector operator as
\begin{align}
\Gamma_i(1,2)\vec{\tau}_1\cdot\vec{\tau}_2.
\end{align}

For the isoscalar operator $\mathbb{I}$, the non-zero matrix elements are
\begin{align}
\la nn|\mathbb{I}|nn\ra = \la pp|\mathbb{I}|pp\ra = \la np|\mathbb{I}|np\ra = \la pn|\mathbb{I}|pn\ra = 1.
\end{align}
In contrast, for the isovector operator $\vec{\tau}_1\cdot\vec{\tau}_2$, the non-zero matrix elements are
\begin{subequations}
\begin{align}
&\la nn|\vec{\tau}_1\cdot\vec{\tau}_2|nn\ra = \la pp|\vec{\tau}_1\cdot\vec{\tau}_2|pp\ra = 1, \\
&\la np|\vec{\tau}_1\cdot\vec{\tau}_2|np\ra = \la pn|\vec{\tau}_1\cdot\vec{\tau}_2|pn\ra = -1, \\
&\la np|\vec{\tau}_1\cdot\vec{\tau}_2|pn\ra = \la pn|\vec{\tau}_1\cdot\vec{\tau}_2|np\ra = 2,
\end{align}
\end{subequations}
with the conventions $\tau_z|n\ra = |n\ra$ and $\tau_z|p\ra = -|p\ra$ for neutron and proton states.

\section{Center-of-mass Motion}\label{sec:app2}

\subsection{Matrix element for center-of-mass motion}

In second quantization the operator for the center of mass correction is given by
\begin{align}
H_{cm} = \frac{1}{2MA}\mathbf{P}^2
= \frac{1}{2MA} \sum_{ab}\mathbf{p}_{ab}^2 b_a^\dagger b_b + \frac{1}{2MA}\sum_{abcd}\mathbf{p}_{ac}\cdot\mathbf{p}_{bd} b_a^\dagger b_b^\dagger b_d b_c = T_{cm}+V_{cm}.
\end{align}
It contains a one-body operator $T_{cm}$ and a two-body operator $V_{cm}$.

The matrix element of the one-body term $T_{cm}$ is similar to the non-relativistic kinetic energy. Because of spherical symmetry and parity conservation, it is diagonal in the block index $\kappa=(l,j)$ and can be derived as
\begin{equation}\label{eq:Tcm}
\la a|T_{cm}|b\ra = -\frac{1}{2MA}\la a|\nb^2|b\ra
= -\frac{1}{2MA}\int dr \Lb F^*_a \ls \frac{\pt^2}{\pt r^2}
- \frac{\k(\k+1)}{r^2}\rs F_{b}
+ G^*_a \ls \frac{\pt^2}{\pt r^2} +\frac{\k(1-\k)}{r^2}\rs G_{b} \Rb.
\end{equation}

The matrix element of the two-body term is
\begin{align}\label{eq:Vcm}
\la ab|V_{cm}|cd\ra =& -\frac{1}{2MA} \la a|\nb|c\ra \cdot \la b|\nb|d\ra\nt\\
=& \frac{1}{2MA} (-1)^{j_c+j_d} \hat{j}_c \hat{j}_d \sum_\mu (-1)^\mu
C_{j_cm_c1\mu}^{j_am_a} C_{j_dm_d1-\mu}^{j_bm_b}\sum_{\eta\zeta=\pm}
(-1)^{l_a^{(\eta)}+l_b^{(\zeta)}} \nt \times \\
&\ti \jb{ l_c^{(\eta)} & \frac{1}{2} & j_c \\ j_a & 1 & l_a^{(\eta)} }
\ls \sqrt{l_c^{(\eta)}+1} A_{ac}^{(\eta)} \d_{l_a^{(\eta)},l_c^{(\eta)}+1}
- \sqrt{l_c^{(\eta)}} B_{ac}^{(\eta)} \d_{l_a^{(\eta)},l_c^{(\eta)}-1} \rs \nt\\
&\ti \jb{ l_d^{(\zeta)} & \frac{1}{2} & j_d \\ j_b & 1 & l_b^{(\zeta)} }
\ls \sqrt{l_d^{(\zeta)}+1} A_{bd}^{(\zeta)} \d_{l_b^{(\zeta)},l_d^{(\zeta)}+1}
- \sqrt{l_d^{(\zeta)}} B_{bd}^{(\zeta)} \d_{l_b^{(\zeta)},l_d^{(\zeta)}-1} \rs.
\end{align}
In this equation, summation indices $\eta,\,\zeta$ sum over the upper ($+$) and lower ($-$) components, and $A,\, B$ are defined as
\begin{align}
A_{ac}^{(+)} =& \int_0^\infty dr F_a^* \lb \frac{d}{dr} - \frac{l_c+1}{r} \rb F_c,~~~~~~~~~~~
A_{ac}^{(-)} = \int_0^\infty dr G_a^* \lb \frac{d}{dr} - \frac{\tilde{l}_c+1}{r} \rb G_c, \\
B_{ac}^{(+)} =& \int_0^\infty dr F_a^* \lb \frac{d}{dr} + \frac{l_c}{r} \rb F_c, ~~~~~~~~~~~~~~~~
B_{ac}^{(-)} = \int_0^\infty dr G_a^* \lb \frac{d}{dr} + \frac{\tilde l_c}{r} \rb G_c.
\end{align}

Here we also give the antisymmetrized $jj$-coupled form of $V_{cm}$.
In the spherical RHF equation, only $ph$ coupling $I=0$ matrix elements in Eq.~(\ref{eq:VphI}) are needed,
\begin{align}
&\la ab|\bar{V}_{cm}|cd\ra_{ph}^{I=0} = \frac{1}{2MA} \hat{j}_a\hat{j}_b
\sum_{\eta\zeta=\pm} (-1)^{l_b^{(\eta)}+l_a^{(\zeta)}}
 \jb{ l_a^{(\eta)} & \frac{1}{2} & j_a \\ j_b & 1 & l_b^{(\eta)} }
 \jb{ l_b^{(\zeta)} & \frac{1}{2} & j_b \\ j_a & 1 & l_a^{(\zeta)} }\ti \nt \\
&~~~~~~\ti
\ls \sqrt{l_a^{(\eta)}+1} A_{bc}^{(\eta)} \d_{l_b^{(\eta)},l_a^{(\eta)}+1}
- \sqrt{l_a^{(\eta)}} B_{bc}^{(\eta)} \d_{l_b^{(\eta)},l_a^{(\eta)}-1} \rs
\ls \sqrt{l_b^{(\zeta)}+1} A_{ad}^{(\zeta)} \d_{l_a^{(\zeta)},l_b^{(\zeta)}+1}
- \sqrt{l_b^{(\zeta)}} B_{ad}^{(\zeta)} \d_{l_a^{(\zeta)},l_b^{(\zeta)}-1} \rs.
\end{align}

\subsection{Center-of-mass correction for total energy}

The energy of the center-of-mass motion is
\begin{align}
E_{cm} =& \la H_{cm} \ra = \sum_{a}^A \la a|T_{cm}|a\ra + \sum_{ab}^A \la ab|\bar{V}_{cm}|ab\ra.
\end{align}
The first term can be obtained from Eq.~(\ref{eq:Tcm}),
\begin{align}
\sum_{a}^A \la a|T_{cm}|a\ra = -\frac{1}{2MA} \sum_a^A \int dr
\Lb F_a\ls \frac{\pt^2}{\pt r^2} - \frac{\k(\k+1)}{r^2}\rs F_a
+ G_a\ls \frac{\pt^2}{\pt r^2} - \frac{\k(\k-1)}{r^2}\rs G_a \Rb.
\end{align}
The second term can be obtained from Eq.~(\ref{eq:Vcm}),
\begin{align}
&\sum_{ab}^A \la ab|\bar{V}_{cm}|ab\ra
= \frac{1}{2MA} {\sum_{ab}^{A}}' \hat{j}_a^2 \hat{j}_b^2
\sum_{\eta\zeta=\pm} (-1)^{l_a^{(\eta)}+l_b^{(\zeta)}}
\jb{ l_b^{(\eta)} & \frac{1}{2} & j_b \\ j_a & 1 & l_a^{(\eta)} }
\jb{ l_a^{(\zeta)} & \frac{1}{2} & j_a \\ j_b & 1 & l_b^{(\zeta)} }\ti \nt \\
&~~~~~~~\ti \ls \sqrt{l_b^{(\eta)}+1} A_{ab}^{(\eta)} \d_{l_a^{(\eta)},l_b^{(\eta)}+1}
- \sqrt{l_b^{(\eta)}} B_{ab}^{(\eta)} \d_{l_a^{(\eta)},l_b^{(\eta)}-1} \rs
\ls \sqrt{l_a^{(\zeta)}+1} A_{ba}^{(\zeta)} \d_{l_b^{(\zeta)},l_a^{(\zeta)}+1}
- \sqrt{l_a^{(\zeta)}} B_{ba}^{(\zeta)} \d_{l_b^{(\zeta)},l_a^{(\zeta)}-1} \rs.
\end{align}
Indeed, the direct contribution in this expression vanishes.
Note that the $\sum'$ does not sum over magnetic quantum number $m$.
This energy $E_{cm}$ should be subtracted in the total energy expression in Eq.~(\ref{eq:finalE}).

\end{widetext}

\section{Coulomb Interaction}
\label{sec:app3}
The Coulomb interaction can be derived from the interaction vertex,
\begin{equation}
\Gamma_{A}(1,2) = \frac{e}{2} \gamma^\mu(1-\tau_z) \frac{e}{2}\gamma_\mu(1-\tau_z).
\end{equation}
In the R(B)HF framework, its contribution to the single-particle potential is
\begin{equation}
\la a|U_{\rm A}|b\ra,
\end{equation}
which is not included in the G-matrix and calculated separately.

\subsection{Direct term}

The direct term of Coulomb field in the spherical case reads
\begin{equation}
U_{\rm Adir}(r) = e \int {r'}^{2}dr  \frac{\rho_p(r')}{r\theta(r-r')+r'\theta(r'-r)},
\end{equation}
with $\theta$ the step functions.
The contribution to total energy is
\begin{equation}
E_{\rm Adir} = \frac{1}{2} 4\pi e \int r^2 U_{\rm Adir}(r) \rho_p(r) dr.
\end{equation}

\subsection{Exchange term}

In principle, the exchange term of Coulomb field is non-local. For simplicity, in the present calculations, we use the relativistic local density approximation (RLDA) for the Coulomb exchange term \cite{Gu2013}, which in the spherical case reads
\begin{equation}
U_{\rm Aex}^{(\rm RLDA)}(r) = -\lb\frac{3}{\pi}\rb^{1/3} e^2 \rho_p^{1/3}(r) + \frac{3\pi}{M^2} e^2 \rho_p(r).
\end{equation}
This is the well-known Slater approximation \cite{Slater1951} plus a relativistic correction.
The contribution to total energy is
\begin{equation}
E_{\rm Aex}^{(\rm RLDA)} = -\frac{3}{4}\lb\frac{3}{\pi}\rb^{1/3} e^2 \int r^2 dr \rho_p^{4/3} \ls 1 - \frac{2}{3}\frac{(3\pi^2\rho_p)^{2/3}}{M^2}\rs.
\end{equation}

\section{Extrapolation with Respect to Box Size}\label{sec:app4}

For fixed $l_{\rm cut}$ and $\varepsilon_{\rm cut}$, the basis space increases dramatically as $R$ increases.
As shown in the coupled-cluster calculations with the harmonic oscillator basis, when the ultraviolet condition is fulfilled, the total energy and radius have specific relations with the box size $R$ due to the infrared cutoffs induced by the box boundary condition \cite{Furnstahl2012}
\begin{subequations}\label{eq:fit}
\begin{align}
E_R =& E_\infty + a_0 e^{-2k_\infty R}, \\
\la r^2\ra_R =& \la r^2\ra_\infty \ls 1-\lb c_0\beta^3+c_1\beta\rb e^{-\beta}\rs,
\end{align}
\end{subequations}
where $\beta=2k_\infty R$, and $E_\infty$ and $\la r^2\ra_\infty$ are the expectation values of total energy and radius when the box size $R$ is extrapolated to infinity.
Together with
$k_\infty,a_0,c_0,c_1$, they are fitted to several calculated points.
Now we can evaluate these relations within the RBHF framework.

\begin{figure}
\includegraphics[width=8cm]{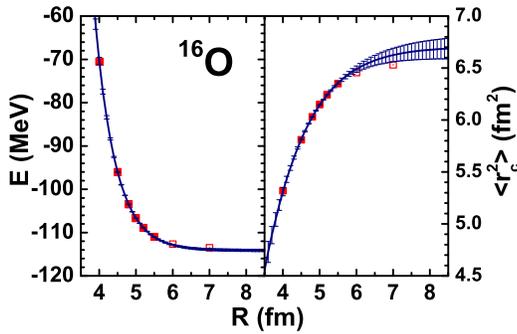}
\caption{(Color online) Extrapolation of total energy and charge radius with respect to the box size.
Solid points are included in the fit of Eq.~(\ref{eq:fit}), whereas open points are not.
The uncertainties are evaluated by the Jackknife resampling method.
\label{fig13}}
\end{figure}

We perform the RBHF calculations for $^{16}$O with several different box sizes from $R=4$ to $R=7$~fm.
The results are shown in Fig.~\ref{fig13}, where squares symbols represent the results calculated by RBHF theory with Bonn A interaction.
The extrapolation of total energy and charge radius is carried out by using the calculated points from $R=4$ to $5.5$~fm indicated with the solid symbols.
The corresponding uncertainties are evaluated by the Jackknife resampling method \cite{Efron1982}.
It is clearly seen that the calculated points with open symbols, which are not included in the fit of Eq.~(\ref{eq:fit}), are well located on the lines of extrapolation.

The extrapolated values of total energy and charge radius of $^{16}$O by the RBHF theory with Bonn A interaction are
\begin{subequations}
\begin{align}
E = & -114.138 \pm 0.265 ~{\rm MeV},\\
\la r_c\ra = & 2.589 \pm 0.023 ~{\rm fm},
\end{align}
\end{subequations}
where the uncertainty comes from Jackknife resampling.

\end{CJK}

\begin{thebibliography}{114}
\expandafter\ifx\csname natexlab\endcsname\relax\def\natexlab#1{#1}\fi
\expandafter\ifx\csname bibnamefont\endcsname\relax
  \def\bibnamefont#1{#1}\fi
\expandafter\ifx\csname bibfnamefont\endcsname\relax
  \def\bibfnamefont#1{#1}\fi
\expandafter\ifx\csname citenamefont\endcsname\relax
  \def\citenamefont#1{#1}\fi
\expandafter\ifx\csname url\endcsname\relax
  \def\url#1{\texttt{#1}}\fi
\expandafter\ifx\csname urlprefix\endcsname\relax\def\urlprefix{URL }\fi
\providecommand{\bibinfo}[2]{#2}
\providecommand{\eprint}[2][]{\url{#2}}

\bibitem[{\citenamefont{Jastrow}(1951)}]{Jastrow1951}
\bibinfo{author}{\bibfnamefont{R.}~\bibnamefont{Jastrow}},
  \bibinfo{journal}{Phys. Rev.} \textbf{\bibinfo{volume}{81}},
  \bibinfo{pages}{165} (\bibinfo{year}{1951}).

\bibitem[{\citenamefont{Brueckner et~al.}(1954)\citenamefont{Brueckner,
  Levinson, and Mahmoud}}]{Brueckner1954}
\bibinfo{author}{\bibfnamefont{K.}~\bibnamefont{Brueckner}},
  \bibinfo{author}{\bibfnamefont{C.}~\bibnamefont{Levinson}}, \bibnamefont{and}
  \bibinfo{author}{\bibfnamefont{H.}~\bibnamefont{Mahmoud}},
  \bibinfo{journal}{Phys. Rev.} \textbf{\bibinfo{volume}{95}},
  \bibinfo{pages}{217} (\bibinfo{year}{1954}).

\bibitem[{\citenamefont{Jastrow}(1955)}]{Jastrow1955}
\bibinfo{author}{\bibfnamefont{R.}~\bibnamefont{Jastrow}},
  \bibinfo{journal}{Phys. Rev.} \textbf{\bibinfo{volume}{98}},
  \bibinfo{pages}{1479} (\bibinfo{year}{1955}).

\bibitem[{\citenamefont{Suzuki and Lee}(1980)}]{Suzuki1980}
\bibinfo{author}{\bibfnamefont{K.}~\bibnamefont{Suzuki}} \bibnamefont{and}
  \bibinfo{author}{\bibfnamefont{S.~Y.} \bibnamefont{Lee}},
  \bibinfo{journal}{Prog. Theor. Phys.} \textbf{\bibinfo{volume}{64}},
  \bibinfo{pages}{2091} (\bibinfo{year}{1980}).

\bibitem[{\citenamefont{Feldmeier et~al.}(1998)\citenamefont{Feldmeier, Neff,
  Roth, and Schnack}}]{Feldmeier1998}
\bibinfo{author}{\bibfnamefont{H.}~\bibnamefont{Feldmeier}},
  \bibinfo{author}{\bibfnamefont{T.}~\bibnamefont{Neff}},
  \bibinfo{author}{\bibfnamefont{R.}~\bibnamefont{Roth}}, \bibnamefont{and}
  \bibinfo{author}{\bibfnamefont{J.}~\bibnamefont{Schnack}},
  \bibinfo{journal}{Nucl. Phys. A} \textbf{\bibinfo{volume}{632}},
  \bibinfo{pages}{61} (\bibinfo{year}{1998}).

\bibitem[{\citenamefont{Bogner et~al.}(2002)\citenamefont{Bogner, Kuo,
  Coraggio, Covello, and Itaco}}]{Bogner2002}
\bibinfo{author}{\bibfnamefont{S.}~\bibnamefont{Bogner}},
  \bibinfo{author}{\bibfnamefont{T.~T.~S.} \bibnamefont{Kuo}},
  \bibinfo{author}{\bibfnamefont{L.}~\bibnamefont{Coraggio}},
  \bibinfo{author}{\bibfnamefont{A.}~\bibnamefont{Covello}}, \bibnamefont{and}
  \bibinfo{author}{\bibfnamefont{N.}~\bibnamefont{Itaco}},
  \bibinfo{journal}{Phys. Rev. C} \textbf{\bibinfo{volume}{65}},
  \bibinfo{pages}{051301} (\bibinfo{year}{2002}).

\bibitem[{\citenamefont{Bogner et~al.}(2007)\citenamefont{Bogner, Furnstahl,
  and Perry}}]{Bogner2007}
\bibinfo{author}{\bibfnamefont{S.~K.} \bibnamefont{Bogner}},
  \bibinfo{author}{\bibfnamefont{R.~J.} \bibnamefont{Furnstahl}},
  \bibnamefont{and} \bibinfo{author}{\bibfnamefont{R.~J.} \bibnamefont{Perry}},
  \bibinfo{journal}{Phys. Rev. C} \textbf{\bibinfo{volume}{75}},
  \bibinfo{pages}{061001} (\bibinfo{year}{2007}).

\bibitem[{\citenamefont{Stoks et~al.}(1994)\citenamefont{Stoks, Klomp,
  Terheggen, and de~Swart}}]{Stoks1994}
\bibinfo{author}{\bibfnamefont{V.}~\bibnamefont{Stoks}},
  \bibinfo{author}{\bibfnamefont{R.}~\bibnamefont{Klomp}},
  \bibinfo{author}{\bibfnamefont{C.}~\bibnamefont{Terheggen}},
  \bibnamefont{and} \bibinfo{author}{\bibfnamefont{J.}~\bibnamefont{de~Swart}},
  \bibinfo{journal}{Phys. Rev. C} \textbf{\bibinfo{volume}{49}},
  \bibinfo{pages}{2950} (\bibinfo{year}{1994}).

\bibitem[{\citenamefont{Wiringa et~al.}(1995)\citenamefont{Wiringa, Stoks, and
  Schiavilla}}]{Wiringa1995}
\bibinfo{author}{\bibfnamefont{R.}~\bibnamefont{Wiringa}},
  \bibinfo{author}{\bibfnamefont{V.}~\bibnamefont{Stoks}}, \bibnamefont{and}
  \bibinfo{author}{\bibfnamefont{R.}~\bibnamefont{Schiavilla}},
  \bibinfo{journal}{Phys. Rev. C} \textbf{\bibinfo{volume}{51}},
  \bibinfo{pages}{38} (\bibinfo{year}{1995}).

\bibitem[{\citenamefont{Machleidt}(2001)}]{Machleidt2001}
\bibinfo{author}{\bibfnamefont{R.}~\bibnamefont{Machleidt}},
  \bibinfo{journal}{Phys. Rev. C} \textbf{\bibinfo{volume}{63}},
  \bibinfo{pages}{024001} (\bibinfo{year}{2001}).

\bibitem[{\citenamefont{Epelbaum
  et~al.}(2009{\natexlab{a}})\citenamefont{Epelbaum, Hammer, and
  Mei{\ss}ner}}]{Epelbaum2009}
\bibinfo{author}{\bibfnamefont{E.}~\bibnamefont{Epelbaum}},
  \bibinfo{author}{\bibfnamefont{H.-W.} \bibnamefont{Hammer}},
  \bibnamefont{and} \bibinfo{author}{\bibfnamefont{U.-G.}
  \bibnamefont{Mei{\ss}ner}}, \bibinfo{journal}{Rev. Mod. Phys.}
  \textbf{\bibinfo{volume}{81}}, \bibinfo{pages}{1773}
  (\bibinfo{year}{2009}{\natexlab{a}}).

\bibitem[{\citenamefont{Machleidt and Entem}(2011)}]{Machleidt2011}
\bibinfo{author}{\bibfnamefont{R.}~\bibnamefont{Machleidt}} \bibnamefont{and}
  \bibinfo{author}{\bibfnamefont{D.}~\bibnamefont{Entem}},
  \bibinfo{journal}{Phys. Rep.} \textbf{\bibinfo{volume}{503}},
  \bibinfo{pages}{1} (\bibinfo{year}{2011}).

\bibitem[{\citenamefont{Carlson et~al.}(2015)\citenamefont{Carlson, Gandolfi,
  Pederiva, Pieper, Schiavilla, Schmidt, and Wiringa}}]{Carlson2015}
\bibinfo{author}{\bibfnamefont{J.}~\bibnamefont{Carlson}},
  \bibinfo{author}{\bibfnamefont{S.}~\bibnamefont{Gandolfi}},
  \bibinfo{author}{\bibfnamefont{F.}~\bibnamefont{Pederiva}},
  \bibinfo{author}{\bibfnamefont{S.~C.} \bibnamefont{Pieper}},
  \bibinfo{author}{\bibfnamefont{R.}~\bibnamefont{Schiavilla}},
  \bibinfo{author}{\bibfnamefont{K.~E.} \bibnamefont{Schmidt}},
  \bibnamefont{and} \bibinfo{author}{\bibfnamefont{R.~B.}
  \bibnamefont{Wiringa}}, \bibinfo{journal}{Rev. Mod. Phys.}
  \textbf{\bibinfo{volume}{87}}, \bibinfo{pages}{1067} (\bibinfo{year}{2015}).

\bibitem[{\citenamefont{Hagen et~al.}(2014)\citenamefont{Hagen, Papenbrock,
  Hjorth-Jensen, and Dean}}]{Hagen2014}
\bibinfo{author}{\bibfnamefont{G.}~\bibnamefont{Hagen}},
  \bibinfo{author}{\bibfnamefont{T.}~\bibnamefont{Papenbrock}},
  \bibinfo{author}{\bibfnamefont{M.}~\bibnamefont{Hjorth-Jensen}},
  \bibnamefont{and} \bibinfo{author}{\bibfnamefont{D.~J.} \bibnamefont{Dean}},
  \bibinfo{journal}{Reports Prog. Phys.} \textbf{\bibinfo{volume}{77}},
  \bibinfo{pages}{096302} (\bibinfo{year}{2014}).

\bibitem[{\citenamefont{Barrett et~al.}(2013)\citenamefont{Barrett,
  Navr{\'{a}}til, and Vary}}]{Barrett2013}
\bibinfo{author}{\bibfnamefont{B.~R.} \bibnamefont{Barrett}},
  \bibinfo{author}{\bibfnamefont{P.}~\bibnamefont{Navr{\'{a}}til}},
  \bibnamefont{and} \bibinfo{author}{\bibfnamefont{J.~P.} \bibnamefont{Vary}},
  \bibinfo{journal}{Prog. Part. Nucl. Phys.} \textbf{\bibinfo{volume}{69}},
  \bibinfo{pages}{131} (\bibinfo{year}{2013}).

\bibitem[{\citenamefont{Dickhoff and Barbieri}(2004)}]{Dickhoff2004}
\bibinfo{author}{\bibfnamefont{W.}~\bibnamefont{Dickhoff}} \bibnamefont{and}
  \bibinfo{author}{\bibfnamefont{C.}~\bibnamefont{Barbieri}},
  \bibinfo{journal}{Prog. Part. Nucl. Phys.} \textbf{\bibinfo{volume}{52}},
  \bibinfo{pages}{377} (\bibinfo{year}{2004}).

\bibitem[{\citenamefont{Lee}(2009)}]{Lee2009}
\bibinfo{author}{\bibfnamefont{D.}~\bibnamefont{Lee}}, \bibinfo{journal}{Prog.
  Part. Nucl. Phys.} \textbf{\bibinfo{volume}{63}}, \bibinfo{pages}{117}
  (\bibinfo{year}{2009}).

\bibitem[{\citenamefont{Hergert et~al.}(2016)\citenamefont{Hergert, Bogner,
  Morris, Schwenk, and Tsukiyama}}]{Hergert2016}
\bibinfo{author}{\bibfnamefont{H.}~\bibnamefont{Hergert}},
  \bibinfo{author}{\bibfnamefont{S.}~\bibnamefont{Bogner}},
  \bibinfo{author}{\bibfnamefont{T.}~\bibnamefont{Morris}},
  \bibinfo{author}{\bibfnamefont{A.}~\bibnamefont{Schwenk}}, \bibnamefont{and}
  \bibinfo{author}{\bibfnamefont{K.}~\bibnamefont{Tsukiyama}},
  \bibinfo{journal}{Phys. Rep.} \textbf{\bibinfo{volume}{621}},
  \bibinfo{pages}{165} (\bibinfo{year}{2016}).

\bibitem[{\citenamefont{Liu et~al.}(2012)\citenamefont{Liu, Otsuka, Shimizu,
  Utsuno, and Roth}}]{Liu2012}
\bibinfo{author}{\bibfnamefont{L.}~\bibnamefont{Liu}},
  \bibinfo{author}{\bibfnamefont{T.}~\bibnamefont{Otsuka}},
  \bibinfo{author}{\bibfnamefont{N.}~\bibnamefont{Shimizu}},
  \bibinfo{author}{\bibfnamefont{Y.}~\bibnamefont{Utsuno}}, \bibnamefont{and}
  \bibinfo{author}{\bibfnamefont{R.}~\bibnamefont{Roth}},
  \bibinfo{journal}{Phys. Rev. C} \textbf{\bibinfo{volume}{86}},
  \bibinfo{pages}{014302} (\bibinfo{year}{2012}).

\bibitem[{\citenamefont{Day}(1967)}]{Day1967}
\bibinfo{author}{\bibfnamefont{B.~D.} \bibnamefont{Day}},
  \bibinfo{journal}{Rev. Mod. Phys.} \textbf{\bibinfo{volume}{39}},
  \bibinfo{pages}{719} (\bibinfo{year}{1967}).

\bibitem[{\citenamefont{Goldstone}(1957)}]{Goldstone1957}
\bibinfo{author}{\bibfnamefont{J.}~\bibnamefont{Goldstone}},
  \bibinfo{journal}{Proc. R. Soc. A Math. Phys. Eng. Sci.}
  \textbf{\bibinfo{volume}{239}}, \bibinfo{pages}{267} (\bibinfo{year}{1957}).

\bibitem[{\citenamefont{Bethe and Goldstone}(1957)}]{Bethe1957}
\bibinfo{author}{\bibfnamefont{H.~A.} \bibnamefont{Bethe}} \bibnamefont{and}
  \bibinfo{author}{\bibfnamefont{J.}~\bibnamefont{Goldstone}},
  \bibinfo{journal}{Proc. R. Soc. A Math. Phys. Eng. Sci.}
  \textbf{\bibinfo{volume}{238}}, \bibinfo{pages}{551} (\bibinfo{year}{1957}).

\bibitem[{\citenamefont{Bethe et~al.}(1963)\citenamefont{Bethe, Brandow, and
  Petschek}}]{Bethe1963}
\bibinfo{author}{\bibfnamefont{H.}~\bibnamefont{Bethe}},
  \bibinfo{author}{\bibfnamefont{B.}~\bibnamefont{Brandow}}, \bibnamefont{and}
  \bibinfo{author}{\bibfnamefont{A.}~\bibnamefont{Petschek}},
  \bibinfo{journal}{Phys. Rev.} \textbf{\bibinfo{volume}{129}},
  \bibinfo{pages}{225} (\bibinfo{year}{1963}).

\bibitem[{\citenamefont{Rajaraman and Bethe}(1967)}]{Rajaraman1967}
\bibinfo{author}{\bibfnamefont{R.}~\bibnamefont{Rajaraman}} \bibnamefont{and}
  \bibinfo{author}{\bibfnamefont{H.}~\bibnamefont{Bethe}},
  \bibinfo{journal}{Rev. Mod. Phys.} \textbf{\bibinfo{volume}{39}},
  \bibinfo{pages}{745} (\bibinfo{year}{1967}).

\bibitem[{\citenamefont{Baranger}(1969)}]{Baranger1969}
\bibinfo{author}{\bibfnamefont{M.}~\bibnamefont{Baranger}}, in
  \emph{\bibinfo{booktitle}{Nuclear Structure and Nuclear Reactions,
  Proceedings of the International School of Physics "Enrico Fermi", Course
  XL}}, edited by \bibinfo{editor}{\bibfnamefont{M.}~\bibnamefont{Jean}}
  (\bibinfo{publisher}{Academic Press}, \bibinfo{address}{New York},
  \bibinfo{year}{1969}), pp. \bibinfo{pages}{511--614}.

\bibitem[{\citenamefont{Coester et~al.}(1970)\citenamefont{Coester, Cohen, Day,
  and Vincent}}]{Coester1970}
\bibinfo{author}{\bibfnamefont{F.}~\bibnamefont{Coester}},
  \bibinfo{author}{\bibfnamefont{S.}~\bibnamefont{Cohen}},
  \bibinfo{author}{\bibfnamefont{B.}~\bibnamefont{Day}}, \bibnamefont{and}
  \bibinfo{author}{\bibfnamefont{C.}~\bibnamefont{Vincent}},
  \bibinfo{journal}{Phys. Rev. C} \textbf{\bibinfo{volume}{1}},
  \bibinfo{pages}{769} (\bibinfo{year}{1970}).

\bibitem[{\citenamefont{Fujita and Miyazawa}(1957)}]{Fujita1957}
\bibinfo{author}{\bibfnamefont{J.-i.} \bibnamefont{Fujita}} \bibnamefont{and}
  \bibinfo{author}{\bibfnamefont{H.}~\bibnamefont{Miyazawa}},
  \bibinfo{journal}{Prog. Theor. Phys.} \textbf{\bibinfo{volume}{17}},
  \bibinfo{pages}{360} (\bibinfo{year}{1957}).

\bibitem[{\citenamefont{Brown and Green}(1969)}]{Brown1969a}
\bibinfo{author}{\bibfnamefont{G.}~\bibnamefont{Brown}} \bibnamefont{and}
  \bibinfo{author}{\bibfnamefont{A.}~\bibnamefont{Green}},
  \bibinfo{journal}{Nucl. Phys. A} \textbf{\bibinfo{volume}{137}},
  \bibinfo{pages}{1} (\bibinfo{year}{1969}).

\bibitem[{\citenamefont{Zuo et~al.}(2002)\citenamefont{Zuo, Lejeune, Lombardo,
  and Mathiot}}]{Zuo2002}
\bibinfo{author}{\bibfnamefont{W.}~\bibnamefont{Zuo}},
  \bibinfo{author}{\bibfnamefont{A.}~\bibnamefont{Lejeune}},
  \bibinfo{author}{\bibfnamefont{U.}~\bibnamefont{Lombardo}}, \bibnamefont{and}
  \bibinfo{author}{\bibfnamefont{J.}~\bibnamefont{Mathiot}},
  \bibinfo{journal}{Nucl. Phys. A} \textbf{\bibinfo{volume}{706}},
  \bibinfo{pages}{418} (\bibinfo{year}{2002}).

\bibitem[{\citenamefont{Pieper and Wiringa}(2001)}]{Pieper2001}
\bibinfo{author}{\bibfnamefont{S.~C.} \bibnamefont{Pieper}} \bibnamefont{and}
  \bibinfo{author}{\bibfnamefont{R.~B.} \bibnamefont{Wiringa}},
  \bibinfo{journal}{Annu. Rev. Nucl. Part. Sci.} \textbf{\bibinfo{volume}{51}},
  \bibinfo{pages}{53} (\bibinfo{year}{2001}).

\bibitem[{\citenamefont{Johnson and Teller}(1955)}]{Johnson1955}
\bibinfo{author}{\bibfnamefont{M.~H.} \bibnamefont{Johnson}} \bibnamefont{and}
  \bibinfo{author}{\bibfnamefont{E.}~\bibnamefont{Teller}},
  \bibinfo{journal}{Phys. Rev.} \textbf{\bibinfo{volume}{98}},
  \bibinfo{pages}{783} (\bibinfo{year}{1955}).

\bibitem[{\citenamefont{Duerr}(1956)}]{Duerr1956}
\bibinfo{author}{\bibfnamefont{H.-P.} \bibnamefont{Duerr}},
  \bibinfo{journal}{Phys. Rev.} \textbf{\bibinfo{volume}{103}},
  \bibinfo{pages}{469} (\bibinfo{year}{1956}).

\bibitem[{\citenamefont{Goeppert-Mayer and Jensen}(1955)}]{GoeppertMayer1955}
\bibinfo{author}{\bibfnamefont{M.}~\bibnamefont{Goeppert-Mayer}}
  \bibnamefont{and} \bibinfo{author}{\bibfnamefont{J.~H.~D.}
  \bibnamefont{Jensen}}, \emph{\bibinfo{title}{Elementary Theory of Nuclear
  Shell Structure}} (\bibinfo{publisher}{John Wiley \& Sons},
  \bibinfo{address}{New York}, \bibinfo{year}{1955}).

\bibitem[{\citenamefont{Rozsnyai}(1961)}]{Rozsnyai1961}
\bibinfo{author}{\bibfnamefont{B.}~\bibnamefont{Rozsnyai}},
  \bibinfo{journal}{Phys. Rev.} \textbf{\bibinfo{volume}{124}},
  \bibinfo{pages}{860} (\bibinfo{year}{1961}).

\bibitem[{\citenamefont{Green and Sawada}(1967)}]{Green1967}
\bibinfo{author}{\bibfnamefont{A.~E.~S.} \bibnamefont{Green}} \bibnamefont{and}
  \bibinfo{author}{\bibfnamefont{T.}~\bibnamefont{Sawada}},
  \bibinfo{journal}{Rev. Mod. Phys.} \textbf{\bibinfo{volume}{39}},
  \bibinfo{pages}{594} (\bibinfo{year}{1967}).

\bibitem[{\citenamefont{Miller and Green}(1972)}]{Miller1972}
\bibinfo{author}{\bibfnamefont{L.}~\bibnamefont{Miller}} \bibnamefont{and}
  \bibinfo{author}{\bibfnamefont{A.}~\bibnamefont{Green}},
  \bibinfo{journal}{Phys. Rev. C} \textbf{\bibinfo{volume}{5}},
  \bibinfo{pages}{241} (\bibinfo{year}{1972}).

\bibitem[{\citenamefont{Walecka}(1974)}]{Walecka1974}
\bibinfo{author}{\bibfnamefont{J.}~\bibnamefont{Walecka}},
  \bibinfo{journal}{Ann. Phys.} \textbf{\bibinfo{volume}{83}},
  \bibinfo{pages}{491} (\bibinfo{year}{1974}).

\bibitem[{\citenamefont{Meng}(2016)}]{Meng2016}
\bibinfo{editor}{\bibfnamefont{J.}~\bibnamefont{Meng}}, ed.,
  \emph{\bibinfo{title}{{Relativistic Density Functional for Nuclear
  Structure}}} (\bibinfo{publisher}{World Scientific Pub.},
  \bibinfo{year}{2016}).

\bibitem[{\citenamefont{Anastasio et~al.}(1980)\citenamefont{Anastasio,
  Celenza, and Shakin}}]{Anastasio1980}
\bibinfo{author}{\bibfnamefont{M.}~\bibnamefont{Anastasio}},
  \bibinfo{author}{\bibfnamefont{L.}~\bibnamefont{Celenza}}, \bibnamefont{and}
  \bibinfo{author}{\bibfnamefont{C.}~\bibnamefont{Shakin}},
  \bibinfo{journal}{Phys. Rev. Lett.} \textbf{\bibinfo{volume}{45}},
  \bibinfo{pages}{2096} (\bibinfo{year}{1980}).

\bibitem[{\citenamefont{Anastasio et~al.}(1983)\citenamefont{Anastasio,
  Celenza, Pong, and Shakin}}]{Anastasio1983}
\bibinfo{author}{\bibfnamefont{M.}~\bibnamefont{Anastasio}},
  \bibinfo{author}{\bibfnamefont{L.}~\bibnamefont{Celenza}},
  \bibinfo{author}{\bibfnamefont{W.}~\bibnamefont{Pong}}, \bibnamefont{and}
  \bibinfo{author}{\bibfnamefont{C.}~\bibnamefont{Shakin}},
  \bibinfo{journal}{Phys. Rep.} \textbf{\bibinfo{volume}{100}},
  \bibinfo{pages}{327} (\bibinfo{year}{1983}).

\bibitem[{\citenamefont{Horowitz and Serot}(1984)}]{Horowitz1984}
\bibinfo{author}{\bibfnamefont{C.}~\bibnamefont{Horowitz}} \bibnamefont{and}
  \bibinfo{author}{\bibfnamefont{B.~D.} \bibnamefont{Serot}},
  \bibinfo{journal}{Phys. Lett. B} \textbf{\bibinfo{volume}{137}},
  \bibinfo{pages}{287} (\bibinfo{year}{1984}).

\bibitem[{\citenamefont{Horowitz and Serot}(1987)}]{Horowitz1987}
\bibinfo{author}{\bibfnamefont{C.}~\bibnamefont{Horowitz}} \bibnamefont{and}
  \bibinfo{author}{\bibfnamefont{B.~D.} \bibnamefont{Serot}},
  \bibinfo{journal}{Nucl. Phys. A} \textbf{\bibinfo{volume}{464}},
  \bibinfo{pages}{613} (\bibinfo{year}{1987}).

\bibitem[{\citenamefont{Brockmann and Machleidt}(1984)}]{Brockmann1984}
\bibinfo{author}{\bibfnamefont{R.}~\bibnamefont{Brockmann}} \bibnamefont{and}
  \bibinfo{author}{\bibfnamefont{R.}~\bibnamefont{Machleidt}},
  \bibinfo{journal}{Phys. Lett. B} \textbf{\bibinfo{volume}{149}},
  \bibinfo{pages}{283} (\bibinfo{year}{1984}).

\bibitem[{\citenamefont{Brockmann and Machleidt}(1990)}]{Brockmann1990}
\bibinfo{author}{\bibfnamefont{R.}~\bibnamefont{Brockmann}} \bibnamefont{and}
  \bibinfo{author}{\bibfnamefont{R.}~\bibnamefont{Machleidt}},
  \bibinfo{journal}{Phys. Rev. C} \textbf{\bibinfo{volume}{42}},
  \bibinfo{pages}{1965} (\bibinfo{year}{1990}).

\bibitem[{\citenamefont{ter Haar and Malfliet}(1986)}]{TerHaar1986}
\bibinfo{author}{\bibfnamefont{B.}~\bibnamefont{ter Haar}} \bibnamefont{and}
  \bibinfo{author}{\bibfnamefont{R.}~\bibnamefont{Malfliet}},
  \bibinfo{journal}{Phys. Rev. Lett.} \textbf{\bibinfo{volume}{56}},
  \bibinfo{pages}{1237} (\bibinfo{year}{1986}).

\bibitem[{\citenamefont{Haar and Malfliet}(1987)}]{Haar1987}
\bibinfo{author}{\bibfnamefont{B.~T.} \bibnamefont{Haar}} \bibnamefont{and}
  \bibinfo{author}{\bibfnamefont{R.}~\bibnamefont{Malfliet}},
  \bibinfo{journal}{Phys. Rep.} \textbf{\bibinfo{volume}{149}},
  \bibinfo{pages}{207} (\bibinfo{year}{1987}).

\bibitem[{\citenamefont{Brown et~al.}(1987)\citenamefont{Brown, Weise, Baym,
  and Speth}}]{Brown1987}
\bibinfo{author}{\bibfnamefont{G.~E.} \bibnamefont{Brown}},
  \bibinfo{author}{\bibfnamefont{W.}~\bibnamefont{Weise}},
  \bibinfo{author}{\bibfnamefont{G.}~\bibnamefont{Baym}}, \bibnamefont{and}
  \bibinfo{author}{\bibfnamefont{J.}~\bibnamefont{Speth}},
  \bibinfo{journal}{Comments Nucl. Part. Phys.} \textbf{\bibinfo{volume}{17}},
  \bibinfo{pages}{39} (\bibinfo{year}{1987}).

\bibitem[{\citenamefont{Machleidt}(1989)}]{Machleidt1989}
\bibinfo{author}{\bibfnamefont{R.}~\bibnamefont{Machleidt}}, in
  \emph{\bibinfo{booktitle}{Adv. Nucl. Phys.}}, edited by
  \bibinfo{editor}{\bibfnamefont{J.~W.} \bibnamefont{Negele}} \bibnamefont{and}
  \bibinfo{editor}{\bibfnamefont{E.}~\bibnamefont{Vogt}}
  (\bibinfo{publisher}{Springer US}, \bibinfo{year}{1989}),
  vol.~\bibinfo{volume}{19} of \emph{\bibinfo{series}{Advances in Nuclear
  Physics}}, pp. \bibinfo{pages}{189--376}.

\bibitem[{\citenamefont{Nuppenau et~al.}(1990)\citenamefont{Nuppenau,
  Mackellar, and Lee}}]{Nuppenau1990}
\bibinfo{author}{\bibfnamefont{C.}~\bibnamefont{Nuppenau}},
  \bibinfo{author}{\bibfnamefont{A.}~\bibnamefont{Mackellar}},
  \bibnamefont{and} \bibinfo{author}{\bibfnamefont{Y.}~\bibnamefont{Lee}},
  \bibinfo{journal}{Nucl. Phys. A} \textbf{\bibinfo{volume}{511}},
  \bibinfo{pages}{525} (\bibinfo{year}{1990}).

\bibitem[{\citenamefont{Xu et~al.}(2016)\citenamefont{Xu, Ma, Zhang, Tian, van
  Dalen, and M{\"{u}}ther}}]{Xu2016}
\bibinfo{author}{\bibfnamefont{R.}~\bibnamefont{Xu}},
  \bibinfo{author}{\bibfnamefont{Z.}~\bibnamefont{Ma}},
  \bibinfo{author}{\bibfnamefont{Y.}~\bibnamefont{Zhang}},
  \bibinfo{author}{\bibfnamefont{Y.}~\bibnamefont{Tian}},
  \bibinfo{author}{\bibfnamefont{E.~N.~E.} \bibnamefont{van Dalen}},
  \bibnamefont{and}
  \bibinfo{author}{\bibfnamefont{H.}~\bibnamefont{M{\"{u}}ther}},
  \bibinfo{journal}{Phys. Rev. C} \textbf{\bibinfo{volume}{94}},
  \bibinfo{pages}{034606} (\bibinfo{year}{2016}).

\bibitem[{\citenamefont{Huber et~al.}(1995)\citenamefont{Huber, Weber, and
  Weigel}}]{Huber1995}
\bibinfo{author}{\bibfnamefont{H.}~\bibnamefont{Huber}},
  \bibinfo{author}{\bibfnamefont{F.}~\bibnamefont{Weber}}, \bibnamefont{and}
  \bibinfo{author}{\bibfnamefont{M.~K.} \bibnamefont{Weigel}},
  \bibinfo{journal}{Phys. Rev. C} \textbf{\bibinfo{volume}{51}},
  \bibinfo{pages}{1790} (\bibinfo{year}{1995}).

\bibitem[{\citenamefont{van Dalen et~al.}(2004)\citenamefont{van Dalen, Fuchs,
  and Faessler}}]{VanDalen2004}
\bibinfo{author}{\bibfnamefont{E.}~\bibnamefont{van Dalen}},
  \bibinfo{author}{\bibfnamefont{C.}~\bibnamefont{Fuchs}}, \bibnamefont{and}
  \bibinfo{author}{\bibfnamefont{A.}~\bibnamefont{Faessler}},
  \bibinfo{journal}{Nucl. Phys. A} \textbf{\bibinfo{volume}{744}},
  \bibinfo{pages}{227} (\bibinfo{year}{2004}).

\bibitem[{\citenamefont{Huber et~al.}(1996)\citenamefont{Huber, Weber, and
  Weigel}}]{Huber1996}
\bibinfo{author}{\bibfnamefont{H.}~\bibnamefont{Huber}},
  \bibinfo{author}{\bibfnamefont{F.}~\bibnamefont{Weber}}, \bibnamefont{and}
  \bibinfo{author}{\bibfnamefont{M.}~\bibnamefont{Weigel}},
  \bibinfo{journal}{Nucl. Phys. A} \textbf{\bibinfo{volume}{596}},
  \bibinfo{pages}{684} (\bibinfo{year}{1996}).

\bibitem[{\citenamefont{Katayama and Saito}(2013)}]{Katayama2013}
\bibinfo{author}{\bibfnamefont{T.}~\bibnamefont{Katayama}} \bibnamefont{and}
  \bibinfo{author}{\bibfnamefont{K.}~\bibnamefont{Saito}},
  \bibinfo{journal}{Phys. Rev. C} \textbf{\bibinfo{volume}{88}},
  \bibinfo{pages}{035805} (\bibinfo{year}{2013}).

\bibitem[{\citenamefont{M{\"{u}}ther et~al.}(2017)\citenamefont{M{\"{u}}ther,
  Sammarruca, and Ma}}]{Muther2017}
\bibinfo{author}{\bibfnamefont{H.}~\bibnamefont{M{\"{u}}ther}},
  \bibinfo{author}{\bibfnamefont{F.}~\bibnamefont{Sammarruca}},
  \bibnamefont{and} \bibinfo{author}{\bibfnamefont{Z.}~\bibnamefont{Ma}},
  \bibinfo{journal}{Int. J. Mod. Phys. E} \textbf{\bibinfo{volume}{26}},
  \bibinfo{pages}{1730001} (\bibinfo{year}{2017}).

\bibitem[{\citenamefont{M{\"{u}}ther et~al.}(1988)\citenamefont{M{\"{u}}ther,
  Machleidt, and Brockmann}}]{Muether1988}
\bibinfo{author}{\bibfnamefont{H.}~\bibnamefont{M{\"{u}}ther}},
  \bibinfo{author}{\bibfnamefont{R.}~\bibnamefont{Machleidt}},
  \bibnamefont{and}
  \bibinfo{author}{\bibfnamefont{R.}~\bibnamefont{Brockmann}},
  \bibinfo{journal}{Phys. Lett. B} \textbf{\bibinfo{volume}{202}},
  \bibinfo{pages}{483} (\bibinfo{year}{1988}).

\bibitem[{\citenamefont{Celenza et~al.}(1990)\citenamefont{Celenza, Gao, and
  Shakin}}]{Celenza1990}
\bibinfo{author}{\bibfnamefont{L.~S.} \bibnamefont{Celenza}},
  \bibinfo{author}{\bibfnamefont{S.-F.} \bibnamefont{Gao}}, \bibnamefont{and}
  \bibinfo{author}{\bibfnamefont{C.~M.} \bibnamefont{Shakin}},
  \bibinfo{journal}{Phys. Rev. C} \textbf{\bibinfo{volume}{41}},
  \bibinfo{pages}{1768} (\bibinfo{year}{1990}).

\bibitem[{\citenamefont{Marcos et~al.}(1991)\citenamefont{Marcos,
  L{\'{o}}pez-Quelle, and {Van Giai}}}]{Marcos1991}
\bibinfo{author}{\bibfnamefont{S.}~\bibnamefont{Marcos}},
  \bibinfo{author}{\bibfnamefont{M.}~\bibnamefont{L{\'{o}}pez-Quelle}},
  \bibnamefont{and} \bibinfo{author}{\bibfnamefont{N.}~\bibnamefont{{Van
  Giai}}}, \bibinfo{journal}{Phys. Lett. B} \textbf{\bibinfo{volume}{257}},
  \bibinfo{pages}{5} (\bibinfo{year}{1991}).

\bibitem[{\citenamefont{Gmuca}(1991)}]{Gmuca1991}
\bibinfo{author}{\bibfnamefont{S.}~\bibnamefont{Gmuca}}, \bibinfo{journal}{J.
  Phys. G Nucl. Part. Phys.} \textbf{\bibinfo{volume}{17}},
  \bibinfo{pages}{1115} (\bibinfo{year}{1991}).

\bibitem[{\citenamefont{Brockmann and Toki}(1992)}]{Brockmann1992}
\bibinfo{author}{\bibfnamefont{R.}~\bibnamefont{Brockmann}} \bibnamefont{and}
  \bibinfo{author}{\bibfnamefont{H.}~\bibnamefont{Toki}},
  \bibinfo{journal}{Phys. Rev. Lett.} \textbf{\bibinfo{volume}{68}},
  \bibinfo{pages}{3408} (\bibinfo{year}{1992}).

\bibitem[{\citenamefont{Fritz et~al.}(1993)\citenamefont{Fritz, M{\"{u}}ther,
  and Machleidt}}]{Fritz1993}
\bibinfo{author}{\bibfnamefont{R.}~\bibnamefont{Fritz}},
  \bibinfo{author}{\bibfnamefont{H.}~\bibnamefont{M{\"{u}}ther}},
  \bibnamefont{and}
  \bibinfo{author}{\bibfnamefont{R.}~\bibnamefont{Machleidt}},
  \bibinfo{journal}{Phys. Rev. Lett.} \textbf{\bibinfo{volume}{71}},
  \bibinfo{pages}{46} (\bibinfo{year}{1993}).

\bibitem[{\citenamefont{{Van Giai} et~al.}(2010)\citenamefont{{Van Giai},
  Carlson, Ma, and Wolter}}]{VanGiai2010}
\bibinfo{author}{\bibfnamefont{N.}~\bibnamefont{{Van Giai}}},
  \bibinfo{author}{\bibfnamefont{B.~V.} \bibnamefont{Carlson}},
  \bibinfo{author}{\bibfnamefont{Z.}~\bibnamefont{Ma}}, \bibnamefont{and}
  \bibinfo{author}{\bibfnamefont{H.}~\bibnamefont{Wolter}},
  \bibinfo{journal}{J. Phys. G Nucl. Part. Phys.}
  \textbf{\bibinfo{volume}{37}}, \bibinfo{pages}{064043}
  (\bibinfo{year}{2010}).

\bibitem[{\citenamefont{Shen et~al.}(2016)\citenamefont{Shen, Hu, Liang, Meng,
  Ring, and Zhang}}]{Shen2016}
\bibinfo{author}{\bibfnamefont{S.-H.} \bibnamefont{Shen}},
  \bibinfo{author}{\bibfnamefont{J.-N.} \bibnamefont{Hu}},
  \bibinfo{author}{\bibfnamefont{H.-Z.} \bibnamefont{Liang}},
  \bibinfo{author}{\bibfnamefont{J.}~\bibnamefont{Meng}},
  \bibinfo{author}{\bibfnamefont{P.}~\bibnamefont{Ring}}, \bibnamefont{and}
  \bibinfo{author}{\bibfnamefont{S.-Q.} \bibnamefont{Zhang}},
  \bibinfo{journal}{Chin. Phys. Lett.} \textbf{\bibinfo{volume}{33}},
  \bibinfo{pages}{102103} (\bibinfo{year}{2016}).

\bibitem[{\citenamefont{Zhou et~al.}(2003)\citenamefont{Zhou, Meng, and
  Ring}}]{Zhou2003}
\bibinfo{author}{\bibfnamefont{S.-G.} \bibnamefont{Zhou}},
  \bibinfo{author}{\bibfnamefont{J.}~\bibnamefont{Meng}}, \bibnamefont{and}
  \bibinfo{author}{\bibfnamefont{P.}~\bibnamefont{Ring}},
  \bibinfo{journal}{Phys. Rev. C} \textbf{\bibinfo{volume}{68}},
  \bibinfo{pages}{034323} (\bibinfo{year}{2003}).

\bibitem[{\citenamefont{Schiller et~al.}(1999)\citenamefont{Schiller,
  M{\"{u}}ther, and Czerski}}]{Schiller1999}
\bibinfo{author}{\bibfnamefont{E.}~\bibnamefont{Schiller}},
  \bibinfo{author}{\bibfnamefont{H.}~\bibnamefont{M{\"{u}}ther}},
  \bibnamefont{and} \bibinfo{author}{\bibfnamefont{P.}~\bibnamefont{Czerski}},
  \bibinfo{journal}{Phys. Rev. C} \textbf{\bibinfo{volume}{59}},
  \bibinfo{pages}{2934} (\bibinfo{year}{1999}).

\bibitem[{\citenamefont{Suzuki et~al.}(2000)\citenamefont{Suzuki, Okamoto,
  Kohno, and Nagata}}]{Suzuki2000}
\bibinfo{author}{\bibfnamefont{K.}~\bibnamefont{Suzuki}},
  \bibinfo{author}{\bibfnamefont{R.}~\bibnamefont{Okamoto}},
  \bibinfo{author}{\bibfnamefont{M.}~\bibnamefont{Kohno}}, \bibnamefont{and}
  \bibinfo{author}{\bibfnamefont{S.}~\bibnamefont{Nagata}},
  \bibinfo{journal}{Nucl. Phys. A} \textbf{\bibinfo{volume}{665}},
  \bibinfo{pages}{92} (\bibinfo{year}{2000}).

\bibitem[{\citenamefont{Brockmann}(1978)}]{Brockmann1978}
\bibinfo{author}{\bibfnamefont{R.}~\bibnamefont{Brockmann}},
  \bibinfo{journal}{Phys. Rev. C} \textbf{\bibinfo{volume}{18}},
  \bibinfo{pages}{1510} (\bibinfo{year}{1978}).

\bibitem[{\citenamefont{Itzykson and Zuber}(1980)}]{Itzykson1980}
\bibinfo{author}{\bibfnamefont{C.}~\bibnamefont{Itzykson}} \bibnamefont{and}
  \bibinfo{author}{\bibfnamefont{J.-B.} \bibnamefont{Zuber}},
  \emph{\bibinfo{title}{Quantum Field Theory}}
  (\bibinfo{publisher}{McGraw-Hill}, \bibinfo{address}{New York},
  \bibinfo{year}{1980}).

\bibitem[{\citenamefont{Koepf and Ring}(1991)}]{Koepf1991}
\bibinfo{author}{\bibfnamefont{W.}~\bibnamefont{Koepf}} \bibnamefont{and}
  \bibinfo{author}{\bibfnamefont{P.}~\bibnamefont{Ring}},
  \bibinfo{journal}{Zeitschrift f\"ur Phys. A Hadron. Nucl.}
  \textbf{\bibinfo{volume}{339}}, \bibinfo{pages}{81} (\bibinfo{year}{1991}).

\bibitem[{\citenamefont{Long et~al.}(2006)\citenamefont{Long, {Van Giai}, and
  Meng}}]{Long2006}
\bibinfo{author}{\bibfnamefont{W.-H.} \bibnamefont{Long}},
  \bibinfo{author}{\bibfnamefont{N.}~\bibnamefont{{Van Giai}}},
  \bibnamefont{and} \bibinfo{author}{\bibfnamefont{J.}~\bibnamefont{Meng}},
  \bibinfo{journal}{Phys. Lett. B} \textbf{\bibinfo{volume}{640}},
  \bibinfo{pages}{150} (\bibinfo{year}{2006}).

\bibitem[{\citenamefont{Davies et~al.}(1969)\citenamefont{Davies, Baranger,
  Tarbutton, and Kuo}}]{Davies1969}
\bibinfo{author}{\bibfnamefont{K.}~\bibnamefont{Davies}},
  \bibinfo{author}{\bibfnamefont{M.}~\bibnamefont{Baranger}},
  \bibinfo{author}{\bibfnamefont{R.}~\bibnamefont{Tarbutton}},
  \bibnamefont{and} \bibinfo{author}{\bibfnamefont{T.}~\bibnamefont{Kuo}},
  \bibinfo{journal}{Phys. Rev.} \textbf{\bibinfo{volume}{177}},
  \bibinfo{pages}{1519} (\bibinfo{year}{1969}).

\bibitem[{\citenamefont{Song et~al.}(1998)\citenamefont{Song, Baldo,
  Giansiracusa, and Lombardo}}]{Song1998}
\bibinfo{author}{\bibfnamefont{H.~Q.} \bibnamefont{Song}},
  \bibinfo{author}{\bibfnamefont{M.}~\bibnamefont{Baldo}},
  \bibinfo{author}{\bibfnamefont{G.}~\bibnamefont{Giansiracusa}},
  \bibnamefont{and} \bibinfo{author}{\bibfnamefont{U.}~\bibnamefont{Lombardo}},
  \bibinfo{journal}{Phys. Rev. Lett.} \textbf{\bibinfo{volume}{81}},
  \bibinfo{pages}{1584} (\bibinfo{year}{1998}).

\bibitem[{\citenamefont{Beck et~al.}(1970)\citenamefont{Beck, Mang, and
  Ring}}]{Beck1970}
\bibinfo{author}{\bibfnamefont{R.}~\bibnamefont{Beck}},
  \bibinfo{author}{\bibfnamefont{H.~J.} \bibnamefont{Mang}}, \bibnamefont{and}
  \bibinfo{author}{\bibfnamefont{P.}~\bibnamefont{Ring}}, \bibinfo{journal}{Z.
  Phys.} \textbf{\bibinfo{volume}{231}}, \bibinfo{pages}{26}
  (\bibinfo{year}{1970}).

\bibitem[{\citenamefont{Ring and Schuck}(1980)}]{Ring1980}
\bibinfo{author}{\bibfnamefont{P.}~\bibnamefont{Ring}} \bibnamefont{and}
  \bibinfo{author}{\bibfnamefont{P.}~\bibnamefont{Schuck}},
  \emph{\bibinfo{title}{{The nuclear many-body problem}}}
  (\bibinfo{publisher}{Springer}, \bibinfo{year}{1980}).

\bibitem[{\citenamefont{Long et~al.}(2004)\citenamefont{Long, Meng, Giai, and
  Zhou}}]{Long2004}
\bibinfo{author}{\bibfnamefont{W.}~\bibnamefont{Long}},
  \bibinfo{author}{\bibfnamefont{J.}~\bibnamefont{Meng}},
  \bibinfo{author}{\bibfnamefont{N.}~\bibnamefont{Giai}}, \bibnamefont{and}
  \bibinfo{author}{\bibfnamefont{S.-g.} \bibnamefont{Zhou}},
  \bibinfo{journal}{Phys. Rev. C} \textbf{\bibinfo{volume}{69}},
  \bibinfo{pages}{034319} (\bibinfo{year}{2004}).

\bibitem[{\citenamefont{Zeh}(1965)}]{Zeh1965}
\bibinfo{author}{\bibfnamefont{H.~D.} \bibnamefont{Zeh}},
  \bibinfo{journal}{Zeitschrift f{\"{u}}r Phys.}
  \textbf{\bibinfo{volume}{188}}, \bibinfo{pages}{361} (\bibinfo{year}{1965}).

\bibitem[{\citenamefont{Chabanat et~al.}(1998)\citenamefont{Chabanat, Bonche,
  Haensel, Meyer, and Schaeffer}}]{Chabanat1998}
\bibinfo{author}{\bibfnamefont{E.}~\bibnamefont{Chabanat}},
  \bibinfo{author}{\bibfnamefont{P.}~\bibnamefont{Bonche}},
  \bibinfo{author}{\bibfnamefont{P.}~\bibnamefont{Haensel}},
  \bibinfo{author}{\bibfnamefont{J.}~\bibnamefont{Meyer}}, \bibnamefont{and}
  \bibinfo{author}{\bibfnamefont{R.}~\bibnamefont{Schaeffer}},
  \bibinfo{journal}{Nucl. Phys. A} \textbf{\bibinfo{volume}{635}},
  \bibinfo{pages}{231} (\bibinfo{year}{1998}).

\bibitem[{\citenamefont{Becker et~al.}(1974)\citenamefont{Becker, Davies, and
  Patterson}}]{Becker1974}
\bibinfo{author}{\bibfnamefont{R.}~\bibnamefont{Becker}},
  \bibinfo{author}{\bibfnamefont{K.}~\bibnamefont{Davies}}, \bibnamefont{and}
  \bibinfo{author}{\bibfnamefont{M.}~\bibnamefont{Patterson}},
  \bibinfo{journal}{Phys. Rev. C} \textbf{\bibinfo{volume}{9}},
  \bibinfo{pages}{1221} (\bibinfo{year}{1974}).

\bibitem[{\citenamefont{Tarbutton and Davies}(1968)}]{Tarbutton1968}
\bibinfo{author}{\bibfnamefont{R.}~\bibnamefont{Tarbutton}} \bibnamefont{and}
  \bibinfo{author}{\bibfnamefont{K.}~\bibnamefont{Davies}},
  \bibinfo{journal}{Nucl. Phys. A} \textbf{\bibinfo{volume}{120}},
  \bibinfo{pages}{1} (\bibinfo{year}{1968}).

\bibitem[{\citenamefont{Campi and Sprung}(1972)}]{Campi1972}
\bibinfo{author}{\bibfnamefont{X.}~\bibnamefont{Campi}} \bibnamefont{and}
  \bibinfo{author}{\bibfnamefont{D.}~\bibnamefont{Sprung}},
  \bibinfo{journal}{Nucl. Phys. A} \textbf{\bibinfo{volume}{194}},
  \bibinfo{pages}{401} (\bibinfo{year}{1972}).

\bibitem[{\citenamefont{Lalazissis et~al.}(2009)\citenamefont{Lalazissis,
  Karatzikos, Serra, Otsuka, and Ring}}]{Lalazissis2009}
\bibinfo{author}{\bibfnamefont{G.~A.} \bibnamefont{Lalazissis}},
  \bibinfo{author}{\bibfnamefont{S.}~\bibnamefont{Karatzikos}},
  \bibinfo{author}{\bibfnamefont{M.}~\bibnamefont{Serra}},
  \bibinfo{author}{\bibfnamefont{T.}~\bibnamefont{Otsuka}}, \bibnamefont{and}
  \bibinfo{author}{\bibfnamefont{P.}~\bibnamefont{Ring}},
  \bibinfo{journal}{Phys. Rev. C} \textbf{\bibinfo{volume}{80}},
  \bibinfo{pages}{041301} (\bibinfo{year}{2009}).

\bibitem[{\citenamefont{Haftel and Tabakin}(1970)}]{Haftel1970}
\bibinfo{author}{\bibfnamefont{M.~I.} \bibnamefont{Haftel}} \bibnamefont{and}
  \bibinfo{author}{\bibfnamefont{F.}~\bibnamefont{Tabakin}},
  \bibinfo{journal}{Nucl. Phys. A} \textbf{\bibinfo{volume}{158}},
  \bibinfo{pages}{1} (\bibinfo{year}{1970}).

\bibitem[{\citenamefont{Gambhir et~al.}(1990)\citenamefont{Gambhir, Ring, and
  Thimet}}]{Gambhir1990}
\bibinfo{author}{\bibfnamefont{Y.}~\bibnamefont{Gambhir}},
  \bibinfo{author}{\bibfnamefont{P.}~\bibnamefont{Ring}}, \bibnamefont{and}
  \bibinfo{author}{\bibfnamefont{A.}~\bibnamefont{Thimet}},
  \bibinfo{journal}{Ann. Phys.} \textbf{\bibinfo{volume}{198}},
  \bibinfo{pages}{132} (\bibinfo{year}{1990}).

\bibitem[{\citenamefont{Wang et~al.}(2012)\citenamefont{Wang, Audi, Wapstra,
  Kondev, MacCormick, Xu, and Pfeiffer}}]{Wang2012}
\bibinfo{author}{\bibfnamefont{M.}~\bibnamefont{Wang}},
  \bibinfo{author}{\bibfnamefont{G.}~\bibnamefont{Audi}},
  \bibinfo{author}{\bibfnamefont{A.}~\bibnamefont{Wapstra}},
  \bibinfo{author}{\bibfnamefont{F.}~\bibnamefont{Kondev}},
  \bibinfo{author}{\bibfnamefont{M.}~\bibnamefont{MacCormick}},
  \bibinfo{author}{\bibfnamefont{X.}~\bibnamefont{Xu}}, \bibnamefont{and}
  \bibinfo{author}{\bibfnamefont{B.}~\bibnamefont{Pfeiffer}},
  \bibinfo{journal}{Chinese Phys. C} \textbf{\bibinfo{volume}{36}},
  \bibinfo{pages}{1603} (\bibinfo{year}{2012}).

\bibitem[{\citenamefont{Angeli and Marinova}(2013)}]{Angeli2013}
\bibinfo{author}{\bibfnamefont{I.}~\bibnamefont{Angeli}} \bibnamefont{and}
  \bibinfo{author}{\bibfnamefont{K.}~\bibnamefont{Marinova}},
  \bibinfo{journal}{At. Data Nucl. Data Tables} \textbf{\bibinfo{volume}{99}},
  \bibinfo{pages}{69} (\bibinfo{year}{2013}).

\bibitem[{\citenamefont{Ozawa et~al.}(2001)\citenamefont{Ozawa, Bochkarev,
  Chulkov, Cortina, Geissel, Hellstr{\"{o}}m, Ivanov, Janik, Kimura, Kobayashi
  et~al.}}]{Ozawa2001}
\bibinfo{author}{\bibfnamefont{A.}~\bibnamefont{Ozawa}},
  \bibinfo{author}{\bibfnamefont{O.}~\bibnamefont{Bochkarev}},
  \bibinfo{author}{\bibfnamefont{L.}~\bibnamefont{Chulkov}},
  \bibinfo{author}{\bibfnamefont{D.}~\bibnamefont{Cortina}},
  \bibinfo{author}{\bibfnamefont{H.}~\bibnamefont{Geissel}},
  \bibinfo{author}{\bibfnamefont{M.}~\bibnamefont{Hellstr{\"{o}}m}},
  \bibinfo{author}{\bibfnamefont{M.}~\bibnamefont{Ivanov}},
  \bibinfo{author}{\bibfnamefont{R.}~\bibnamefont{Janik}},
  \bibinfo{author}{\bibfnamefont{K.}~\bibnamefont{Kimura}},
  \bibinfo{author}{\bibfnamefont{T.}~\bibnamefont{Kobayashi}},
  \bibnamefont{et~al.}, \bibinfo{journal}{Nucl. Phys. A}
  \textbf{\bibinfo{volume}{691}}, \bibinfo{pages}{599} (\bibinfo{year}{2001}).

\bibitem[{\citenamefont{Coraggio et~al.}(2003)\citenamefont{Coraggio, Itaco,
  Covello, Gargano, and Kuo}}]{Coraggio2003}
\bibinfo{author}{\bibfnamefont{L.}~\bibnamefont{Coraggio}},
  \bibinfo{author}{\bibfnamefont{N.}~\bibnamefont{Itaco}},
  \bibinfo{author}{\bibfnamefont{A.}~\bibnamefont{Covello}},
  \bibinfo{author}{\bibfnamefont{A.}~\bibnamefont{Gargano}}, \bibnamefont{and}
  \bibinfo{author}{\bibfnamefont{T.~T.~S.} \bibnamefont{Kuo}},
  \bibinfo{journal}{Phys. Rev. C} \textbf{\bibinfo{volume}{68}},
  \bibinfo{pages}{034320} (\bibinfo{year}{2003}).

\bibitem[{\citenamefont{Long et~al.}(2007)\citenamefont{Long, Sagawa, Giai, and
  Meng}}]{Long2007}
\bibinfo{author}{\bibfnamefont{W.~H.} \bibnamefont{Long}},
  \bibinfo{author}{\bibfnamefont{H.}~\bibnamefont{Sagawa}},
  \bibinfo{author}{\bibfnamefont{N.~V.} \bibnamefont{Giai}}, \bibnamefont{and}
  \bibinfo{author}{\bibfnamefont{J.}~\bibnamefont{Meng}},
  \bibinfo{journal}{Phys. Rev. C} \textbf{\bibinfo{volume}{76}},
  \bibinfo{pages}{034314} (\bibinfo{year}{2007}).

\bibitem[{\citenamefont{Hu et~al.}(2017)\citenamefont{Hu, Xu, Wu, Ma, and
  Sun}}]{Hu2017a}
\bibinfo{author}{\bibfnamefont{B.~S.} \bibnamefont{Hu}},
  \bibinfo{author}{\bibfnamefont{F.~R.} \bibnamefont{Xu}},
  \bibinfo{author}{\bibfnamefont{Q.}~\bibnamefont{Wu}},
  \bibinfo{author}{\bibfnamefont{Y.~Z.} \bibnamefont{Ma}}, \bibnamefont{and}
  \bibinfo{author}{\bibfnamefont{Z.~H.} \bibnamefont{Sun}},
  \bibinfo{journal}{Phys. Rev. C} \textbf{\bibinfo{volume}{95}},
  \bibinfo{pages}{034321} (\bibinfo{year}{2017}).

\bibitem[{\citenamefont{Hagen et~al.}(2009)\citenamefont{Hagen, Papenbrock,
  Dean, Hjorth-Jensen, and Asokan}}]{Hagen2009}
\bibinfo{author}{\bibfnamefont{G.}~\bibnamefont{Hagen}},
  \bibinfo{author}{\bibfnamefont{T.}~\bibnamefont{Papenbrock}},
  \bibinfo{author}{\bibfnamefont{D.~J.} \bibnamefont{Dean}},
  \bibinfo{author}{\bibfnamefont{M.}~\bibnamefont{Hjorth-Jensen}},
  \bibnamefont{and} \bibinfo{author}{\bibfnamefont{B.~V.}
  \bibnamefont{Asokan}}, \bibinfo{journal}{Phys. Rev. C}
  \textbf{\bibinfo{volume}{80}}, \bibinfo{pages}{021306}
  (\bibinfo{year}{2009}).

\bibitem[{\citenamefont{Roth et~al.}(2011)\citenamefont{Roth, Langhammer,
  Calci, Binder, and Navr{\'{a}}til}}]{Roth2011}
\bibinfo{author}{\bibfnamefont{R.}~\bibnamefont{Roth}},
  \bibinfo{author}{\bibfnamefont{J.}~\bibnamefont{Langhammer}},
  \bibinfo{author}{\bibfnamefont{A.}~\bibnamefont{Calci}},
  \bibinfo{author}{\bibfnamefont{S.}~\bibnamefont{Binder}}, \bibnamefont{and}
  \bibinfo{author}{\bibfnamefont{P.}~\bibnamefont{Navr{\'{a}}til}},
  \bibinfo{journal}{Phys. Rev. Lett.} \textbf{\bibinfo{volume}{107}},
  \bibinfo{pages}{072501} (\bibinfo{year}{2011}).

\bibitem[{\citenamefont{Entem and Machleidt}(2003)}]{Entem2003}
\bibinfo{author}{\bibfnamefont{D.~R.} \bibnamefont{Entem}} \bibnamefont{and}
  \bibinfo{author}{\bibfnamefont{R.}~\bibnamefont{Machleidt}},
  \bibinfo{journal}{Phys. Rev. C} \textbf{\bibinfo{volume}{68}},
  \bibinfo{pages}{041001} (\bibinfo{year}{2003}).

\bibitem[{\citenamefont{L{\"{a}}hde et~al.}(2014)\citenamefont{L{\"{a}}hde,
  Epelbaum, Krebs, Lee, Mei{\ss}ner, and Rupak}}]{Lahde2014}
\bibinfo{author}{\bibfnamefont{T.~A.} \bibnamefont{L{\"{a}}hde}},
  \bibinfo{author}{\bibfnamefont{E.}~\bibnamefont{Epelbaum}},
  \bibinfo{author}{\bibfnamefont{H.}~\bibnamefont{Krebs}},
  \bibinfo{author}{\bibfnamefont{D.}~\bibnamefont{Lee}},
  \bibinfo{author}{\bibfnamefont{U.-G.} \bibnamefont{Mei{\ss}ner}},
  \bibnamefont{and} \bibinfo{author}{\bibfnamefont{G.}~\bibnamefont{Rupak}},
  \bibinfo{journal}{Phys. Lett. B} \textbf{\bibinfo{volume}{732}},
  \bibinfo{pages}{110} (\bibinfo{year}{2014}).

\bibitem[{\citenamefont{Epelbaum
  et~al.}(2009{\natexlab{b}})\citenamefont{Epelbaum, Krebs, Lee, and
  Mei{\ss}ner}}]{Epelbaum2009a}
\bibinfo{author}{\bibfnamefont{E.}~\bibnamefont{Epelbaum}},
  \bibinfo{author}{\bibfnamefont{H.}~\bibnamefont{Krebs}},
  \bibinfo{author}{\bibfnamefont{D.}~\bibnamefont{Lee}}, \bibnamefont{and}
  \bibinfo{author}{\bibfnamefont{U.~G.} \bibnamefont{Mei{\ss}ner}},
  \bibinfo{journal}{Eur. Phys. J. A} \textbf{\bibinfo{volume}{41}},
  \bibinfo{pages}{125} (\bibinfo{year}{2009}{\natexlab{b}}).

\bibitem[{\citenamefont{Lapoux et~al.}(2016)\citenamefont{Lapoux, Som{\`{a}},
  Barbieri, Hergert, Holt, and Stroberg}}]{Lapoux2016}
\bibinfo{author}{\bibfnamefont{V.}~\bibnamefont{Lapoux}},
  \bibinfo{author}{\bibfnamefont{V.}~\bibnamefont{Som{\`{a}}}},
  \bibinfo{author}{\bibfnamefont{C.}~\bibnamefont{Barbieri}},
  \bibinfo{author}{\bibfnamefont{H.}~\bibnamefont{Hergert}},
  \bibinfo{author}{\bibfnamefont{J.~D.} \bibnamefont{Holt}}, \bibnamefont{and}
  \bibinfo{author}{\bibfnamefont{S.~R.} \bibnamefont{Stroberg}},
  \bibinfo{journal}{Phys. Rev. Lett.} \textbf{\bibinfo{volume}{117}},
  \bibinfo{pages}{052501} (\bibinfo{year}{2016}).

\bibitem[{\citenamefont{Ekstr\"om et~al.}(2015)\citenamefont{Ekstr\"om, Jansen,
  Wendt, Hagen, Papenbrock, Carlsson, Forss\'en, Hjorth-Jensen, Navr\'atil, and
  Nazarewicz}}]{Ekstrom2015}
\bibinfo{author}{\bibfnamefont{A.}~\bibnamefont{Ekstr\"om}},
  \bibinfo{author}{\bibfnamefont{G.~R.} \bibnamefont{Jansen}},
  \bibinfo{author}{\bibfnamefont{K.~A.} \bibnamefont{Wendt}},
  \bibinfo{author}{\bibfnamefont{G.}~\bibnamefont{Hagen}},
  \bibinfo{author}{\bibfnamefont{T.}~\bibnamefont{Papenbrock}},
  \bibinfo{author}{\bibfnamefont{B.~D.} \bibnamefont{Carlsson}},
  \bibinfo{author}{\bibfnamefont{C.}~\bibnamefont{Forss\'en}},
  \bibinfo{author}{\bibfnamefont{M.}~\bibnamefont{Hjorth-Jensen}},
  \bibinfo{author}{\bibfnamefont{P.}~\bibnamefont{Navr\'atil}},
  \bibnamefont{and}
  \bibinfo{author}{\bibfnamefont{W.}~\bibnamefont{Nazarewicz}},
  \bibinfo{journal}{Phys. Rev. C} \textbf{\bibinfo{volume}{91}},
  \bibinfo{pages}{051301} (\bibinfo{year}{2015}).

\bibitem[{\citenamefont{M{\"{u}}ther et~al.}(1990)\citenamefont{M{\"{u}}ther,
  Machleidt, and Brockmann}}]{Muether1990}
\bibinfo{author}{\bibfnamefont{H.}~\bibnamefont{M{\"{u}}ther}},
  \bibinfo{author}{\bibfnamefont{R.}~\bibnamefont{Machleidt}},
  \bibnamefont{and}
  \bibinfo{author}{\bibfnamefont{R.}~\bibnamefont{Brockmann}},
  \bibinfo{journal}{Phys. Rev. C} \textbf{\bibinfo{volume}{42}},
  \bibinfo{pages}{1981} (\bibinfo{year}{1990}).

\bibitem[{\citenamefont{{De Vries} et~al.}(1987)\citenamefont{{De Vries}, {De
  Jager}, and {De Vries}}}]{DeVries1987}
\bibinfo{author}{\bibfnamefont{H.}~\bibnamefont{{De Vries}}},
  \bibinfo{author}{\bibfnamefont{C.}~\bibnamefont{{De Jager}}},
  \bibnamefont{and} \bibinfo{author}{\bibfnamefont{C.}~\bibnamefont{{De
  Vries}}}, \bibinfo{journal}{At. Data Nucl. Data Tables}
  \textbf{\bibinfo{volume}{36}}, \bibinfo{pages}{495} (\bibinfo{year}{1987}).

\bibitem[{\citenamefont{Lu et~al.}(2013)\citenamefont{Lu, Mueller, Drake,
  N{\"{o}}rtersh{\"{a}}user, Pieper, and Yan}}]{Lu2013}
\bibinfo{author}{\bibfnamefont{Z.-T.} \bibnamefont{Lu}},
  \bibinfo{author}{\bibfnamefont{P.}~\bibnamefont{Mueller}},
  \bibinfo{author}{\bibfnamefont{G.~W.~F.} \bibnamefont{Drake}},
  \bibinfo{author}{\bibfnamefont{W.}~\bibnamefont{N{\"{o}}rtersh{\"{a}}user}},
  \bibinfo{author}{\bibfnamefont{S.~C.} \bibnamefont{Pieper}},
  \bibnamefont{and} \bibinfo{author}{\bibfnamefont{Z.-C.} \bibnamefont{Yan}},
  \bibinfo{journal}{Rev. Mod. Phys.} \textbf{\bibinfo{volume}{85}},
  \bibinfo{pages}{1383} (\bibinfo{year}{2013}).

\bibitem[{\citenamefont{Nogga et~al.}(2000)\citenamefont{Nogga, Kamada, and
  Gl{\"{o}}ckle}}]{Nogga2000}
\bibinfo{author}{\bibfnamefont{A.}~\bibnamefont{Nogga}},
  \bibinfo{author}{\bibfnamefont{H.}~\bibnamefont{Kamada}}, \bibnamefont{and}
  \bibinfo{author}{\bibfnamefont{W.}~\bibnamefont{Gl{\"{o}}ckle}},
  \bibinfo{journal}{Phys. Rev. Lett.} \textbf{\bibinfo{volume}{85}},
  \bibinfo{pages}{944} (\bibinfo{year}{2000}).

\bibitem[{\citenamefont{Binder et~al.}(2016)\citenamefont{Binder, Calci,
  Epelbaum, Furnstahl, Golak, Hebeler, Kamada, Krebs, Langhammer, Liebig
  et~al.}}]{Binder2016}
\bibinfo{author}{\bibfnamefont{S.}~\bibnamefont{Binder}},
  \bibinfo{author}{\bibfnamefont{A.}~\bibnamefont{Calci}},
  \bibinfo{author}{\bibfnamefont{E.}~\bibnamefont{Epelbaum}},
  \bibinfo{author}{\bibfnamefont{R.~J.} \bibnamefont{Furnstahl}},
  \bibinfo{author}{\bibfnamefont{J.}~\bibnamefont{Golak}},
  \bibinfo{author}{\bibfnamefont{K.}~\bibnamefont{Hebeler}},
  \bibinfo{author}{\bibfnamefont{H.}~\bibnamefont{Kamada}},
  \bibinfo{author}{\bibfnamefont{H.}~\bibnamefont{Krebs}},
  \bibinfo{author}{\bibfnamefont{J.}~\bibnamefont{Langhammer}},
  \bibinfo{author}{\bibfnamefont{S.}~\bibnamefont{Liebig}},
  \bibnamefont{et~al.}, \bibinfo{journal}{Phys. Rev. C}
  \textbf{\bibinfo{volume}{93}}, \bibinfo{pages}{044002}
  (\bibinfo{year}{2016}).

\bibitem[{\citenamefont{Navr{\'{a}}til}(2007)}]{Navratil2007a}
\bibinfo{author}{\bibfnamefont{P.}~\bibnamefont{Navr{\'{a}}til}},
  \bibinfo{journal}{Few-Body Syst.} \textbf{\bibinfo{volume}{41}},
  \bibinfo{pages}{117} (\bibinfo{year}{2007}).

\bibitem[{\citenamefont{Machleidt et~al.}(1996)\citenamefont{Machleidt,
  Sammarruca, and Song}}]{Machleidt1996}
\bibinfo{author}{\bibfnamefont{R.}~\bibnamefont{Machleidt}},
  \bibinfo{author}{\bibfnamefont{F.}~\bibnamefont{Sammarruca}},
  \bibnamefont{and} \bibinfo{author}{\bibfnamefont{Y.}~\bibnamefont{Song}},
  \bibinfo{journal}{Phys. Rev. C} \textbf{\bibinfo{volume}{53}},
  \bibinfo{pages}{R1483} (\bibinfo{year}{1996}).

\bibitem[{\citenamefont{Epelbaum et~al.}(2015)\citenamefont{Epelbaum, Krebs,
  and Mei{\ss}ner}}]{Epelbaum2015}
\bibinfo{author}{\bibfnamefont{E.}~\bibnamefont{Epelbaum}},
  \bibinfo{author}{\bibfnamefont{H.}~\bibnamefont{Krebs}}, \bibnamefont{and}
  \bibinfo{author}{\bibfnamefont{U.-G.} \bibnamefont{Mei{\ss}ner}},
  \bibinfo{journal}{Phys. Rev. Lett.} \textbf{\bibinfo{volume}{115}},
  \bibinfo{pages}{122301} (\bibinfo{year}{2015}).

\bibitem[{\citenamefont{Roth and Navr{\'{a}}til}(2007)}]{Roth2007}
\bibinfo{author}{\bibfnamefont{R.}~\bibnamefont{Roth}} \bibnamefont{and}
  \bibinfo{author}{\bibfnamefont{P.}~\bibnamefont{Navr{\'{a}}til}},
  \bibinfo{journal}{Phys. Rev. Lett.} \textbf{\bibinfo{volume}{99}},
  \bibinfo{pages}{092501} (\bibinfo{year}{2007}).

\bibitem[{\citenamefont{Hagen et~al.}(2007)\citenamefont{Hagen, Dean,
  Hjorth-Jensen, Papenbrock, and Schwenk}}]{Hagen2007}
\bibinfo{author}{\bibfnamefont{G.}~\bibnamefont{Hagen}},
  \bibinfo{author}{\bibfnamefont{D.~J.} \bibnamefont{Dean}},
  \bibinfo{author}{\bibfnamefont{M.}~\bibnamefont{Hjorth-Jensen}},
  \bibinfo{author}{\bibfnamefont{T.}~\bibnamefont{Papenbrock}},
  \bibnamefont{and} \bibinfo{author}{\bibfnamefont{A.}~\bibnamefont{Schwenk}},
  \bibinfo{journal}{Phys. Rev. C} \textbf{\bibinfo{volume}{76}},
  \bibinfo{pages}{044305} (\bibinfo{year}{2007}).

\bibitem[{\citenamefont{Hagen et~al.}(2010)\citenamefont{Hagen, Papenbrock,
  Dean, and Hjorth-Jensen}}]{Hagen2010}
\bibinfo{author}{\bibfnamefont{G.}~\bibnamefont{Hagen}},
  \bibinfo{author}{\bibfnamefont{T.}~\bibnamefont{Papenbrock}},
  \bibinfo{author}{\bibfnamefont{D.~J.} \bibnamefont{Dean}}, \bibnamefont{and}
  \bibinfo{author}{\bibfnamefont{M.}~\bibnamefont{Hjorth-Jensen}},
  \bibinfo{journal}{Phys. Rev. C} \textbf{\bibinfo{volume}{82}},
  \bibinfo{pages}{034330} (\bibinfo{year}{2010}).

\bibitem[{\citenamefont{Becker}(1970)}]{Becker1970}
\bibinfo{author}{\bibfnamefont{R.}~\bibnamefont{Becker}},
  \bibinfo{journal}{Phys. Rev. Lett.} \textbf{\bibinfo{volume}{24}},
  \bibinfo{pages}{400} (\bibinfo{year}{1970}).

\bibitem[{\citenamefont{Ren et~al.}(2016)\citenamefont{Ren, Li, Geng, Long,
  Ring, and Meng}}]{Ren2016}
\bibinfo{author}{\bibfnamefont{X.-L.} \bibnamefont{Ren}},
  \bibinfo{author}{\bibfnamefont{K.-W.} \bibnamefont{Li}},
  \bibinfo{author}{\bibfnamefont{L.-S.} \bibnamefont{Geng}},
  \bibinfo{author}{\bibfnamefont{B.}~\bibnamefont{Long}},
  \bibinfo{author}{\bibfnamefont{P.}~\bibnamefont{Ring}}, \bibnamefont{and}
  \bibinfo{author}{\bibfnamefont{J.}~\bibnamefont{Meng}},
  \bibinfo{journal}{arXiv}:\bibinfo{pages}{1611.08475}.

\bibitem[{\citenamefont{Li et~al.}(2016)\citenamefont{Li, Ren, Geng, and
  Long}}]{Li2016}
\bibinfo{author}{\bibfnamefont{K.-W.} \bibnamefont{Li}},
  \bibinfo{author}{\bibfnamefont{X.-L.} \bibnamefont{Ren}},
  \bibinfo{author}{\bibfnamefont{L.-S.} \bibnamefont{Geng}}, \bibnamefont{and}
  \bibinfo{author}{\bibfnamefont{B.} \bibnamefont{Long}},
  \bibinfo{journal}{arXiv}:\bibinfo{pages}{1612.08482}.

\bibitem[{\citenamefont{de~Shalit and Talmi}(1963)}]{Shalit1963}
\bibinfo{author}{\bibfnamefont{A.}~\bibnamefont{de~Shalit}} \bibnamefont{and}
  \bibinfo{author}{\bibfnamefont{I.}~\bibnamefont{Talmi}},
  \emph{\bibinfo{title}{Nuclear Shell Theory}} (\bibinfo{publisher}{Academic
  Press}, \bibinfo{address}{New York}, \bibinfo{year}{1963}).

\bibitem[{\citenamefont{Varshalovich et~al.}(1988)\citenamefont{Varshalovich,
  Moskalev, and Khersonskii}}]{Varshalovich1988}
\bibinfo{author}{\bibfnamefont{D.~A.} \bibnamefont{Varshalovich}},
  \bibinfo{author}{\bibfnamefont{A.~N.} \bibnamefont{Moskalev}},
  \bibnamefont{and} \bibinfo{author}{\bibfnamefont{V.~K.}
  \bibnamefont{Khersonskii}}, \emph{\bibinfo{title}{{Quantum Theory of Angular
  Momentum}}} (\bibinfo{publisher}{World Scientific}, \bibinfo{year}{1988}).

\bibitem[{\citenamefont{Gu et~al.}(2013)\citenamefont{Gu, Liang, Long, {Van
  Giai}, and Meng}}]{Gu2013}
\bibinfo{author}{\bibfnamefont{H.-Q.} \bibnamefont{Gu}},
  \bibinfo{author}{\bibfnamefont{H.}~\bibnamefont{Liang}},
  \bibinfo{author}{\bibfnamefont{W.~H.} \bibnamefont{Long}},
  \bibinfo{author}{\bibfnamefont{N.}~\bibnamefont{{Van Giai}}},
  \bibnamefont{and} \bibinfo{author}{\bibfnamefont{J.}~\bibnamefont{Meng}},
  \bibinfo{journal}{Phys. Rev. C} \textbf{\bibinfo{volume}{87}},
  \bibinfo{pages}{041301} (\bibinfo{year}{2013}).

\bibitem[{\citenamefont{Slater}(1951)}]{Slater1951}
\bibinfo{author}{\bibfnamefont{J.}~\bibnamefont{Slater}},
  \bibinfo{journal}{Phys. Rev.} \textbf{\bibinfo{volume}{81}},
  \bibinfo{pages}{385} (\bibinfo{year}{1951}).

\bibitem[{\citenamefont{Furnstahl et~al.}(2012)\citenamefont{Furnstahl, Hagen,
  and Papenbrock}}]{Furnstahl2012}
\bibinfo{author}{\bibfnamefont{R.~J.} \bibnamefont{Furnstahl}},
  \bibinfo{author}{\bibfnamefont{G.}~\bibnamefont{Hagen}}, \bibnamefont{and}
  \bibinfo{author}{\bibfnamefont{T.}~\bibnamefont{Papenbrock}},
  \bibinfo{journal}{Phys. Rev. C} \textbf{\bibinfo{volume}{86}},
  \bibinfo{pages}{031301} (\bibinfo{year}{2012}).

\bibitem[{\citenamefont{Efron}(1982)}]{Efron1982}
\bibinfo{author}{\bibfnamefont{B.}~\bibnamefont{Efron}},
  \emph{\bibinfo{title}{{The Jackknife, the Bootstrap and Other Resampling
  Plans}}} (\bibinfo{publisher}{Stanford University}, \bibinfo{year}{1982}).

\end{thebibliography}
\end{document}